\documentclass[a4paper,fleqn,usenatbib]{mnras}

\usepackage[T1]{fontenc}
\usepackage{ae,aecompl}

\usepackage{graphicx}	
\usepackage{amssymb}	
\usepackage{multirow,arydshln,tabu}   
\usepackage[table]{xcolor}

\title[Bulge-disk decomposition of GAMA galaxies]{Galaxy And Mass Assembly (GAMA): Bulge-disk decomposition of KiDS data in the nearby universe}

\author[Casura et al.]{
Sarah Casura,$^{1}$\thanks{E-mail: sarah.casura@uni-hamburg.de}
Jochen Liske,$^{1}$
Aaron S. G. Robotham,$^{2}$
Sarah Brough,$^{3}$
Simon P. Driver,$^{2}$
\newauthor
Alister W. Graham,$^{4}$
Boris H\"au\ss ler,$^{5}$
Benne W. Holwerda,$^{6}$
Andrew M. Hopkins,$^{7}$
Lee S. Kelvin,$^{8}$
\newauthor
Amanda J. Moffett,$^{9}$
Dan S. Taranu,$^{8,2}$
Edward N. Taylor$^{4}$
\\
$^{1}$Hamburger Sternwarte, Universit\"at Hamburg, Gojenbergsweg 112, 21029 Hamburg, Germany\\
$^{2}$ ICRAR, M468, University of Western Australia, Crawley, WA 6009, Australia \\
$^{3}$School of Physics, University of New South Wales, NSW 2052, Australia\\
$^{4}$Centre for Astrophysics and Supercomputing, Swinburne University of Technology, Hawthorn, VIC 3122, Australia \\
$^{5}$European Southern Observatory, Alonso de Cordova 3107, Casilla 19001, Santiago, Chile\\
$^{6}$Department of Physics and Astronomy, 102 Natural Science Building, University of Louisville, Louisville KY 40292, USA \\
$^{7}$Australian Astronomical Optics, Macquarie University, 105 Delhi Rd, North Ryde, NSW 2113, Australia \\ 
$^{8}$Department of Astrophysical Sciences, Princeton University, 4 Ivy Lane, Princeton, NJ 08544, USA\\
$^{9}$Department of Physics and Astronomy, University of North Georgia, 3820 Mundy Mill Rd., Oakwood GA 30566, USA 
}

\date{Accepted XXX. Received YYY; in original form ZZZ}

\pubyear{2020}

\begin{document}
\label{firstpage}
\pagerange{\pageref{firstpage}--\pageref{lastpage}}
\maketitle

\begin{abstract}

We derive single S\'ersic fits and bulge-disk decompositions for 13096 galaxies at redshifts $z<0.08$ in the GAMA II equatorial survey regions in the Kilo-Degree Survey (KiDS) $g$, $r$ and $i$ bands. The surface brightness fitting is performed using the Bayesian two-dimensional profile fitting code \texttt{ProFit}. We fit three models to each galaxy in each band independently with a fully automated Markov-chain Monte Carlo analysis: a single S\'ersic model, a S\'ersic plus exponential and a point source plus exponential. After fitting the galaxies, we perform model selection and flag galaxies for which none of our models are appropriate (mainly mergers/Irregular galaxies). The fit quality is assessed by visual inspections, comparison to previous works, comparison of independent fits of galaxies in the overlap regions between KiDS tiles and bespoke simulations. The latter two are also used for a detailed investigation of systematic error sources. We find that our fit results are robust across various galaxy types and image qualities with minimal biases. Errors given by the MCMC underestimate the true errors typically by factors 2-3. Automated model selection criteria are accurate to $>90\%$ as calibrated by visual inspection of a subsample of galaxies. We also present $g-r$ component colours and the corresponding colour-magnitude diagram, consistent with previous works despite our increased fit flexibility. Such reliable structural parameters for the components of a diverse sample of galaxies across multiple bands will be integral to various studies of galaxy properties and evolution. All results are integrated into the GAMA database.
\end{abstract}

\begin{keywords}
galaxies: structure -- galaxies: fundamental parameters -- methods: statistical -- catalogues -- galaxies: photometry
\end{keywords}



\section{Introduction}

The quantitative modelling of galaxy surface brightness distributions has a long history dating back to \citet{deVaucouleurs1948}, \citet{Sersic1963} and even earlier works; see \citet{Graham2013b} for a review of the development of light profile models. While the early works focussed on azimuthally averaged galaxy profiling with just a single functional form \citep[e.g.][]{Kormendy1977}, modern codes allow users to decompose galaxies into several distinct components and to take into account the full two-dimensional information. To this end, there are many different techniques, methods and code packages, all of which have become increasingly sophisticated as the quality and quantity of available astronomical data have grown. Broadly, they can be divided into parametric and non-parametric modelling as well as one-dimensional and two-dimensional methods. Which of these is most appropriate to use depends on the science case and the available data. This work falls into the regime of large-scale automated analyses of galaxies with often barely resolved components, for which we want to obtain structural parameters that are easily comparable between galaxies. Hence, two-dimensional parametric analysis is most appropriate (see also the discussion in \citealt{Robotham2017} and references therein). 
Examples of such two-dimensional, parametric fitting tools used for large-scale automated analyses include \texttt{GIM2D} \citep{Simard2002}, \texttt{BUDDA} \citep{deSouza2004}, \texttt{GALFIT3} \citep{Peng2010}, \texttt{GALFITM} \citep{Vika2013}, \texttt{IMFIT} \citep{Erwin2015}, \texttt{ProFit} \citep{Robotham2017} and \texttt{PHI} \citep{Argyle2018}. 
Each of these tools comes with its own advantages and disadvantages, which goes to show how difficult the problem of galaxy modelling is, especially when automated for large samples of a very diverse galaxy population. Usually, some form of post-processing is needed to assess the influence of systematic errors, judge the convergence, exclude bad fits and identify the most appropriate model to use for each galaxy. This can be achieved via visual inspection (for small enough samples), logical filters, frequentist statistics such as the $F$-test, Bayesian inference or similar methods \citep[see, e.g.][]{Allen2006, Gadotti2009, Simard2011, Vika2014, Meert2015, Lange2016, Mendez-Abreu2017}. 

Despite the associated difficulties (e.g. convergence and quality of fit metrics), many authors have performed two-dimensional surface brightness profile fitting for large numbers of galaxies, modelling the radial light profile as a simple functional form, most often a S\'ersic function \citep[][to name just a few]{Blanton2003, Blanton2005, Barden2005, Trujillo2006, Hyde2009, LaBarbera2010, Kelvin2012, vanderWel2012, Haeussler2013, Shibuya2015, Sanchez-Janssen2016}. 
The results of such analyses have been used to derive a number of key relations between different galaxy properties, their formation and evolutionary history, and interactions with the environment. 
For example, many works have studied the distribution of, and relation between, size and mass or luminosity for different galaxy types (split by e.g. S\'ersic index or colour), sometimes including morphology, surface brightness, internal velocity, environment, wavelength, colour, or redshift effects \citep[e.g.][]{Shen2003, Barden2005, Blanton2005, Trujillo2006, Hyde2009, LaBarbera2010, Kelvin2014, vanderWel2014, Lange2015, Shibuya2015, Kawinwanichakij2021, Nedkova2021}.\\ 

With improving data quality of surveys, the galaxy fitting community has increasingly shifted towards fitting more than one component, i.e. to perform bulge-disk decomposition. While some authors, such as \citet{Gadotti2009}, \citet{Salo2015} or \citet{Gao2017} also account for bars, central point sources, spiral arms or other additional morphological features, most works focus on the bulge and disk. The focus on only two components is especially true when running automated analyses of large samples, since in many cases the data quality is not sufficient to meaningfully constrain more than one or two components, or it would require extensive manual tuning based on visual inspection. From a more physical point of view, the majority of the stellar mass in the local universe resides in ellipticals, disks and classical bulges, with pseudo-bulges and bars only contributing a few percent \citep{Gadotti2009}. Hence, for automated analyses it is common practice to fit only two components, where the term ``bulge" is used to describe the central component, irrespective of whether it is a classical bulge, pseudo-bulge, bar, lens, active galactic nucleus (AGN), or a mixture thereof, while ``disk" refers to a more extended component with typically lower surface brightness and potential additional structure such as spiral arms, breaks, flares or rings. 
 
Examples of large bulge-disk decomposition studies include \citet{Simard2002, Simard2011, Allen2006, Benson2007, Gadotti2009, Lackner2012, Fernandez-Lorenzo2014, Head2014, Mendel2014, Vika2014, Meert2015, Meert2016, Kennedy2016, Kim2016, Lange2016, Dimauro2018, Bottrell2019, Cook2019, Barsanti2021}; and \citet{Haeussler2022}. 
Such catalogues can then be used to determine the relative numbers of different galaxy components as well as their luminosity or stellar mass functions, size-mass or size-luminosity relations, including their redshift evolution and dependence on other properties of the galaxy and its environment (similar to the studies of entire galaxies mentioned earlier). This has been done by \citet{Driver2007b, Dutton2011, Tasca2014, Kennedy2016, Lange2016, Moffett2016, Dimauro2019}, and many others.

In addition, quantitative measures for the components of galaxies aid the comparison of observational data to theory and simulations. Bulges and disks are often decisively different not only in their visual appearance but also in their structure, dynamics, stellar populations, gas and dust content and are thought to have different formation pathways \citep{Cole2000, Cook2009, Driver2013, Lange2016, Dimauro2018, Lagos2018, Oh2020}. Consequently, bulge-disk decomposition studies provide stringent constraints on the formation and evolutionary histories of galaxies and their physical properties that are not easily measured directly such as the dark matter halo, the build-up of stellar mass (in different components) over time, or merger histories 
\citep[examples include][]{Driver2013, Bottrell2017b, Bluck2019, Rodriguez-Gomez2019, deGraaff2022}. 
Hence, consistently measuring the structure of the stellar components is essential to make full use of current and future large-scale observational surveys such as the Kilo-Degree Survey \citep[KiDS;][]{deJong2013} or the Legacy Survey of Space and Time \citep[LSST;][]{Ivezic2019}, 
and of cosmological hydrodynamical simulations such as Illustris \citep{Vogelsberger2014} and IllustrisTNG \citep[The Next Generation;][]{Pillepich2018} or Evolution and Assembly of GaLaxies and their Environments \citep[EAGLE;][]{Schaye2015}.\\ 

In the present study, we obtain single S\'ersic fits and bulge-disk decompositions for 13096 GAMA galaxies in the KiDS $g$, $r$ and $i$ bands.
We choose \texttt{ProFit} as our modelling software due to its Bayesian nature (allowing full MCMC treatment including more realistic error estimates), its suitability to large-scale automated analyses and its ability, in combination with \texttt{ProFound}, to serve as a fully self-contained package covering all steps of the analysis from image segmentation through to model fitting. We supplement this functionality with our own routines for the rejection of unsuitable fits, model selection, and a characterisation of systematic uncertainties. The resulting catalogue has already been used to aid the kinematic bulge-disk decomposition of a sample of galaxies in the Sydney-AAO Multi-object Integral-field spectroscopy (SAMI) Galaxy Survey \citep{Oh2020} and to examine the properties of galaxy groups \citep{Cluver2020}, with many more studies in progress. 

Our own plans include deriving the stellar mass functions of bulges and disks, studying component colours and trends of other S\'ersic parameters with wavelength, and constraining the nature and distribution of dust in galaxy disks. The latter can be achieved by comparing the distribution of bulges and disks in the luminosity-size plane to dust radiative transfer models such as those presented in \citet{Popescu2011} and preceding papers of this series (similar to the analysis performed by \citealt{Driver2007} albeit with more and better data and at several wavelengths). For these science aims, we are most interested in obtaining structural parameters that are directly comparable amongst each other, i.e. consistent within the dataset; and correctly represent the statistical properties of the entire sample, with less emphasis placed on capturing all aspects of the detailed structure of individual galaxies. Correspondingly, we choose to model a maximum of two components for each galaxy and use the terms ``bulge" and ``disk" in their widest senses, in line with previous automated decompositions of large samples. In particular the ``bulges" we obtain are often mixtures of classical or pseudo-bulges, bars, lenses and AGN. Similarly, we place more emphasis on the central, high surface brightness regions of galaxies by modelling only a relatively tight region around each galaxy of interest. While most of the fits we obtain are not perfect (because galaxies are more complex than two simple components), they do achieve the above named aims and are comparable to similar studies. \\

In Section~\ref{sec:datasamplecode} we describe our data (GAMA and KiDS), sample selection, and code (\texttt{ProFit} and \texttt{ProFound}), and discuss in detail the distinguishing features of this study compared to previous work. Section~\ref{sec:pipeline} then presents the pipeline we developed for the bulge-disk decomposition, including preparatory work and post-processing, before we show our main results in Section~\ref{sec:results}. Sections~\ref{sec:prevwork} and \ref{sec:simulations} focus on the quality control of the fits by comparison to previous work and a detailed investigation into systematic uncertainties and biases from simulations and the overlap sample. We conclude with a summary and information on catalogue access in Section~\ref{sec:conclusions}. We assume a standard cosmology of $H_0 = 70$\,km\,s$^{-1}$\,Mpc$^{-1}$, $\Omega_m=0.3$ and $\Omega_{\lambda}=0.7$ throughout.

\section{Data, sample and code}
\label{sec:datasamplecode}

\subsection{GAMA}
The Galaxy and Mass Assembly (GAMA)\footnote{\label{foot:gama}\url{http://www.gama-survey.org}} survey is a large low-redshift spectroscopic survey covering $\sim$\,238\,000 galaxies in 286\,deg$^2$ of sky (split into 5 survey regions) out to a redshift of approximately 0.6 and a depth of $r$<19.8 mag. The observations were taken using the AAOmega spectrograph on the Anglo-Australian Telescope and were completed in 2014. The survey strategy and spectroscopic data reduction are described in detail in \citet{Driver2009, Baldry2010, Robotham2010, Driver2011, Hopkins2013, Baldry2014} and \citet{Liske2015}. 

In addition to the spectroscopic data, the GAMA team collected imaging data on the same galaxies from a number of independent surveys in more than 20 bands with wavelengths between 1\,nm and 1\,m. Details of the imaging surveys and the photometric data reduction are given in \citet{Liske2015, Driver2016}; \citet{Driver2022} and references therein. The combined spectroscopic and multiwavelength photometric data at this depth, resolution and completeness provide a unique opportunity to study a variety of properties of the low-redshift galaxy population.

In this work, we focus on the KiDS $g$-band, $r$-band and $i$-band imaging data (see Section~\ref{sec:kids}) in the GAMA II equatorial survey regions, which are 3 regions of size $12\degr\times5\degr$ located along the equator at 9, 12 and approximately 14.5 hours in right ascencion (the G09, G12 and G15 regions). For our sample selection, we make use of the equatorial input catalogue\footnote{For the sake of reproducibility, we always give the exact designation of a catalogue on the GAMA database in parentheses: the data management unit (DMU) that produced the catalogue (e.g. \texttt{EqInputCat}) followed by the catalogue name (e.g. \texttt{TilingCat}) and the version used (e.g. \texttt{v46}).} \citep[\texttt{EqInputCat:TilingCatv46},][]{Baldry2010} and the most recent version of the redshifts originally described by \citet{Baldry2012} (\texttt{LocalFlowCorrection:DistancesFramesv14}), see details in Section~\ref{sec:sampleselection}. For the stellar mass-size relation (Section~\ref{sec:sizemass}), we also use the Data Release (DR) 3 version of the stellar mass catalogue first presented in \citet{Taylor2011} (\texttt{StellarMasses:StellarMassesv19}); for the comparison to previous work (Section~\ref{sec:comparelee}) we use the single S\'ersic fits of \citet{Kelvin2012} (\texttt{SersicPhotometry:SersicCatSDSSv09}); and in order to correct galaxy colours for Galactic extinction, we use the corresponding table provided along with the equatorial input catalogue (\texttt{EqInputCat:GalacticExtinctionv03}). All of these catalogues can be obtained from the GAMA database$^{\ref{foot:gama}}$. 

\subsection{KiDS}
\label{sec:kids}
The Kilo-Degree Survey \citep[KiDS,][]{deJong2013} is a wide-field imaging survey in the Southern sky using the VLT Survey Telescope (VST) at the ESO Paranal Observatory. 1350\,deg$^2$ are mapped in the optical broad-band filters $u, g, r, i$; while the VISTA Kilo-degree INfrared Galaxy (VIKING) Survey \citep{Edge2013} provides the corresponding near-infrared data in the $Z, Y, J, H, K_s$ bands. The GAMA II equatorial survey regions have been covered as of DR3.0.  

KiDS provides $\sim$\,1\degr$\times$1\degr\ science tiles calibrated to absolute values of flux with associated weight maps (inverse variance) and binary masks. The science tiles are composed of 5 dithers (4 in $u$) totalling 1000, 900, 1800 and 1200\,s exposure time in $u,g,r,i$, with all dithers aligned in the right ascenscion and declination axes (i.e. no rotational dithers). The $r$-band observations are performed during the best seeing conditions in dark time; while $g$, $u$ and $i$ have progressively worse seeing and $i$ is additionally taken during grey time or bright moon. During co-addition, the dithers across all 4 bands are re-gridded onto a common pixel scale of $0\farcs$2. The magnitude zeropoint of the science tiles is close to zero with small corrections given in the image headers. The $r$-band point spread function (PSF) size is typically $0\farcs$7 and the limiting magnitudes in $u,g,r,i$ are $\sim$\,24.2, 25.1, 25.0, 23.7\,mag respectively (5$\sigma$ in a 2\arcsec aperture). This high image quality, depth, survey size and wide wavelength coverage in combination with VIKING make KiDS data unique. For details, see \citet{Kuijken2019}. 

For this work, we use the $g$-band, $r$-band and $i$-band science tiles, weight maps and masks from KiDS DR4.0 \citep{Kuijken2019}, which are publicly available\footnote{\url{http://kids.strw.leidenuniv.nl/DR4/index.php}} for our selected sample of galaxies (Section~\ref{sec:sampleselection}). We plan to extend the analysis to include the KiDS $u$ and the VIKING $Z, Y, J, H,$ and $K_s$ bands in a future work.

\subsection{Sample selection}
\label{sec:sampleselection}
Our main sample consists of all GAMA II equatorial region main survey targets with a reliable redshift in the range $0.005<z<0.08$, which are a total of 12958 objects.\footnote{In detail, we select all targets with NQ$\geq$3, SURVEY\textunderscore CLASS$\geq$4 and 0.005$<$Z\textunderscore CMB$<$0.08 from \texttt{EqInputCat:TilingCatv46} joined to \texttt{LocalFlowCorrection:DistancesFramesv14} on CATAID.} In addition, we include all 2404 targets of the ``GAMA sample" of the SAMI Galaxy Survey\footnote{Taking the CATIDs listed in sami\textunderscore sel\textunderscore 20140413\textunderscore v1.9\textunderscore publiclist from \url{https://sami-survey.org/data/target_catalogue}} \citep{Bryant2015}, the majority of which are already in our main sample. The combination of both results in the full sample of 13096 unique physical objects, which were imaged a total of 14966 times in each of the KiDS $g$, $r$ and $i$ bands due to small overlap regions between the tiles. 11301, 1742, 31 and 22 objects were imaged once, twice, three and four times respectively. We keep these multiple data matches to the same physical object separate during all processing steps to serve as an internal consistency check. 

\subsection{ProFit}
\label{sec:profit}
We fit the surface brightness distributions of our sample of galaxies using \texttt{ProFit}\footnote{https://github.com/ICRAR/ProFit} (v1.3.2) which is a free and open-source, fully Bayesian two-dimensional profile fitting code \citep{Robotham2017}. \texttt{ProFit} offers great flexibility: there are several built-in profiles to choose from, it is easy to add several components of the same or different profiles, there is a choice of likelihood calculations and optimisation algorithms that can be used (various downhill gradient options, genetic algorithms, over 60 variants of Markov Chain Monte Carlo methods, MCMC), parameters can be fitted in linear or logarithmic space, it is possible to add complex priors for each, as well as constraints relating several parameters; and much more. The pixel integrations are performed using a standalone C++ library (\texttt{libprofit}), making it both faster and more accurate than other commonly used algorithms such as \texttt{GALFIT} (\citeauthor{Peng2010} \citeyear{Peng2010}; see detailed comparison in \citeauthor{Robotham2017} \citeyear{Robotham2017}). This allows us to fit galaxies with the computationally more expensive MCMC algorithms, overcoming the main problems of downhill gradient based optimisers: their susceptibility to initial guesses and their inability to easily derive realistic error estimates \citep[e.g.][]{Lange2016}. This makes \texttt{ProFit} highly suitable for the decomposition of large sets of galaxies with little user intervention.

\subsection{ProFound}
\label{sec:profound} 
\texttt{ProFit} (Section~\ref{sec:profit}) requires a number of inputs apart from the (sky-subtracted) science image and the chosen model to fit, most importantly initial parameter guesses, a segmentation map specifying which pixels to fit, a sigma (error) map and a PSF image. To provide these inputs in a robust and consistent manner, the sister package \texttt{ProFound}\footnote{https://github.com/asgr/ProFound/} \citep{Robotham2018} was developed, which also serves as a stand-alone source finding and image analysis tool. The main novelties of \texttt{ProFound} compared to other commonly used free and open-source packages such as \texttt{Source Extractor} \citep{Bertin1996} are that, rather than elliptical apertures, \texttt{ProFound} uses dilated ``segments" (collections of pixels of arbitrary shape) with watershed de-blending across saddle-points in flux. This means that the flux from each pixel is attributed to exactly one source (or the background) and apertures are never overlapping or nested. It also allows for extracting more complex object shapes than ellipses while still capturing the total flux due to the segment dilation (expansion) process. This makes it less prone to catastrophic segmentation failures (such as fragmentation of bright sources or blending of several sources into one aperture), reducing the need for manual intervention or multiple runs with ``hot" and ``cold" deblending settings, hence making \texttt{ProFound} particularly suitable for large-scale automated analysis of deep extragalactic surveys \citep{Robotham2018, Davies2018, Bellstedt2020}.

Apart from the segmentation map, the main function of the package, \texttt{profoundProFound}, also returns estimated sky and sky-RMS maps (if not given as inputs) and a wealth of ancillary data including a list of segments and their properties such as their size, ellipticity and the flux contained. The latter is particularly useful to obtain reasonable initial parameter guesses for galaxy fitting; or for identifying certain types of sources (e.g. stars for PSF estimation). The package also contains many additional functions for further image analysis and processing, all within the same framework. In addition, combining \texttt{ProFound} with \texttt{ProFit} allows the user to estimate a PSF (see Section~\ref{sec:psfestimation}), hence entirely removing any dependence on external tools. Finally, both packages come with comprehensive documentation and many extended examples and vignettes which serve as great resources for newcomers to the fields of source extraction and galaxy fitting. 

We use \texttt{ProFound} (v1.9.2) along with \texttt{ProFit} (v1.3.2) for all preparatory steps (image segmentation/source identification, sky subtraction, initial parameter estimates and PSF determination; see Section~\ref{sec:preparatorysteps} for details) producing the inputs needed for the galaxy fitting with \texttt{ProFit}.

\section{Bulge-disk decomposition pipeline}
\label{sec:pipeline}

We use the free and open-source \texttt{R} programming language \citep{Rv3.6} for all scripting.

\subsection{Preparatory steps} 
\label{sec:preparatorysteps}
\subsubsection{Cutouts and masking}
\label{sec:cutoutsandmasking}
KiDS imaging tiles are registered to the same pixel grid across all four bands (with matching weight maps and masks), such that a joint analysis of the bands is straightforward. They are also aligned such that the $x$-axis corresponds to right ascencion (RA) and the $y$-axis to declination (Dec). Hence, we obtain a $400\arcsec\times 400\arcsec$ cutout of the KiDS tile, associated weight map and mask for each object in our sample and for each of the three KiDS bands we used ($g,r,i$). The masks of all 3 bands are then combined and all pixels which have a value greater than 0 in any of the masks are excluded from analysis. This results in approximately 20\% of all pixels being masked out. This large fraction of masking is primarily due to the reflection halos of bright stars that are also clearly visible in the data (see \citealt{deJong2015} for details). We combine the masks in this way to ensure that the pixels used for analysis are exactly the same in all bands and so the results are most directly comparable between bands. Objects for which the central pixel is masked ($\sim20\%$ of all galaxies) are skipped in the galaxy fitting.
\subsubsection{Image segmentation}
\label{sec:imagesegmentation}
We perform image segmentation in order to determine which pixels to fit for each of our objects, identify other nearby sources, improve the background subtraction and obtain reasonable initial guesses for the galaxy parameters. This is performed on the joint $g, r, i$ cutouts with \texttt{ProFound} in several steps.  

First, we add the cutouts in the $g$, $r$, and $i$ bands using inverse variance weighting and compute the joint weight map. We then estimate the (joint $gri$) sky by running the stacked image through \texttt{profoundProFound} passing in the correct magnitude zeropoint, mask and weight, but leaving \texttt{skycut} on its default of 1. This means that all pixels with a flux at least $1\sigma$ above the median are progressively assigned to segments (collections of pixels belonging to an object) using an iterative process: starting with the brightest pixel in the image, segments are grown by adding neighbouring pixels with lower flux; new segments are started when a pixel shows more flux than its neighbours (within some tolerance) or when all neighbouring pixels above the \texttt{skycut} value have been assigned. Once all pixels above \texttt{skycut} have been assigned, the resulting segments are additionally expanded until flux convergence is reached. For more details, see \citet{Robotham2018}. 

\begin{figure}
	\includegraphics[width=\columnwidth]{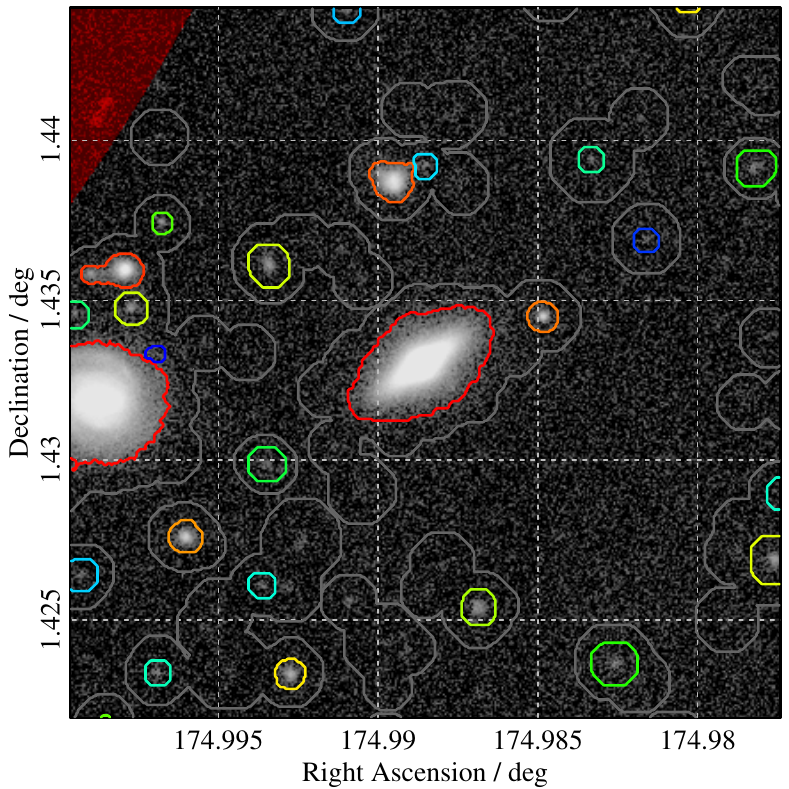}
    \caption{The \texttt{ProFound} segmentation map obtained for the galaxy 396740 overlaid on the KiDS $r$-band image. Note this is only a cutout of the full segmentation map showing the central $100\arcsec\times 100\arcsec$. Identified objects (segments) are shown with contours, coloured from red to blue according to the flux contained. Grey contours indicate the more dilated segmentation map used for the background subtraction. Masked areas are shaded red.} 
    \label{fig:exampleseg}
\end{figure}

Along the way, \texttt{ProFound} estimates the sky background several times since object detection relies on accurate background subtraction and vice versa.\footnote{The sky variance can also be estimated, but in our case this is already provided as the KiDS weight maps and given to \texttt{ProFound} as an input.} For the final sky estimate, the already-dilated segments are expanded even further to ensure that no object flux will bias the background determination. This very aggressive object mask is indicated with grey contours in Figure \ref{fig:exampleseg}. We use it for the joint-$gri$ sky estimate here and also for the band-specific background determination detailed in Section \ref{sec:backgroundsubtraction} (performed in the same way).  

For the galaxy fitting, however, we decided to use tighter segments that do not push that deeply into the sky. Besides speeding up the fit, this naturally results in the best possible fit to the inner, high signal-to-noise regions of the galaxy that we are most interested in and reduces the sensitivity to background subtraction problems, flux from the wings of other objects and features that cannot be captured by our models such as disk breaks and flares, and edge-on disks requiring the inclined disk model \citep{vanderKruit1981}. Note, however, that this choice comes with some trade-offs, most notably that the fit frequently overpredicts the flux outside the segment boundary. We address this in more detail in Section~\ref{sec:tightsegments}. 

To obtain these tighter segmentation maps, we run \texttt{profoundProFound} again with the sky now fixed and a higher \texttt{skycut} value of 2. This means that only pixels with a flux at least $2\sigma$ above the background level are considered in the segmentation, which ensures that fewer noise fluctuations are ``detected" and segment borders are smooth. In order to capture all flux of the galaxy wings, the segment for the object of interest (only) is then expanded further (using \texttt{profoundMakeSegimDilate}) such that its area increases by typically around 30\%. This last step also ensures unbiased smooth borders of the segment since it is entirely independent of noise fluctuations. The resulting segmentation map is indicated with coloured contours in Figure \ref{fig:exampleseg} and is used for galaxy fitting in all bands, so that exactly the same pixels are fitted in each band (the segmentation statistics are of course re-calculated in each band). 
\subsubsection{Background subtraction}
\label{sec:backgroundsubtraction}
KiDS tiles are background-subtracted already, however we opt to use the sky estimated by \texttt{ProFound} to even out inhomogeneities on smaller scales. For this, we split our $400\arcsec\times 400\arcsec$ cutout into 16 square boxes and mask out all objects using the aggressively dilated segmentation map indicated with grey contours in Figure \ref{fig:exampleseg} (cf. Section~\ref{sec:imagesegmentation}). The sky is then estimated as the median of the remaining (background) pixels in each box independently; and the solutions between the boxes interpolated with a bicubic spline.\footnote{This is done by \texttt{profoundProFound} internally; with the box size and the order of the interpolation spline being some of the variables we set.} This is done for each band independently, however the segmentation maps used to mask out objects are the same in all bands.

This procedure for the background subtraction was chosen after extensive testing during pipeline development. In short, we found that the \texttt{ProFound} sky adopted here does not subtract object wings while still homogenising the background well enough to avoid having to fit it along with the object of interest (introducing possible parameter degeneracies). It also decreases the sensitivity of the fit to the chosen segment size. 
\subsubsection{Sigma maps}
Once the image segmentation and background subtraction is completed, we also calculate the sigma (error) map for each cutout (independently in each band). This is a combination of the KiDS weight map (where $\sigma=1/\sqrt{weight}$) and the object shot noise. The latter is estimated as $\sqrt{N}$, where $N$ is the number of photons per pixel (using positive-valued pixels only). This, in turn, is obtained by converting the image into counts using the gain provided in the meta-data associated with each KiDS tile.
\subsubsection{PSF estimation}
\label{sec:psfestimation}
PSF fitting is performed on the background-subtracted $400\arcsec\times400\arcsec$ cutouts with corresponding masks and sigma maps in each band. The segmentation statistics returned by \texttt{ProFound} are used to identify isolated stars (round, bright, small and highly concentrated objects with few nearby segments). More details on the star candidate selection are given in Appendix~\ref{app:psfdetails}. These objects are then fitted with a Moffat function using \texttt{ProFit}; fitting all parameters except boxyness, i.e. the position, magnitude, full width at half maximum (FWHM), concentration index, axial ratio and position angle. Scale parameters are fitted in logarithmic space, a Normal likelihood function is used, initial guesses are taken from the segmentation statistics and we use the \texttt{BFGS} algorithm from \texttt{optim} \citep{Rv3.6}, which is a fast downhill gradient optimisation using a quasi-Newton method published simultaneously by \citet{Broyden1970, Fletcher1970, Goldfarb1970, Shanno1970}. 

Some of the objects fitted above may not actually be suitable for PSF estimation as they can be too faint or bright (close to saturation), have irregular features, bad pixels or additional small objects included in the fitting segment. Unsuitable objects are excluded by a combination of hard cuts in reduced chi-square ($\chi^2_\nu$), position and magnitude relative to the \texttt{ProFound} estimates and an iterative $2\sigma$-clipping in FWHM, concentration index, angle and axial ratio. Again, more details can be found in Appendix~\ref{app:psfdetails}. Finally, we take the median of the Moffat parameters of a maximum of 8 suitable stars (the closest 2 from each quadrant where possible to ensure an even distribution around the position of interest) and use these Moffat parameters to create a model PSF image. The size of the PSF image is adjusted to include at least 99\% of the total flux; or to a maximum of the median segment size within which the stars were fitted, with pixels in the corners of the image set to zero to avoid having a rectangular PSF. 

\begin{figure}
	\includegraphics[width=\columnwidth]{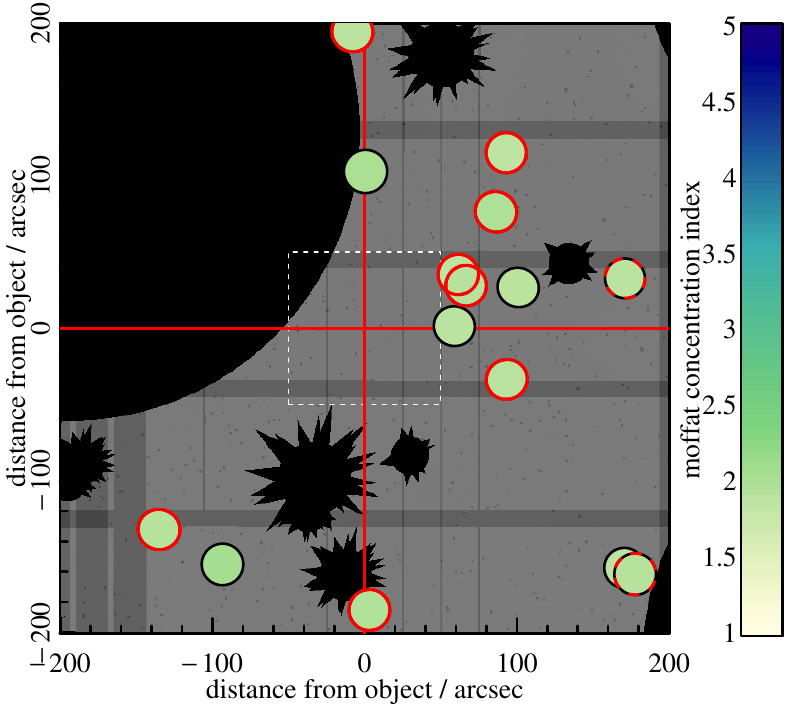}
    \caption{The result of the PSF fitting for the galaxy 396740 in the KiDS $r$-band with the dashed white square indicating the cutout shown in Figure~\ref{fig:exampleseg}. The greyscale image shows the $r$-band weight map with lighter colours meaning higher weight. Masked areas from the stacked $gri$-masks are shown in black (zero weight). The vertical and horizontal red line indicate the position of the object of interest (galaxy 396740) and split the image into its 4 quadrants. All fitted PSFs are shown as coloured ellipses with the size (FWHM multiplied by 20), axial ratio, orientation angle and concentration index (colour) taken from the fitted Moffat parameters. Stars selected for estimating the final model PSF have red borders; dashed red borders mean a fit was classified as suitable, but not selected because the maximum of 8 stars was already reached.} 
    \label{fig:examplepsfoutput}
\end{figure}

Figure \ref{fig:examplepsfoutput} shows an example diagnostic plot of the PSF fitting result. 

\subsubsection{Outputs}
For the fitting, we are only interested in the central galaxy and the closest neighbouring sources (for potential simultaneous fitting and to gain a better overview during visual inspection). Hence we do not save the entire $2000\times2000$\,pixel cutouts used in the preparatory work as that would unnecessarily waste storage space and computational time used on reading and writing files. Instead, the image, corresponding mask, segmentation map, sigma map and sky image are cut down to the smallest possible size that includes the object of interest (centred) and its neighbouring (touching) segments before saving. These 5 files, the model PSF image and some ancillary information such as the segment statistics are the main outputs of the preparatory work pipeline and serve as inputs for the galaxy fitting, which we describe in the next section.

\subsection{Galaxy fitting}
\label{sec:galaxyfitting}

\subsubsection{Inputs and models}
\label{sec:inputsandmodels}
We use the Bayesian code \texttt{ProFit} \citep{Robotham2017} to perform 2-dimensional multi-component surface brightness modelling in each band independently, assuming elliptical geometry and a (combination of) S\'ersic function(s) as the radial profile. The \citet{Sersic1963} function is described by three main parameters: the S\'ersic index $n$ giving the overall shape (with special cases $n=0.5$: Gaussian; $n=1$: exponential and $n=4$: \citet{deVaucouleurs1948} profile), the effective radius $R_e$ including half of the total flux and the overall normalisation which we specify as total magnitude $m$. In addition, in 2 dimensions the axial ratio $b/a$ gives the ratio of the minor to the major axis of the elliptical model and the position angle PA its orientation, while $x$ and $y$ are used to define the position in RA and Dec. Throughout the paper, $R_e$ refers to the effective radius along the major axis of the elliptical model. The S\'ersic model is detailed in \citet{Graham2005}. 

The data inputs for \texttt{ProFit} are a background-subtracted image, corresponding mask, segmentation map, sigma map and PSF. All of these are obtained during the preparatory steps (Section~\ref{sec:preparatorysteps}). On the modelling side, the main choices are the profile(s) to fit with initial parameter guesses and priors, the likelihood function to use, the fitting algorithm and convergence criteria; which are detailed in Sections~\ref{sec:initialguesses} to~\ref{sec:fitandconvergence}. In short, we choose to fit each object with 3 different models in a 4-step procedure: 
\begin{enumerate}
\item{Single component S\'ersic fits with initial guesses from segmentation statistics.}
\item{Double component S\'ersic bulge plus exponential disk fits with initial guesses from single component fits.}
\item{Double component re-fits for a subset of galaxies which seemed to have the bulge and disk components swapped in step (ii), see Section~\ref{sec:componentswapping}.}
\item{``1.5-component" point source bulge plus exponential disk fits with initial guesses from double component fits.}
\end{enumerate} 
Note that, for brevity, we will call the central component ``bulge" throughout this paper, even if it may not be a classical bulge. In particular, we do not distinguish classical bulges from pseudo-bulges, bars, AGNs, nuclear disks, combinations thereof or anything else that may emit light near the centre of a galaxy. Hence, we also use the term ``bulge" for 1.5-component fits where the central component is unresolved and for double component central components with low S\'ersic index and/or low axial ratios. 

To implement our three models, we make use of two of the many models built into \texttt{ProFit}, namely the S\'ersic and point source models. We fit all parameters except \texttt{boxyness} (i.e. we do not allow deviations of components from an elliptical shape) and, for the double and 1.5-component models, tying the positions of the two components together. Exponential disks are implemented using a S\'ersic profile with the S\'ersic index fixed to 1. This leaves 7 free parameters for our single S\'ersic and 1.5-component models and 11 free parameters of the double component fits, which are summarised in Table~\ref{tab:fittingparameters}. Scale parameters (S\'ersic index, effective radius and axial ratio) are treated in logarithmic space throughout, i.e. the actual fitting parameters are $\log_{10}(X)$ for scale parameters $X$. 

\begin{table}
	\centering
	\caption{The fitting parameters for each of our three models.}
	\label{tab:fittingparameters}
	\begin{tabular}{l l c l cc l cc} 
		\hline
		 && single && \multicolumn{2}{c}{double} && \multicolumn{2}{c}{1.5-comp.} \\
		 parameter && && bulge & disk && bulge & disk \\
		 \hline
		$x$-centre && free && \multicolumn{2}{c}{free} && \multicolumn{2}{c}{free} \\
		$y$-centre && free &&  \multicolumn{2}{c}{free} && \multicolumn{2}{c}{free}\\
		$m$ && free && free & free && free & free\\
		$\log_{10}(R_e)$&& free && free & free && N/A & free\\
		$\log_{10}(n)$ && free && free & 1 && N/A & 1 \\
		$\log_{10}(b/a)$&& free && free & free && N/A & free\\
		PA && free && free & free && N/A & free\\
		boxyness && 0 && 0 & 0 && N/A & 0\\
		\hline
	\end{tabular}
\end{table}

The 1.5-component model is needed for around 15-30\% of our double component systems where the bulge is too small relative to the image resolution to meaningfully constrain its S\'ersic parameters (the exact number depends on the band due to the different PSF sizes). With the point source profile, at least we can determine the existence of a second component and constrain its magnitude and hence the bulge-to-total (or AGN-to-total, bar-to-total, etc.) flux ratio. 

If the centre of an object is masked or the PSF estimation failed (which happens if large fractions of the surrounding area are masked), then the object is skipped and no fits are obtained. This affects approximately 20\% of the galaxies. All other objects are fitted with all three models; and the best model is selected subsequently (see Section~\ref{sec:modelselection} on details of the model selection and Section~\ref{sec:statistics} for the corresponding statistics).

\subsubsection{Initial guesses}
\label{sec:initialguesses}
Since we use MCMC algorithms, our fits do not strongly depend on the initial guesses. However, reasonable starting parameters are still required for convergence within finite computing times. 

The initial guesses for the single component S\'ersic parameters are obtained directly from the segmentation statistics output by \texttt{profoundProFound} (Section~\ref{sec:profound}) where we use the position, magnitude, effective radius (\texttt{R50}), axial ratio and angle as given; and the inverse of the concentration (1/\texttt{con}) for the S\'ersic index. 

For the double component fits, we convert the single component fits into initial guesses as follows: the position is taken unchanged, the magnitude of the single component fit is split equally between the two components, the bulge and disk effective radii are taken as 1/2 and 1 times the single component effective radius respectively, the S\'ersic index of the bulge is set to 4 and its axial ratio to 1 (round), the disk axial ratio is set to the axial ratio of the single component fit and the position angles of both components are taken as that of the single component fit.

Initial guesses for the 1.5-component fits are taken from the double component fits (after making sure the components are not swapped, see Section~\ref{sec:componentswapping}), where the bulge magnitude is used as the point source magnitude and the disk parameters are taken unchanged. 
\subsubsection{Priors, intervals, constraints}
\label{sec:priorsintervalsconstraints}
All parameters are limited to fixed intervals. In addition, there can be constraints between parameters (such that, e.g., the bulge and disk positions can be tied together). If a (trial) parameter is outside the bounds of its interval or constraint during any step of the fitting process, \texttt{ProFit} moves it back onto the limit before the likelihood is evaluated.

The limits for single-component fits are given in Table~\ref{tab:singlefitlimits}. In addition, the position angle is constrained such that if it leaves its interval, it is not just moved back onto the limit but jumps back 180\degr (which is the same angle, just more in the centre of the fitting interval). 
\begin{table}
	\centering
	\caption{The fitting limits for single-component fits.}
	\label{tab:singlefitlimits}
	\begin{tabular}{lll} 
		\hline
		parameter(s) & lower limit & upper limit \\
		\hline
		$x$- and $y$-centre & 0 & cutout side length\\
		magnitude & 10 & 25\\ 
		effective radius & 0.5 pixels & $\sqrt{2}$ cutout side length\\
		S\'ersic index & 0.1 & 20\\
		axial ratio & 0.05 & 1\\
		position angle & -90\degr & 270\degr\\
		\hline
	\end{tabular}
\end{table}

There are no additional priors or constraints for single component  fits. This means that in effect, we use unnormalised uniform priors which are 1 everywhere in the respective interval and zero otherwise. For scale  parameters (which are fitted in logarithmic space) the priors are uniform in logarithmic space, corresponding to \citet{Jeffreys1946}, i.e. uninformative, priors. 

The limits and constraints for double and 1.5-component fits are the same as for the single component fits (for both bulges and disks), except for the magnitude where the individual component magnitudes have infinity as their upper limit and instead the \emph{total} magnitude is constrained to be within  the magnitude limits. This is most consistent and also allows the fitting procedure to discard one of the two components for systems which can equally well be fitted with a single S\'ersic function (we then take this into account in the model selection). 

Note that the above procedure results in unnormalised likelihoods. The lack of normalization does not impede our analysis because the only time when we compare likelihoods is during model selection, where we effectively fold the normalisation into the calibration during visual inspection (Section~\ref{sec:modelselection}). 
\subsubsection{Likelihood function}
We use a Normal likelihood function for all fits. We have tested a t-distribution likelihood function which is less sensitive to outliers/unfittable regions; but found that the Normal likelihood function is better suited to our needs for several reasons. 

First of all, the t-distribution fits often preferred to use the freedom of the bulge parameters to fit disk features instead (e.g. rings, bumps, flares, etc. that cannot be captured by the exponential model), treating the bulge as an outlier since the t-distribution prefers a few strong outliers (the bulge pixels) over many weak ones. 

Second, the t-distribution fits fail for galaxies which are perfectly fitted by the model since then the errors truly are distributed Normally. This is a relatively common occurrence.

Hence some galaxies ($\sim20\%$) need to be fitted with a Normal distribution anyway, which, third, makes model selection much harder since the likelihood values obtained with different likelihood functions cannot easily be compared to each other. 
\subsubsection{Fit and convergence}
\label{sec:fitandconvergence}
All fits are performed on the sky-subtracted image within the galaxy segment only using the \texttt{convergeFit} function from the \texttt{AllStarFit} package \citep{AllStarFit}.
This function uses a combination of different downhill gradient algorithms available in the \texttt{nloptr} package \citep{nloptr} followed by several MCMC fits with \texttt{LaplacesDemon} \citep{LaplacesDemon} until convergence is reached.

The downhill gradient algorithms are used first to improve the initial guesses. The MCMC chain is not very sensitive to the initial guesses, but converges much faster if starting closer to the peak of the likelihood. Once the MCMC chains have converged, 2000 further likelihood points are collected to ensure a stationary sample for the subsequent analysis of the galaxy. 

We only fit the primary object of interest. While simultaneously fitting neighbouring sources is possible in \texttt{ProFit} and might have improved the fit on a few objects, the effects are generally small since the galaxies we study are not in highly crowded fields and the segmentation process usually excludes the vast majority of the flux from other sources. This is especially true since we use tight fitting segments within which the galaxy flux is dominant (cf. Section~\ref{sec:imagesegmentation}); and considering that the watershed algorithm of \texttt{ProFound} cleanly separates even overlapping sources, so neighbours are automatically masked (Section~\ref{sec:profound}). Hence we opted for the simpler and computationally cheaper option of just fitting the main objects. We confirm that this does not lead to major biases in Section~\ref{sec:parameterrecovery}).

\begin{figure*}
	\includegraphics[width=\textwidth]{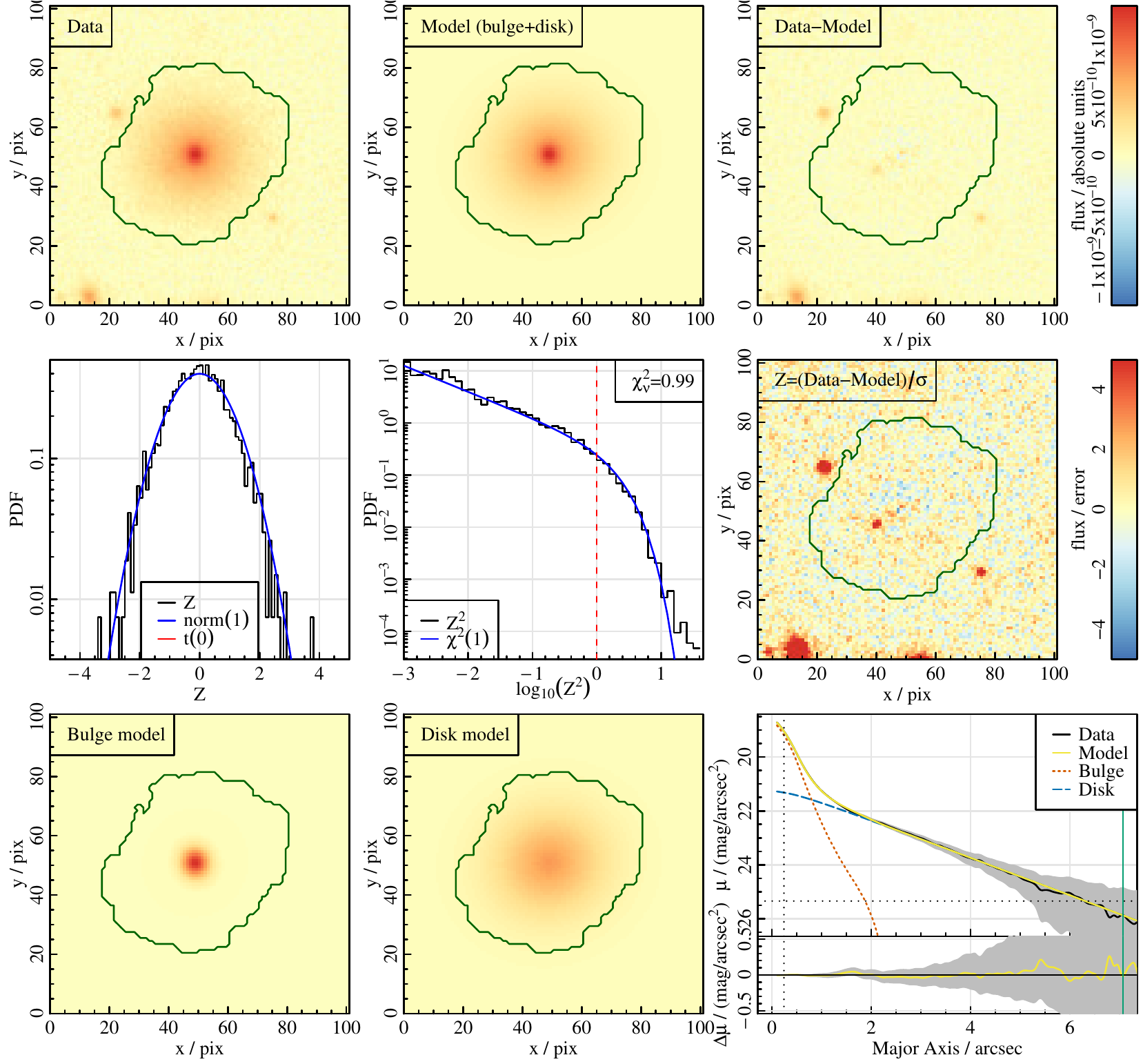}
    \caption{The result of the 2-component (S\'ersic bulge plus exponential disk) fit for the galaxy 611298 in the KiDS $r$-band. \textbf{Top row:} the data, 2-component model and residual between them shown in absolute values of flux given by the colour bar on the right. The green contour indicates the segment used for fitting. Note that the flux scaling here is non-linear and optimised to increase visibility of galaxy features, but is the same in all 3 panels. \textbf{Middle row:} goodness of fit statistics. The right panel is the normalised residual $Z$ (colour bar on the right) capped at $\pm5\sigma$. The left panel is the one-dimensional distribution (measured probability density function, PDF) of $Z$ within the segment; with blue and red curves showing a Normal distribution with a standard deviation of 1 and a Students-t distribution with the relevant degrees of freedom for comparison. The middle panel shows the measured PDF of $Z^2$ compared to a $\chi^2$-distribution of 1 degree of freedom (blue). The reduced chi-square, $\chi^2_\nu$ (sum over $Z^2$ divided by the degrees of freedom of the fit) is given in the top right corner. \textbf{Bottom row:} The bulge and disk models in 2 dimensions on the same flux scale as the top row; and the bulge, disk, and total model compared against the data in one-dimensional form (azimuthally averaged over elliptical annuli). The FWHM of the PSF and the approximate $1\sigma$ surface brightness limits are indicated by vertical and horizontal dotted lines for orientation. The vertical solid green line indicates the segment radius beyond which our model is extrapolated. The pixel scale is $0\farcs$2 for KiDS data, i.e. 1\arcsec\ corresponds to 5\,pix.} 
    \label{fig:examplefit}
\end{figure*}

An example fit for an object which is well-represented by our 2-component model is shown in Figure~\ref{fig:examplefit}. 

\subsubsection{Component swapping}
\label{sec:componentswapping}
Approximately 20-30\% of the double component fits have their bulge and disk components swapped, i.e. the exponential component fitting the central region and the S\'ersic component fitting the wings (this is a common problem in galaxy fitting, first pointed out by \citealt{Allen2006}). In particular, the freedom of the S\'ersic component is often used to fit disks that do not follow pure exponential profiles while at the same time being the dominant component in terms of flux (which is the case for most galaxies). To solve this problem, we devised an empirical swapping procedure guided by the visual inspection of a subsample of our galaxies. 

First, we select the galaxies that are most likely to have swapped components based on a cut in the plane of the ratio of S\'ersic indices and the ratio of effective radii for the single component fits and the bulge of the double component fits. The reasoning for choosing this parameter space to calibrate the cut was that we would generally expect bulges alone to be more concentrated (i.e. smaller effective radius and higher S\'ersic index) than when mixed with their respective disks in the single S\'ersic fits. This results in approximately 30\% of our sample being flagged as possibly swapped, which we then re-fit in a second step.

The re-fit is performed in exactly the same way as the original fit, except that we now use the results of the previous double-component fit as initial guesses, swapping around the bulge and disk components (except for the bulge S\'ersic index for which we use a value of 4). While the MCMC chain is less sensitive to initial guesses than a downhill gradient algorithm, it will still show some dependency for finite run-times. In particular, in our double component model the two components are \emph{nearly} interchangeable with the only difference being the S\'ersic index (fixed to 1 for the disk, free for the bulge). Hence there will always be 2 high maxima in likelihood space, which are far apart in the 11-dimensional parameter space. Moving from one to the other would require changing 9 parameters (all except position) at once in the right direction and hence is statistically unlikely. Therefore, we assist the code in finding the other maximum by manually swapping the initial guesses. 

In approximately 5\% of all re-fits, the code still converges on the same fit as before the swapping, but in most cases we find another likelihood maximum which corresponds to the bulge and disk components being reversed. As a third step we then select between the old and the new fit to obtain the physically more appropriate one. For this we first check whether either of the fits has a bulge S\'ersic index smaller than 2 \emph{and} a bulge effective radius at least 10\% larger than the disk effective radius \emph{and} a bulge-to-total ratio above 0.7 (i.e. the ``bulge" component is close to exponential, larger than the disk and contains the majority of the flux). If this is the case for only one of the fits, we choose the other one. If it is true for both or neither of the fits, then we apply our main criterion, which is that we choose the fit with the higher absolute value of bulge flux in the central pixel. These selection criteria are again based on visual inspection guided by the notion that we expect the bulge to be smaller and steeper than the disk and have proven to work very well. Note that the fit we select in this way is the one that is physically better motivated (i.e. with the bulge at the centre), and not necessarily the one which is statistically better.

After this procedure, the number of galaxies which still have the bulge and disk components swapped (and are classified as double component fits in model selection) is reduced to $\sim1-2\%$. 

\subsection{Post-processing}
\label{sec:postprocessing}

\subsubsection{Flagging of bad fits}
\label{sec:flaggingofbadfits}
After all three models have been fitted to all objects, we run them through our outlier flagging process (separately in each band). Each model is treated separately first; they are then combined during the model selection (Section~\ref{sec:modelselection}). 

The criteria for flagging bad fits (outliers) are: a very irregular fitting segment, an extreme bulge-to-total flux ratio, numerical integration problems, a parameter hitting its fit limits, poor $\chi^2$ statistics, a large distance between the input and fitted positions and a small fraction of model flux within the fitting segment. Additionally, there are some cautionary flags that identify fits which should be treated with extra care. All criteria are derived from and calibrated against visual inspection and described in more detail below. For orientation, we give the percentage of affected $r$-band fits in parentheses for each criterion. Note, however, that bad fits tend to fall into multiple of these categories, so the total number of bad fits is smaller than the sum of flagged objects in each category. Overall, approximately 9\%, 11\% and 10\% of all non-skipped fits are flagged in the $g$, $r$ and $i$ bands respectively (after model selection, Section~\ref{sec:modelselection}).

\begin{description}
\item[Very irregular segment (5.4\%):]{we calculate the difference between the magnitude of the model contained within the segment and the magnitude contained within the ``segment radius", which is defined as the maximum distance between the centre of the fit and the edge of segment. Objects where this magnitude difference is larger than 0.3 are flagged, as this is an indication for irregular segments (shredded, partly masked or cut off by another object for example). Note this criterion, as expected, often shows overlap with the criterion on the fraction of model flux contained within the segment (see description below).}

\item[Extreme bulge-to-total ratio (0.1\%):]{we flag double component and 1.5-component fits with a bulge-to-total ratio smaller than 0.001 or larger than 0.999 because in these cases the second component has negligible flux and a single component fit is better suited.}

\item[Numerical integration problems (0.2\%):]{\texttt{ProFit} includes an oversampling scheme for accurate pixel flux integration where pixels containing steep flux gradients are recursively oversampled up to an oversampling factor of 4096; in the central pixel even up to $\sim10^9$ \citep[for more details see][]{Robotham2017}. However, for very extreme model parameters, even this procedure may not be accurate enough anymore, leading to significant errors in the pixel flux calculations. This could be improved by changing the default oversampling values to achieve higher accuracy (at the cost of increased computational time), however we opted for simply excluding those cases since usually this only happens for unresolved bulges which are better represented by the 1.5-component fits anyway.}

\item[Parameter hitting limit (5.8\%):]{we flag objects where the magnitude, effective radius or S\'ersic index hit either of their limits (cf. Section~\ref{sec:priorsintervalsconstraints}); or the axial ratio hit its lower limit (for double component fits this applies to both components individually). The axial ratio upper limit is not flagged because fits are allowed to be exactly round, but there is a cautionary flag for all objects which hit any of its parameter limits (6.5\%). We also add a cautionary flag for suspiciously small or large errors on any parameter, where ``suspicious" is defined as being an outlier in the respective distribution of errors (2.1\%).}

\item[Poor $\chi^2$ statistics (0.1\%):]{we flag fits with a $\chi_\nu^2$ larger than 80; or where the $\chi^2$ in the central pixel is more than 1000 times larger than the average $\chi^2$ per pixel since that is an indication that the bulge was not fitted.}

\item[Large distance between input and output position (0.3\%):]{we flag fits with a distance between the input and output position of more than 2\arcsec (10\,pix), which are usually highly asymmetrical objects, mergers, objects with very nearby other objects (especially small objects embedded in the wings of much larger objects), or objects in regions of the image with unmasked instrumental effects. Often the fitted object then is not the one that we intended to fit. There is also a cautionary flag for offsets above 1\arcsec (1.3\%).}

\item[Small fraction of model flux within fitting segment (1.4\%):]{we flag fits where the amount of model flux (of any component) that falls within the fitting segment is less than 20\%. With so little flux to work on \texttt{ProFit} cannot constrain the parameters well anymore and these are often objects which are cut off by a masked region (e.g. a bright star) or other nearby objects. There is a cautionary flag for objects where the fraction of model flux (of any component) that falls within the segment is less than 50\% (9.3\%).}
\end{description}

\subsubsection{Model selection}
\label{sec:modelselection}
In Bayesian analysis, model selection is performed by computing the posterior odds ratio between two models, $O_{1:2}$, which is the ratio of the probabilities of the models given the same set of data and background information. With the help of Bayes' theorem and assuming a prior odds ratio of 1, this becomes the Bayes factor \citep[see, e.g.,][for a detailed treatment]{Sivia2006}: 
\begin{equation}
\label{eq:oddsratio}
O_{1:2} = \frac{\mathrm{p}(data \vert M_1, I)}{\mathrm{p}(data \vert M_2, I)} ,
\end{equation}
where $\mathrm{p}(data \vert M_1, I)$ is the probability of the data given Model 1 and any background information $I$ (the marginalised likelihood of Model 1). In practice, these probabilities are often difficult to compute because they require marginalising (i.e. integrating) over all model parameters.  

Hence, many information criteria tests have been developed which are based on the (non-marginalised) likelihood (or $\chi^2$) combined with some penalty term depending on the number of model parameters. This penalty term serves to judge whether a more complicated model is justified and takes the role of Ockham's factor (which is automatically included in equation~\ref{eq:oddsratio} due to the integration over all parameters). Commonly used tests include the Akaike information criterion \citep[AIC,][]{Akaike1974}, the Bayesian information criterion \citep[BIC,][]{Schwarz1978}, or the deviance information criterion \citep[DIC,][]{Spiegelhalter2002}. We choose to use the deviance information criterion, which is usually recommended over the AIC or BIC in Bayesian analysis \citep{Hilbe2017} and straightforward to compute from an MCMC output. Brief tests using the BIC or the estimated log marginal likelihood output by \texttt{LaplacesDemon} showed similar results.

The DIC is a direct output of the \texttt{LaplacesDemon} function (see Section~\ref{sec:fitandconvergence}) and is defined as: 
\begin{equation}
\label{eq:dic}
\mathrm{DIC} = Dev + pD = Dev + \mathrm{var}(Dev)/2,
\end{equation}
where $pD$ is a measure of the number of free parameters in the model and $Dev = -2\times$log-likelihood is the deviance. In theory, then, if the DIC difference $\Delta$DIC between two models is negative, the first model is preferred and if it is positive, then the second model is preferred; with differences larger than approximately 4 being considered meaningful \citep{Hilbe2017}. However, for the case of galaxy fitting where many features are present that cannot be captured by the model (bars, spiral arms, disk breaks or flares, tidal tails, mergers, foreground objects, etc.), we want to choose the model that we consider physically more appropriate rather than better in a strictly statistical sense. This requires visual classification, logical filters, detailed simulations or a manual calibration of the $\Delta$DIC cut (or whichever other chosen diagnostic) by visual inspection of a representative sub-sample \citep[e.g.][]{Allen2006, Simard2011, Vika2014, Argyle2018, Kruk2018}. We choose the latter approach, which has the added advantages that we do not need to worry about normalising our likelihoods (cf. Section~\ref{sec:priorsintervalsconstraints}), hence circumventing dependencies of the results on prior widths; nor the fact that our pixel values are correlated (due to the PSF) -- these effects are simply folded into the visual calibration. 

We use a random sample of $\sim700$ non-skipped objects per band (i.e. $\sim2000$ objects in total) for the calibration; and a further 1000 $r$-band objects that were previously inspected for cross-checking the results. In addition, our model selection procedure takes into account some of the outlier flagging (Section~\ref{sec:flaggingofbadfits}). 
For each of the $\sim700$ objects in each band, SC visually inspected the fits of all three models and classified the object into one of the categories: ``single component", ``1.5-component", ``double component", ``not sure if 1.5- or double component", ``not sure at all", ``unfittable" (outlier). We then calculate the DIC differences between all three models (i.e. $\Delta$DIC$_{1-1.5}$, $\Delta$DIC$_{1-2}$ and $\Delta$DIC$_{1.5-2}$) and calibrate them for model selection in two steps: first, we select between single component fit or not; of the ones that are not single component fits we then select between double component or 1.5-component fits.

For the first step of model selection calibration, the $\Delta$DIC$_{1-1.5}$ and $\Delta$DIC$_{1-2}$ cuts are optimised such that the minimum number of fits is classified wrongly. ``Wrong" in this case means a fit was manually classified as ``single" but is now a double/1.5; or a fit was manually classified as ``1.5", ``double", or ``not sure if 1.5 or double" but is now a single. ``Unfittable" and ``not sure at all" cases are ignored. For the second step of model selection calibration, the $\Delta$DIC$_{1.5-2}$ cut is optimised in the same way; where ``wrong" now means that the fit was manually classified as ``1.5" but is now a double or vice versa, with all other categories being ignored. For the two steps of the calibration, we bootstrap the manual sample 1000 and 500 times respectively and repeat the optimisation to get an estimate of the error on the chosen DIC cuts. These errors are chosen as the $1\sigma$ quantiles (i.e. they contain the central 68\% of DIC cut distributions). Our calibrated DIC cuts hence all have a median, a lower limit and an upper limit. Any object within these limits is flagged as unsure in the model selection, i.e. the DIC differences are not conclusive for this object. 

To perform the actual model selection, the calibrated DIC cuts in each band are then applied to the entire sample, again in a two-step procedure: the single component fit is selected if neither of the 1.5- or double component fits are significantly better (as indicated by the DIC differences). Double component fits need to be significantly better than 1.5-component fits, too. In all cases, if the DIC difference is very clear, we do the model selection first; then flag objects as outliers if needed.\footnote{This means that it is possible (and not uncommon) that a galaxy which is classified as an outlier has a non-flagged fit in another model (but the fit that was chosen was significantly better than the other one, despite it being an outlier).} In the unsure region of the DIC difference, we choose the model that is not flagged as outlier; if neither is flagged, the DIC cut is applied.  

Compared against visual inspection (keeping in mind that visual classification is not free of errors either), roughly 7\%, 9\% and 6\% of the galaxies end up in the wrong category in total in the $g$, $r$ and $i$ bands respectively (in both steps of model selection combined, ignoring cases which were visually classified as ``unsure"). Table~\ref{tab:modelselconfusionr} gives the detailed confusion matrix for the $r$-band. Note that we do not consider the success of the outlier flagging here, so for outliers
we show what the galaxy would have been classified as if it were not flagged (absolute value of
the \texttt{NCOMP} column in our catalogue). We highlight those galaxies that are correctly classified in bold and show those that were ignored during the model selection calibration process in grey font. The remaining (black) numbers add up to the 9\% quoted above. Corresponding confusion matrices for the $g$ and $i$ bands are given in Appendix~\ref{app:modelselconfusion} (Tables~\ref{tab:modelselconfusiong} and~\ref{tab:modelselconfusioni}).
Note that since we minimise the \emph{total} number of fits classified wrongly, there is a slight bias against the rarer categories in the automated model selection. For example, the relative fraction of true 1.5-component objects (as per the visual inspection) that is classified wrongly by the automated selection is higher simply because 1.5-component objects are much rarer than single or double component objects.

\begin{table}
	\centering
	\caption{The confusion matrix for our model selection based on a DIC difference cut compared against visual inspection for the $r$-band. All values are in percent of the total number of visually inspected $r$-band galaxies. Bold font highlights galaxies classified correctly, while grey shows those that were ignored during the calibration.}
	\label{tab:modelselconfusionr}
	\begin{tabu}{lcrrrc} 
		\hline
		 & \multicolumn{5}{r}{number of components} \\
		visual classification & & 1 & 1.5 & 2 &\\
		\hline
		``single" && \textbf{41.6} & 0 & 2.7 &\\
		``1.5" && 2.2 & \textbf{2.4} & 0.9 &\\
		``double" && 3.1 & 0.1 & \textbf{9.2} &\\
		``1.5 or double" && 0.3 & \textbf{0.6} & \textbf{3.0} &\\
		\rowfont{\color{gray}}
		``unsure" && 16.1 & 0.4 & 13.1 &\\
		\rowfont{\color{gray}}
		``unfittable" && 0.9 & 0.6 & 2.7 & \\
        \hline
	\end{tabu}
\end{table}

In addition to this band-specific model selection, we perform a joint model selection for all three bands. For this, we sum the DIC values of all three bands for each model before computing the DIC differences. Then we perform the same optimisation procedure as for the single bands (using all $\sim2000$ visually classified objects across the three bands) to obtain cuts in DIC difference which we subsequently apply for the model selection. Note that the model selected in this way is by necessity a compromise between the different bands, which have different depth and seeing. In this procedure, approximately 9\% of fits are classified wrongly across all bands compared to visual classification. The corresponding confusion matrix is shown in Table~\ref{tab:modelselconfusionjoint}.

The accuracy of the model selection is also confirmed using simulations, to the extent to which our simulations allow us to do so (see Section~\ref{sec:simulationsmodelselection} for details).

\subsubsection{Truncating to segment radii}
\label{sec:tightsegments}
As detailed in Section~\ref{sec:imagesegmentation}, we produce segmentation maps that define the fitting region, meaning that only pixels within the fitting segment are considered during the evaluation of the likelihood of the model (equivalent to giving all pixels outside the segment zero weight in the fit). We choose tight fitting segments (cf. Section~\ref{sec:imagesegmentation}) in order to obtain the best possible fit in the inner, high signal-to-noise ratio regions of the galaxies and be less sensitive to disk breaks, flares, nearby other objects, sky subtraction problems and similar. The disadvantage of this approach is that profiles are not necessarily forced to zero for large radii, i.e. our S\'ersic fits often show unphysically large effective radii combined with high S\'ersic indices. 

To mitigate this effect, we define a ``segment radius" for each galaxy segment, which is simply the maximum distance between the fitted galaxy centre and the edge of the segment and can be understood as the upper limit to within which our model is valid. We then calculate the ``segment magnitude", $m_{seg}$, which is the magnitude of the (intrinsic, not PSF-convolved) profile integrated to the segment radius (rather than infinity); and the ``segment effective radius", $R_{e, seg}$, which is the radius containing half of the flux defined by the segment magnitude. These values (and quantities derived from them, such as segment bulge-to-total flux ratios) are provided in the catalogue (labelled *\_SEGRAD) and we strongly recommend using these instead of the S\'ersic values integrated to infinity whenever they are available. For a direct parameter comparison to other works, the values in those catalogues should also be appropriately truncated.

In the following, we explain this recommendation in more detail; with further points to note in Sections~\ref{sec:comparelee} and \ref{sec:systematics}.  \\

\begin{figure*}
    \includegraphics[width=\columnwidth]{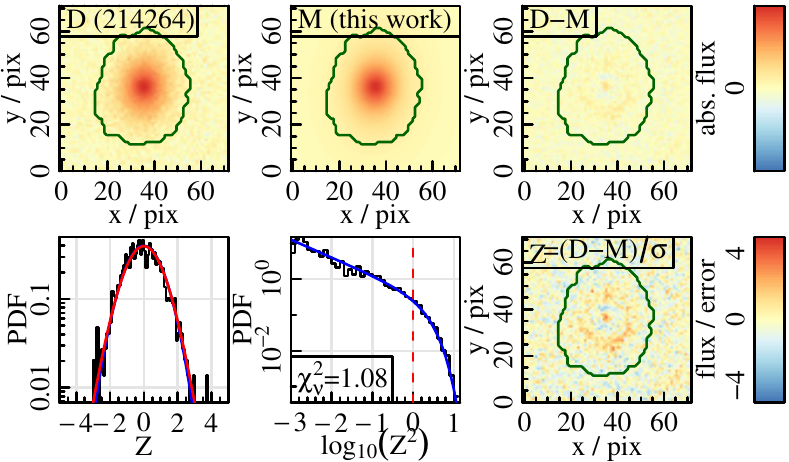}
    \includegraphics[width=\columnwidth]{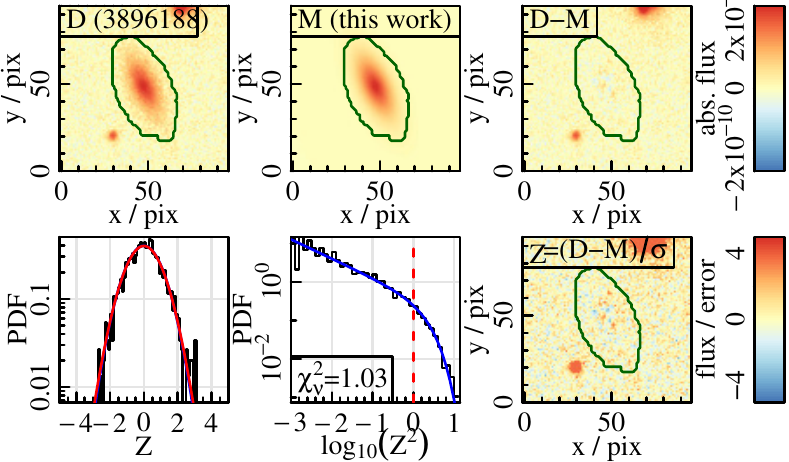}
    \includegraphics[width=\columnwidth]{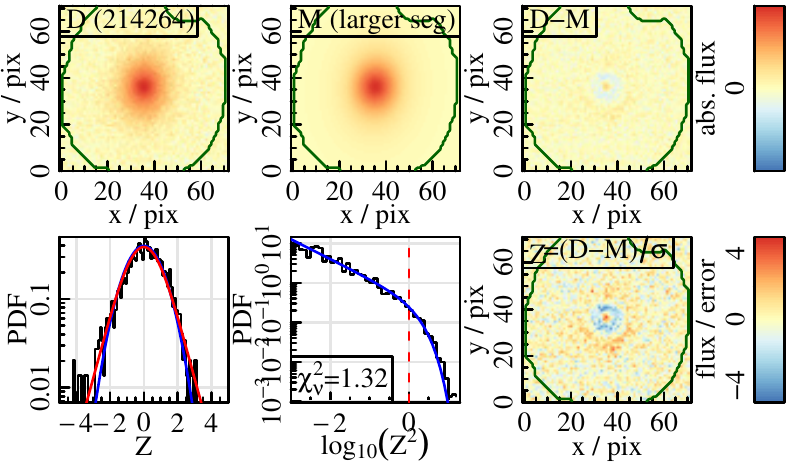}
    \includegraphics[width=\columnwidth]{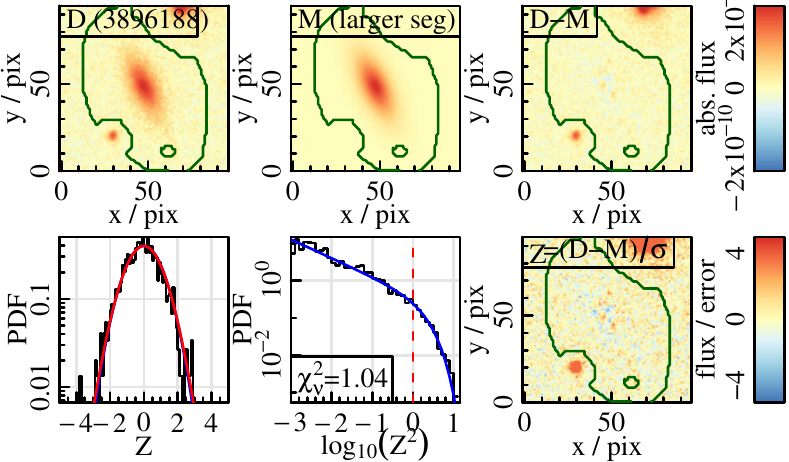}
    \includegraphics[width=\columnwidth]{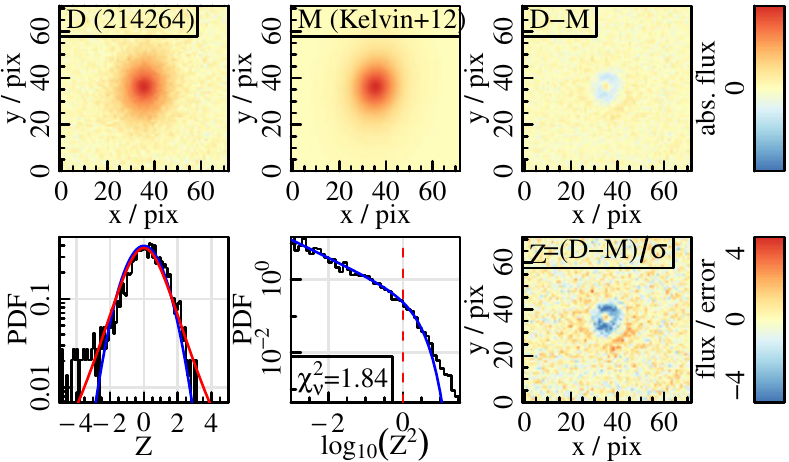}
    \includegraphics[width=\columnwidth]{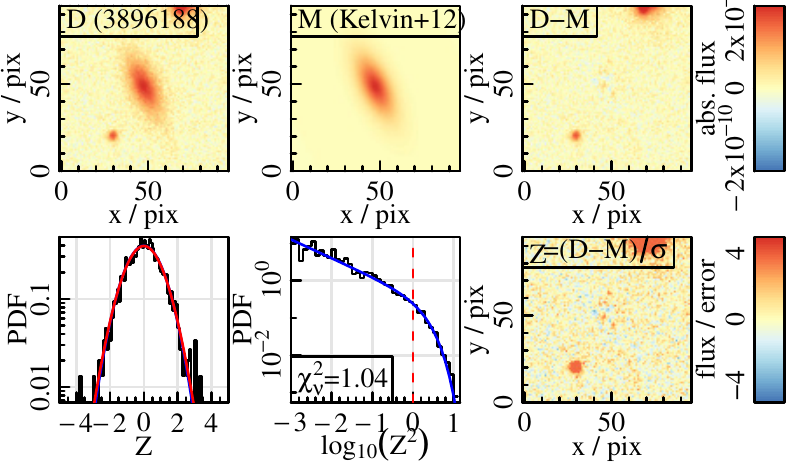}
    \includegraphics[width=\columnwidth]{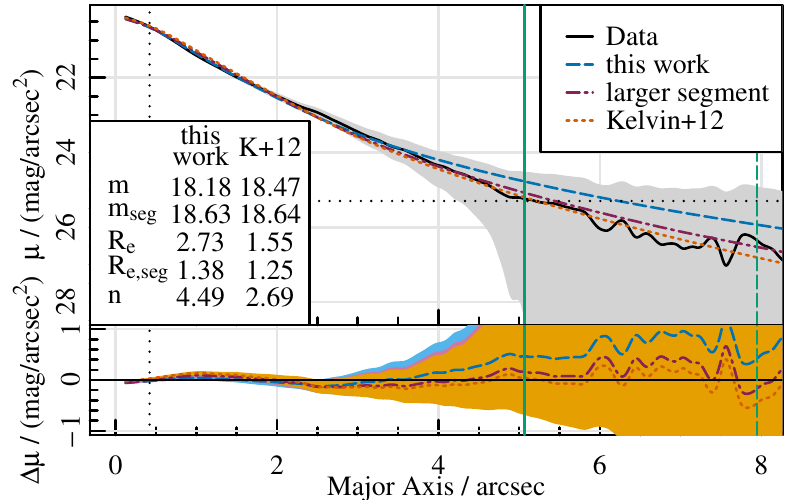}
    \includegraphics[width=\columnwidth]{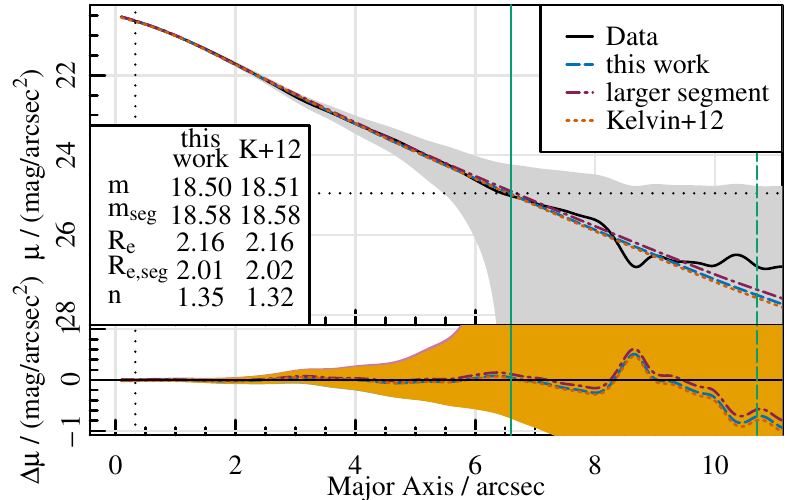}
    \caption{\textbf{Left:} detailed comparison of our single-S\'ersic fit, our fit using a larger segment, and the \citet{Kelvin2012} fit to the galaxy 214264, which in reality is a 1.5-component system. \textbf{Right:} the same for galaxy 3896188, which is well-described by a single S\'ersic component. \textbf{Top two rows:} our fit to the KiDS $r$-band data with panels the same as those in Figure~\ref{fig:examplefit}. \textbf{Rows three and four:} the fit we obtained by using a larger fitting segment as indicated. \textbf{Rows five and six:} the \citet{Kelvin2012} fits (originally performed on SDSS $r$-band data) evaluated on the KiDS $r$-band data, which use a fitting region larger than the cutout shown. \textbf{Bottom two panels:} direct comparison of the one-dimensional profiles, see text for details. The vertical green solid and dotted lines indicate the segment radii for the two segment sizes. The vertical dotted line shows the HWHM of the PSF; the horizontal dotted line is the $1\sigma$ surface brightness limit of the data. The quoted values for the effective radius are in arcseconds.
    }
    \label{fig:tightseg}
\end{figure*}

Figure~\ref{fig:tightseg} illustrates the effects produced by our tight fitting segments, how to mitigate those by truncating the magnitude and effective radius appropriately; and the circumstances under which this correction is necessary. For two example galaxies - 214264 and 3896188 - we show a detailed comparison of our single S\'ersic fit to a fit using a larger segment and to the fit obtained in \citet{Kelvin2012}. We present a more general (statistical) comparison of our fit results to those of \citet{Kelvin2012} in Section~\ref{sec:comparelee}, where we also give more details on how their fits were derived. For the purposes of the analysis in this section it suffices to say that \citet{Kelvin2012} used much larger fitting regions than we do, while the remaining analysis is in many ways analogous to ours (although they use different data, code and procedures in detail). 

Focussing on the left half of Figure~\ref{fig:tightseg} first, the first six panels (top two rows) show the KiDS $r$-band data, our single S\'ersic model, the residual and various goodness of fit statistics as described in the caption of Figure~\ref{fig:examplefit}. Panels seven to twelve (rows three and four) show the same for a larger fitting segment as indicated by the green contour. Panels 13 to 18 again show the same for the \citet{Kelvin2012} fit, where we note that this was originally performed on $r$-band Sloan Digital Sky Survey (SDSS, \citealt{York2000}) data but is now evaluated on the $r$-band KiDS data. The \citet{Kelvin2012} fits were performed on cutouts larger than the size shown here, i.e. they include all visible pixels (and more) in the fit. Note that the reduced chi-square value quoted in the bottom middle panel of each set of plots always is evaluated within the smallest segment so that they can be directly compared. Finally, the bottom two panels show a direct comparison of the one-dimensional profiles of all three fits, which we will now study in detail. 

In the top panel of this one-dimensional plot, we show the surface brightness (azimuthally averaged over elliptical annuli) against the projected major axis for the data (solid black line with grey uncertainty region), our model fit for the fiducial segment (dashed blue line) and the larger segment (dash-dotted pink line) and the \citet{Kelvin2012} model fit (dotted orange line). The vertical green solid and dashed lines indicate the segment radii (for the two segment sizes respectively) beyond which our model is an extrapolation. The vertical dotted line shows the half width at half maximum (HWHM) of the PSF and the horizontal dotted line is the $1\sigma$ surface brightness limit of the data. The inset in the bottom left of this plot shows the fitted magnitude $m$, effective radius $R_e$ in arcseconds and S\'ersic index $n$ values for our and the \citet{Kelvin2012} fits; and the corresponding segment-radius-truncated values for $m$ and $R_e$. Finally, the bottom panel shows the difference between all three models and the data (with errors): our fiducial fit in blue with a dashed line, the fit in the larger segment in pink with a dash-dotted line and the \citet{Kelvin2012} fit in orange with a dotted line.

Clearly, our model is a better fit to the inner regions of the galaxy than the \citet{Kelvin2012} fit (out to about 2\arcsec, also evident from the two-dimensional plots and from the reduced $\chi^2$-value within the segment decreasing from 1.84 to 1.08), owing to the higher S\'ersic index which better represents the steep bulge at the centre. However, it has a large effective radius and considerable amounts of model flux at large radii which are not observed in the data. In particular in the region beyond the segment radius, where our model is merely extrapolated, it is clearly oversubtracting the data (also visible in the 2-dimensional plots). Correnspondingly, the truncated segment quantities differ substantially from the fitted S\'ersic values. The \citet{Kelvin2012} fits, instead, use a larger fitting region and hence follow the data out to larger radii, which results in a worse fit of the central regions but does not contain such large amounts of excess flux beyond the surface brightness limit. Hence, truncating to segment radii has a smaller effect on the parameter values. The truncated values for both models are then in reasonable agreement with each other, except for the S\'ersic index, for which no truncated version exists as it would be unclear how to define such a value. Our fit in the larger segment is in between the two others in all respects, since it has a fitting region intermediate to the other two.

Note that the differences only come about when the model is not (in a formal statistical sense) a good representation of the data, i.e. when there is a need to compromise between fitting different regions. In the case of the left side of Figure~\ref{fig:tightseg}, the galaxy shown is better described by a 1.5-component model ($m^B = 20.47$, $m^D = 18.79$, $R_e^D = 1.89\arcsec$), although in general there are many objects in our sample for which even a two-component model cannot capture all aspects of the data. For comparison, in the right half of Figure~\ref{fig:tightseg}, we show a galaxy that is well-described by a single S\'ersic model: here, both our and the \citet{Kelvin2012} fits arrive at virtually the same solution despite the different fitting regions. In fact, all three models and the data are nearly indistinguishable all the way down to the $1\sigma$ surface brightness limit. 

In short, there is no perfect way to fit a S\'ersic function to an object which intrinsically does \emph{not} have a pure S\'ersic profile. For such objects, which unfortunately comprise the majority of our sample, the fitted parameters will always depend on the exact fitting region used as well as the quality of the data (its depth in particular). Most previous work, including \citet{Kelvin2012}, opted to use large fitting regions in order to include enough sky pixels to ensure that the profiles are constrained to approach zero flux at large radii (although a S\'ersic function technically never reaches zero exactly). Here, we choose a different approach by using smaller fitting segments. This means that the profiles are not constrained to approach zero flux at large radii. Instead more emphasis is placed on adequately representing the inner regions of the galaxies. We choose this approach since it is most appropriate for our science case, where we are primarily interested in comparing the high signal-to-noise regions of galaxies from the same data set amongst each other. 
In addition, it decreases the sensitivity of our fits to deviations from a S\'ersic profile in the low surface brightness wings of objects (arguably no galaxy truly follows a S\'ersic profile to infinity) as well as nearby other objects and inaccuracies in the sky subtraction. 
We stress that this means that our parameters are not directly comparable to other works using larger fitting segments. In particular, our S\'ersic indices tend to be systematically higher (see Section~\ref{sec:comparelee}) since high S\'ersic indices result in high amounts of flux at large radii and are hence suppressed when constraining the models to zero flux at large radii. Magnitudes and effective radii can be compared to those of other studies by truncating to segment radii.

\section{Results}
\label{sec:results}

\subsection{\texttt{BDDecomp} DMU}
Our main result is the \texttt{BDDecomp} DMU in the GAMA database. It contains 8 catalogues: \texttt{BDInputs} (three times, one for each band) with the most important outputs of the preparatory work pipeline (segmentation, PSF estimation, initial guesses), \texttt{BDModelsAll} (three times, one for each band) with the output from the actual galaxy fitting and post-processing (model selection, flagging of bad fits and truncating to segment radii) and \texttt{BDModels} which combines the most important columns of the 3 \texttt{BDModelsAll} tables and has a few additional joint columns (mainly joint model selection). Finally, the table \texttt{BDModelsAlt} presents the same information as \texttt{BDModels} just with the three bands arranged in rows instead of columns. Each table is accompanied by comprehensive documentation including descriptions of all columns, details on the processing steps and practical tips for using the catalogue. The DMU also provides all input data used for the fitting (i.e. image cutouts, masks, error maps, segmentation maps, sky estimates, PSFs) as well as various diagnostic plots of the fit results on the GAMA file server, where detailed descriptions of these files can be found.

In the following sections we present an overview over the contents of the main catalogue, \texttt{BDModels}. 

\subsection{Catalogue statistics}
\label{sec:statistics}
Table~\ref{tab:results} gives an overview of the fit and postprocessing results. Starting with our full sample (13096 galaxies from the combination of our main and SAMI samples, see Section~\ref{sec:sampleselection}), we show how the number of galaxies evolves through all steps of the pipeline. The results are split per-band and per-model where necessary. At some steps, we also include percentages of galaxies lost or remaining (grey font). In short, we lose nearly 20\% of our sample to masking and a further almost 10\% to the flagging of bad fits; where the former is a random subset while the latter preferentially affects certain types of galaxies (e.g. mergers and irregulars). 

\begin{table*}
	\centering
	\caption{Fit results: numbers (black) and percentages (grey) of galaxies remaining or lost at each step in our pipeline, split per-band and per-model where necessary.$^1$ Notes/explanations of each step:\\
	1) The full sample results from the combination of our main and SAMI samples (Section~\ref{sec:sampleselection}). \\
	2) Some galaxies have been imaged more than once due to overlap regions between KiDS tiles. These duplicate observations of the same physical objects are treated independently throughout our pipeline (Section~\ref{sec:sampleselection}).\\
	3) We use the associated KiDS masks, combining the three bands. Images for which the central galaxy pixel is masked ($\sim20\%$) are skipped during the fitting (Section~\ref{sec:cutoutsandmasking}).\\
	4) For each image in each band, a PSF is then estimated by fitting nearby stars. If the PSF estimation fails, the galaxy is skipped during the fit (Section~\ref{sec:psfestimation}). Note that technically, we estimate PSFs also for galaxies that are masked in step 3), but we do not list those here.
	5) For each non-masked image with a successful PSF estimate, we attempt 3 fits: a single S\'ersic (1), a pointsource + exponential (1.5) and a S\'ersic + exponential (2). Very rarely, the fit attempts fail with an error (Section~\ref{sec:fitandconvergence}).\\
	6) Each fit (for each model independently) is passed through our outlier flagging process, identifying bad fits (Section~\ref{sec:flaggingofbadfits}).\\
	7) Of the non-flagged (i.e. good) fits, we then select the most appropriate one during model selection (Section~\ref{sec:modelselection}).\\
	8) Summing up the selected fits for each model (step 7) gives the total number of good fits. The difference between the good and successful fits (step 5) stems from the outlier flagging. Skipped fits are due to masking, PSF or fit fails (steps 3, 4, 5). The sum of good, flagged and skipped fits gives the total number of independent fits (step 2).\\
	9) Removing duplicate observations for the same physical objects gives the number of good, flagged and skipped galaxies, which sum to the number of unique objects (step 1). Here we always use the best available result for each galaxy, i.e. it is counted as ``good" if at least one of the multiple observations was ``good".}
	\label{tab:results}
	\setlength{\tabcolsep}{4pt} 
	\begin{tabu}{l rrr c rrr c rrr c rrr} 
	    \hline
	    band
	    & \multicolumn{3}{c}{$g$}
	    && \multicolumn{3}{c}{$r$}
	    && \multicolumn{3}{c}{$i$}
	    && \multicolumn{3}{c}{joint $gri$}
	    \\
	    model (components)
	    & \multicolumn{1}{c}{1} & \multicolumn{1}{c}{1.5} & \multicolumn{1}{c}{2} 
	    && \multicolumn{1}{c}{1} & \multicolumn{1}{c}{1.5} & \multicolumn{1}{c}{2} 
	    && \multicolumn{1}{c}{1} & \multicolumn{1}{c}{1.5} & \multicolumn{1}{c}{2} 
	    && \multicolumn{1}{c}{1} & \multicolumn{1}{c}{1.5} & \multicolumn{1}{c}{2} 
	    \\
	    \hline
	    number of: 
	    \\
	    1) unique objects (galaxies)
	    & \multicolumn{15}{c}{13096}
	    \\
	    2) images (independent fits)
	    & \multicolumn{15}{c}{14966}
	    \\
	    3) images not masked
	    & \multicolumn{15}{c}{11989}
	    \\
	    \rowfont{\color{gray}}
	    lost due to masking (\%)
	    & \multicolumn{15}{c}{20}
	    \\
	    \\	    
	    4) successful PSFs
	    & \multicolumn{3}{c}{11838}
	    && \multicolumn{3}{c}{11872}
	    && \multicolumn{3}{c}{11946}
	    && \multicolumn{3}{c}{11683}
	    \\
	    \rowfont{\color{gray}}
	    lost due to PSF fails (\%)
	    & \multicolumn{3}{c}{1}
	    && \multicolumn{3}{c}{0.8}
	    && \multicolumn{3}{c}{0.3}
	    && \multicolumn{3}{c}{2}
	    \\
	    \\
		5) successful fits 
		& 11837 & 11837 & 11831 
		&& 11872 & 11870 & 11861 
		&& 11946 & 11943 & 11945
		&& 11682 & 11678 & 11665
		\\		
		\rowfont{\color{gray}}
		lost due to fit fails (\%)
		& <0.01 & <0.01 & 0.05 
		&& 0 & 0.01 & 0.07 
		&& 0 & 0.02 & <0.01
		&& <0.01 & 0.03 & 0.12
		\\		
		6) fits not flagged 
		& 10951 & 7122 & 8022 
		&& 11025 & 8164 & 8759 
		&& 11086 & 7620 & 7775
		&& 10680 & 6446 & 5870
		\\		
		\rowfont{\color{gray}} 
	    not flagged/successful (\%) 
		& 93 & 60 & 68 
		&& 93 & 69 & 74 
		&& 93 & 64 & 65 
		&& 91 & 55 & 50
		\\		
		7) selected fits 
		& 8294 & 740 & 1743 
		&& 7061 & 585 & 2935 
		&& 7411 & 662 & 2663 
		&& 7308 & 621 & 2009
		\\		
		\rowfont{\color{gray}} 
		selected/successful (\%)
		& 70 & 6 & 15 
		&& 59 & 5 & 25 
		&& 62 & 6 & 22 
		&& 63 & 5 & 17
		\\		
		\\
		total number (per band) of:\\
		8) good\,|\,flagged\,|\,skipped fits
		& \multicolumn{3}{c}{10777\,|\,1061\,|\,3128} 
		&& \multicolumn{3}{c}{10581\,|\,1291\,|\,3094} 
		&& \multicolumn{3}{c}{10736\,|\,1210\,|\,3020} 
		&& \multicolumn{3}{c}{9938\,|\,1745\,|\,3283} 
		\\
		\rowfont{\color{gray}} 
		good\,|\,f.\,|\,s./all images (\%)
		& \multicolumn{3}{c}{72\,|\,7\,|\,21} 
		&& \multicolumn{3}{c}{71\,|\,9\,|\,21} 
		&& \multicolumn{3}{c}{72\,|\,8\,|\,20} 
		&& \multicolumn{3}{c}{66\,|\,12\,|\,22}
		\\
		\\
		9) good\,|\,flagged\,|\,skipped gal.
		& \multicolumn{3}{c}{9722\,|\,935\,|\,2439}
		&& \multicolumn{3}{c}{9545\,|\,1145\,|\,2406} 
		&& \multicolumn{3}{c}{9687\,|\,1059\,|\,2350} 
		&& \multicolumn{3}{c}{8998\,|\,1559\,|\,2539} 
		\\
		\rowfont{\color{gray}} 
		good\,|\,f.\,|\,s./unique objects (\%) 
		& \multicolumn{3}{c}{74\,|\,7\,|\,19} 
		&& \multicolumn{3}{c}{73\,|\,9\,|\,18} 
		&& \multicolumn{3}{c}{74\,|\,8\,|\,18} 
		&& \multicolumn{3}{c}{69\,|\,12\,|\,19}
		\\
        \hline
        \multicolumn{16}{l}{$^1$Based on information given in the \texttt{*\_BDQUAL\_FLAG}, \texttt{*\_OUTLIER\_FLAG} and \texttt{*\_NCOMP} columns of the \texttt{BDModels} catalogue.}
        \\
	\end{tabu}
 \end{table*}
Note that we used stacked $gri$ images for segmentation and masking, but then treated the galaxies independently in all bands except for the model selection, where we performed both a per-band and a joint version. Therefore, the column ``joint $gri$" always gives the number of galaxies that were ``good" in all three bands (hence why numbers are generally lower), except for the model selection, where it shows the results of the joint model selection (cf. Section~\ref{sec:modelselection}).

\begin{figure}
	\includegraphics[width=\columnwidth]{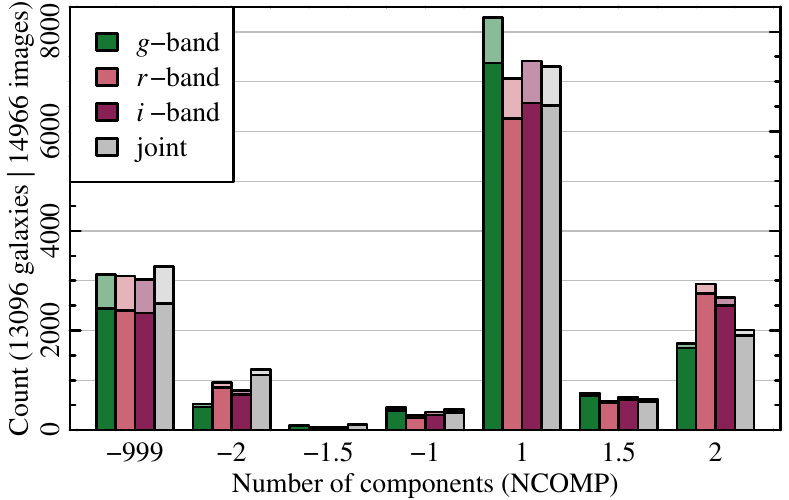}
    \caption{The number of components assigned in our model selection procedure for individual bands and the joint analysis. 1, 1.5 and 2 mean single S\'ersic, point source bulge + exponential disk and S\'ersic bulge + exponential disk models respectively. Negative values indicate that the chosen (best) fit was flagged as unreliable (mostly irregular or partly masked galaxies). -999 is assigned to skipped fits, either because the galaxy centre is masked (most cases) or because the PSF estimation failed. The lighter (higher) bars show the number of images, whereas the shorter bars in the foreground indicate the number of unique objects. See text and Table~\ref{tab:results} for details.}	
    \label{fig:ncompstats}
\end{figure}

Figure~\ref{fig:ncompstats} visualises the most important information given in Table~\ref{tab:results}, namely the final number of objects classified in each category: lighter bars in the background refer to individual fits (total 14966) with the number of unique galaxies (total 13096) overplotted. When several fits to the same galaxy were classified in different categories, we allocate it to the highest of those,\footnote{This means that a galaxy is classified as ``outlier" if \emph{all} fits to it are outliers and it is ``skipped" only if \emph{all} fits are skipped. 
Galaxies with good fits are allocated to the most complex model of the available fits (assuming that one of the images was deeper and allowed to constrain more components than the other(s)), while within the outlier categories we allocate it to the simplest model. Note that in Table~\ref{tab:results} we only show the total number of flagged fits and do not split them into the different outlier categories.} which is consistent with Table~\ref{tab:results}. \texttt{NCOMP}\,=\,$-999$ means the object was skipped (not fitted) because it is masked or the PSF estimation failed (usually because of large masked areas in the immediate vicinity of the object). \texttt{NCOMP}\,=\,1, 1.5 or 2 indicates that this is a good fit classified as single, 1.5- or double component fit. \texttt{NCOMP}\,=\,$-1$, $-1.5$ or $-2$ indicates that this is a bad fit (outlier) which would have been classified as single, 1.5- or double component fit if it were not an outlier (most often these are mergers/irregular galaxies for which our models are not appropriate; or galaxies that are partly masked). We keep these three classes separate since automated outlier identification can never be perfect; and what should be considered a bad fit will depend on the use case. The flagging of fits is hence only intended as a guide and all available information in the catalogue is retained for all fitted objects. 
Figure~\ref{fig:examplefit} shows an example two-component fit. Examples for a single, 1.5-component and outlier can be found in Appendix~\ref{app:exampleplots}. 

\subsection{Parameter distributions}
\begin{figure*}
    \includegraphics[width=\textwidth]{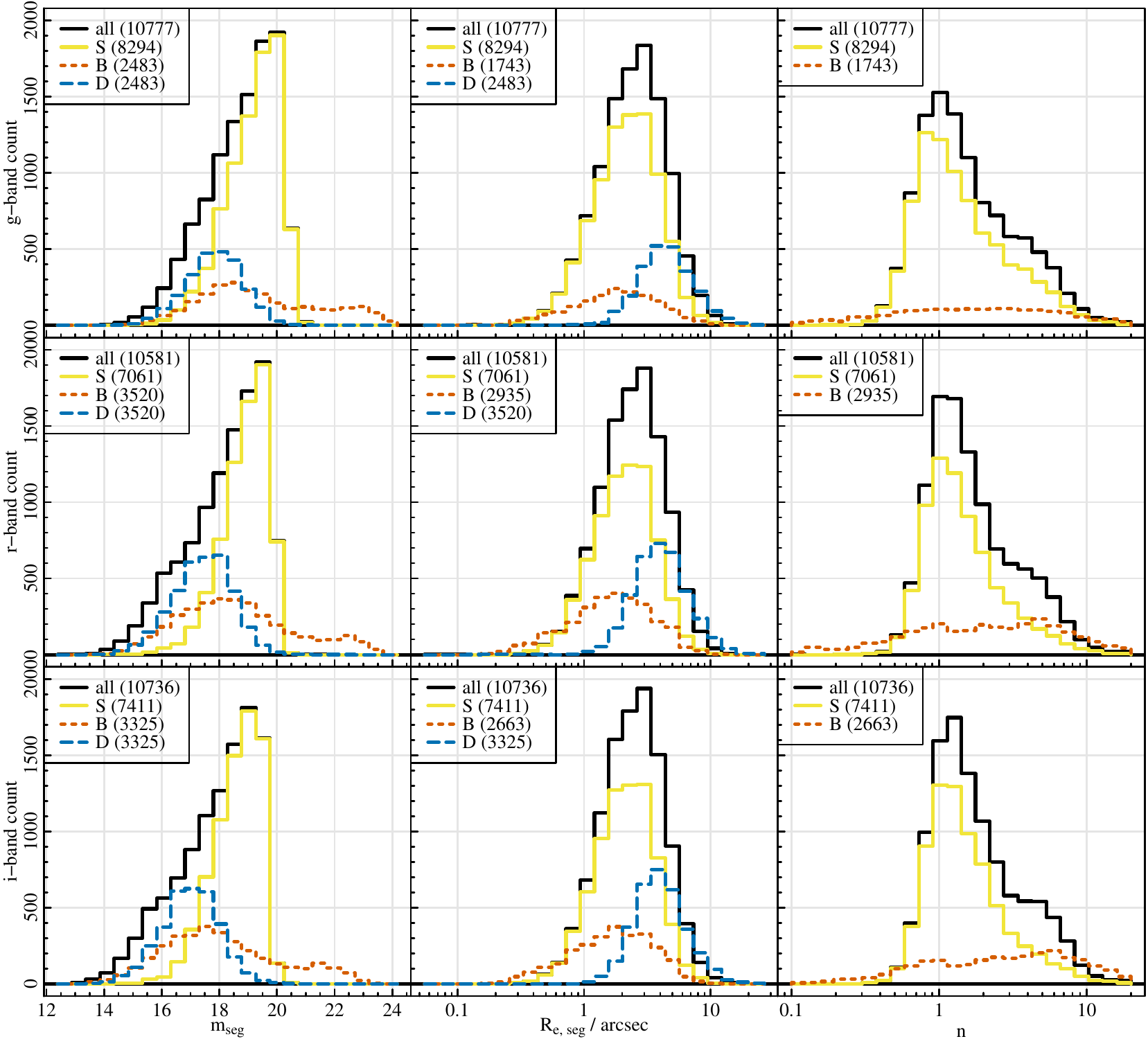}
    \caption{The distribution of the main parameters (limited to segment radii) for all bands and models. Left, middle and right columns show magnitude, effective radius and S\'ersic index while top, middle and bottom rows show the $g$-band, $r$-band and $i$-band respectively. The solid yellow lines are the single S\'ersic values for those galaxies which were classified as single component systems, dotted red and dashed blue lines show bulges and disks, respectively, for those objects classified as 1.5- or double component systems. For reference the solid black line shows the single S\'ersic fits for all galaxies with \texttt{NCOMP}\,>\,0 (i.e. including those classified as 1.5- or double component systems). The number of objects in each histogram is given in the legends, where the number of bulges and disks differs for effective radii and S\'ersic indices because these parameters do not exist for 1.5-component fits (point source bulge). We do not show disk S\'ersic indices since they were fixed to 1 (exponential).}	
    \label{fig:resultshists}
\end{figure*}

Figure~\ref{fig:resultshists} shows the distribution of the main parameters - magnitude, effective radius and S\'ersic index - in all three bands ($g$, $r$ and $i$) for single S\'ersic fits, bulges and disks. The single S\'ersic fit distributions are shown for all galaxies with \texttt{NCOMP}\,>\,0 (i.e. all non-outliers) in black and for those galaxies which were actually classified as single component systems (\texttt{NCOMP}\,=\,1) in yellow. Red dotted and blue dashed lines show bulges and disks, respectively. For disks we show the 1.5-component fits and double component fits combined (i.e. the 1.5-component parameters for objects with \texttt{NCOMP}\,=\,1.5 and double component parameters for those with \texttt{NCOMP}\,=\,2 added into one histogram); the S\'ersic index is not shown since it was fixed to 1. Bulge magnitudes are also shown for 1.5- and double component fits combined; effective radii and S\'ersic indices are only shown for the double component fits since they do not exist in the point source model. The legend indicates the numbers of objects in each histogram, which can also be inferred from Table~\ref{tab:results}. Magnitudes and effective radii are truncated at the segment radii which we found to give more robust results than using the S\'ersic values extrapolated to infinity (see Sections~\ref{sec:tightsegments}, \ref{sec:comparelee} and \ref{sec:systematics}). 

The first thing apparent from Figure~\ref{fig:resultshists} is that the distributions in the three bands are generally very similar, which is reassuring given that the fits were performed independently. Looking at the first column, the single S\'ersic number counts increase up to a sharp drop just before 20\,mag in all bands, which is not surprising given the GAMA survey limit of 19.8\,mag. The faintest of these objects are all classified as single component galaxies (yellow line is on top of black line), while some of the brighter objects are successfully decomposed into bulges and disks. Disks are generally slightly brighter than bulges. The bulges show a second, smaller peak at very faint magnitudes which we found to be the ones from the 1.5-component fits (unresolved, faint bulges). There is a slight trend for magnitudes to become brighter moving from $g$ to $r$ to $i$ for all components, as expected from typical galaxy colours. We investigate the colours further in Section~\ref{sec:colours}. 

From the middle panels it becomes obvious that bulges tend to be smaller than disks by a factor $\sim$2, while single S\'ersic fits span a wide range of sizes. Similar to the trend observed in the magnitudes, the smallest objects are classified as single component systems, while some of the larger galaxies can be successfully decomposed. 

The S\'ersic indices of single component systems show a clear peak around a value of 1 (exponential), a sharp drop-off at lower values and a longer tail towards higher values. Interestingly, the single S\'ersic distributions showing all systems (black lines) have a secondary ``bump" around a value of 4 or 5 (classical de Vaucouleurs bulge), which is not apparent in those galaxies classified as single component systems (yellow line). Hence most of those high S\'ersic index objects were indeed found to contain bulges and were classified as double component systems. The bulges themselves show a wide range of S\'ersic indices with (at least in $r$ and $i$ bands) a slightly double-peaked nature around values of 1 and 4-6. At this point, we would like to remind the reader that we use the term ``bulge" to refer to all kinds of central components of galaxies, including classical bulges, pseudo-bulges, bars and AGN (cf. Section~\ref{sec:inputsandmodels}). Hence the ``bulge" distribution will include a variety of physical components and their combinations, leading to the wide spread of values. In addition, the S\'ersic index tends to be the parameter with the largest uncertainty, with typical galaxies showing relative errors on their bulge S\'ersic index of 1-10\%, adding further scatter to the distribution. \\

\begin{figure}
    \includegraphics[width=\columnwidth]{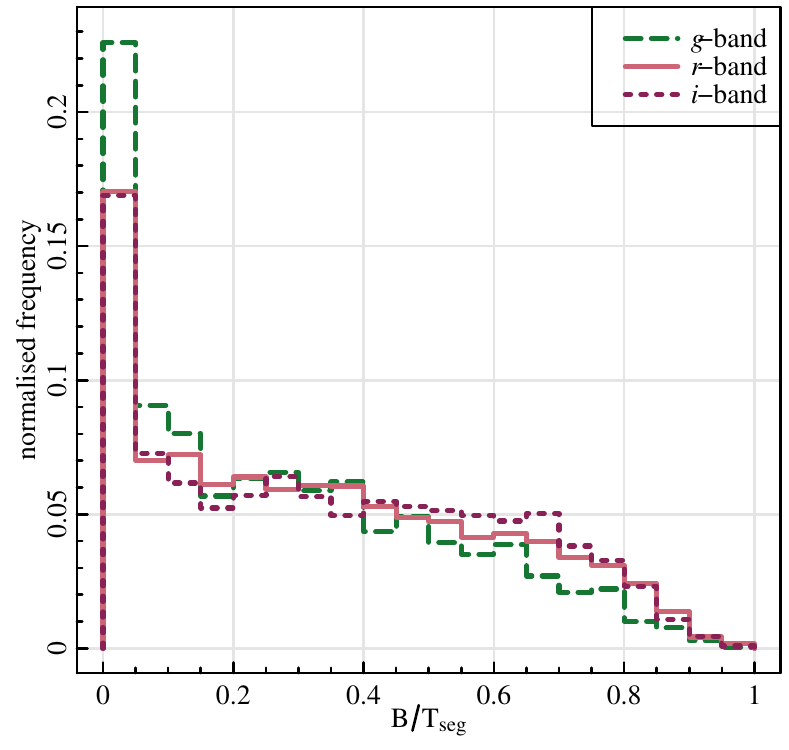}
    \caption{The distribution of the bulge to total flux ratio (limited to segment radii) for the 1.5- and double component fits in all bands. Dashed green, solid light red and dotted dark pink lines refer to the $g$, $r$ and $i$ bands, respectively. The histograms have been normalised by their respective total number of fits (cf. Figure~\ref{fig:resultshists}) to make the bands directly comparable.}	
    \label{fig:BThist}
\end{figure}

Since the bulge to total flux ratio is a derived parameter that is frequently of interest, we additionally show it in Figure~\ref{fig:BThist} for all three bands; for those galaxies that were classified as a 1.5- or double component fit in the respective band. The majority of systems have intermediate values of B/T with only a few percent at the extreme end above 0.8. The secondary peak at very low B/T values around 0.02 stems from the 1.5-component fits. The B/T ratio generally increases from $g$ to $r$ to $i$, as expected (see Section~\ref{sec:prevcols}). 

\begin{figure}
    \includegraphics[width=\columnwidth]{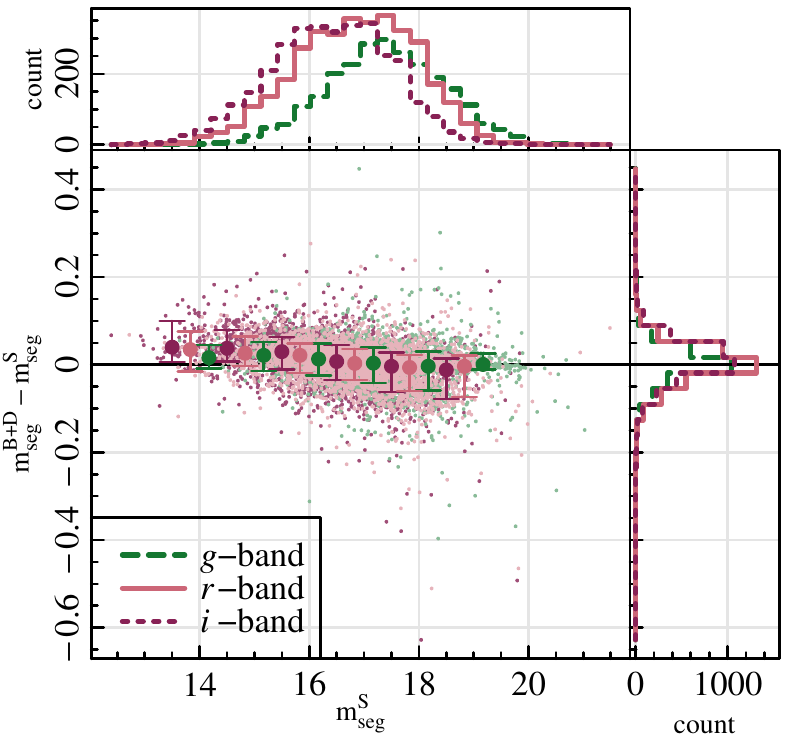}
    \caption{The difference between the single S\'ersic magnitude and the total magnitude derived from the double or 1.5-component fits for those galaxies that were classified as such, for all bands (all magnitudes limited to segment radii). The scatter plot shows the difference between the two magnitudes against the single S\'ersic magnitude for all three bands with the running medians and $1\sigma$-percentiles overplotted. The top and right panels show the respective marginal distributions. Dashed green, solid light red and dotted dark pink lines refer to the $g$, $r$ and $i$ bands respectively.}	
    \label{fig:magrecovery}
\end{figure}

Finally, as a first consistency check, we show the difference between the single S\'ersic magnitude and the total magnitude derived from the double or 1.5-component fits, all limited to segment radii, in Figure~\ref{fig:magrecovery}. The distributions for all three bands are highly peaked around zero, with the vast majority of objects having total magnitudes consistent with the single S\'ersic magnitudes within 0.1\,mag (over the entire magnitude range). We only show galaxies that were classified as 1.5- or double component fits here to ensure reliable bulge and disk magnitudes, but note that the distribution is very similar when including objects classified as single S\'ersic fits. This indicates that the total magnitude is well-constrained even in the case when the individual component magnitudes are not (see also Section~\ref{sec:colours}).

\subsection{Galaxy and component colours}
\label{sec:colours}
\begin{figure*}
    \includegraphics[width=\textwidth]{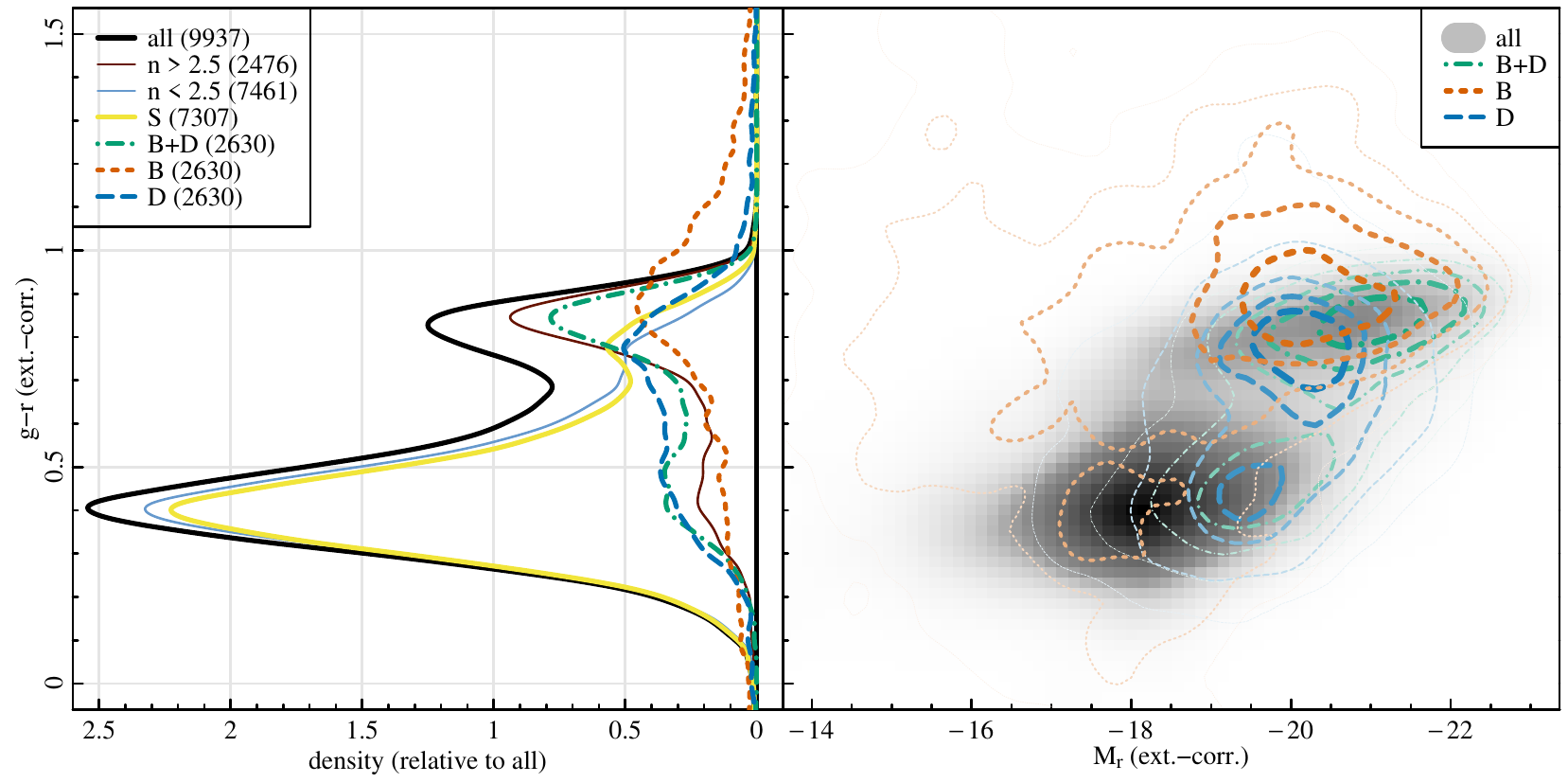}
    \caption{\textbf{Left panel:} The Galactic extinction-corrected $g-r$ colour distributions (limited to segment radii) for galaxies and their components. The colour coding of the lines is the same as for Figure~\ref{fig:resultshists}, although with a few additions: The solid black line shows single S\'ersic fits for all galaxies with \texttt{NCOMP}\,>\,0 in the joint model selection; the thinner dark red and light blue solid lines split this sample into those with $n>2.5$ and $n<2.5$ in the $r$-band. The solid yellow line gives the single S\'ersic values for those galaxies which were classified as single component systems, dotted red and dashed blue lines show bulges and disks, respectively, for those objects classified as 1.5 or double component systems (always in the joint model selection). The dot-dashed green histogram gives the total galaxy colour (derived from the addition of bulge and disk flux) for 1.5 and double component systems. The number of objects in each histogram is given in the legend.
\textbf{Right panel:} The Galactic extinction-corrected $g-r$ vs. $M_r$ colour-magnitude diagram (limited to segment radii) for galaxies and their components. The colour coding of the lines is the same as for the left panel. Contours include 10, 25, 50, 75 and 90\% of the sample.}	
    \label{fig:colourplots}
\end{figure*}

The left panel of Figure~\ref{fig:colourplots} shows the distribution of $g-r$ colours for galaxies and their components. The colours are corrected for Galactic extinction, but not for dust attenuation in the emitting galaxy. The Galactic extinction was obtained from v03 of the \texttt{GalacticExtinction} catalogue accompanying the equatorial input catalogue on the GAMA database. We focus on $g-r$ colours here since those bands are the deepest KiDS exposures. Results for $g-i$ are qualitatively similar, albeit a bit more noisy. 

The solid black line shows the colour distribution for all single S\'ersic fits that were not classified as outliers in the joint model selection (\texttt{NCOMP}\,>\,0). It is clearly bimodal, with redder colours typically belonging to higher S\'ersic index objects as indicated by the thinner dark red and light blue lines splitting the distribution at $n=2.5$ (in the $r$-band). Not entirely surprisingly (given the distribution of S\'ersic indices in Figure~\ref{fig:resultshists}), the distribution of single S\'ersic objects actually classified as such (\texttt{NCOMP}\,=\,1, solid yellow line) mostly follows the distribution of low S\'ersic index objects; while the high S\'ersic index objects tend to be classified as double component systems. For the latter, we show total colours with a dash-dotted green line, bulge colours with a dotted red and disk colours with a dashed blue line. As expected, bulges tend to be redder than disks, although the scatter is large. \\

The right panel of Figure~\ref{fig:colourplots} shows the corresponding colour-magnitude diagram. Colours and absolute magnitudes are both corrected for Galactic extinction but not for dust attenuation in the emitting galaxy. The absolute magnitude was calculated using the distance modulus provided in v14 of the \texttt{DistancesFrames} catalogue from the GAMA database which we also used to obtain redshifts for the sample selection. Again, we focus on $g-r$ colour and absolute $r$-band magnitude, $M_r$, but note that results using $g-i$ and/or $M_i$ are qualitatively similar.

The grey density plot in the background shows the single S\'ersic fits for all non-outlier (\texttt{NCOMP}\,>\,0) galaxies, corresponding to the black line in the left panel of Figure~\ref{fig:colourplots}. The bimodality of the distribution is even clearer here, with the red sequence and blue cloud being well-separated. The green contours indicate the part of the sample that was classified as 1.5 or double component object\footnote{To be precise, the green contours were derived by adding the respective bulge and disk fluxes of the 1.5 or double component objects (for consistency with the bulge and disk contours), while the grey density plot is based on the single S\'ersic fits (for robustness at low magnitudes). These two versions of the total galaxy magnitude are generally very similar as evidenced by Figure~\ref{fig:magrecovery}.}: as expected, this is concentrated towards the bright end of the galaxy distribution and hence encompasses mostly galaxies located in the red sequence. Correspondingly, bulges and disks are both relatively red, with bulges on average slightly redder than the total galaxies and disks slightly bluer (while both components - obviously - are fainter than the total galaxy). However, both components show a large scatter and overlap with each other: both faint blue bulges exist as well as bright red disks. 

A detailed study of component colours and the different populations in the right panel of Figure~\ref{fig:colourplots} is beyond the scope of this work. However, we note that the total galaxy colours show much less scatter; indicating that the scatter results from a different splitting of the light into bulge and disk components in the $g$ and $r$ bands, while the total amount of light is well-constrained (cf. also Figure~\ref{fig:magrecovery}). A further brief investigation into extreme systems (blue bulges with red disks and also excessively red bulges with very blue disks) suggests that they are caused by a variety of remaining uncertainties in our analysis, e.g. swapped components in one of the two bands (Section~\ref{sec:componentswapping}), small faint bulges that are barely detected in the $r$-band and missed in the $g$-band, the ``bulge" component dominating both small and large radii in one of the two bands (cf. Section~\ref{sec:modelselectioncaveats}) or failures in the flagging of bad fits, all combined with model selection uncertainties and the necessity of joint model selection to compromise between the bands. While each of these processes by itself only affects a small number of galaxies, in sum across both bands they do reach the 10-20\% level. Still, on average our colours do follow the expected trends, as we show in Section~\ref{sec:prevcols} with an overview of similar studies in the literature. We will study the colours of galaxies and their components in more detail in forthcoming work, also including further bands (KiDS $u$ and VIKING $Z, Y, J, H, K_s$) and taking full account of inclination effects due to dust in the emitting galaxies \citep[see, e.g.][]{Driver2008}. We will then also assess trends in other parameters, such as the component effective radii, with wavelength.

\subsection{Catalogue limitations}
\label{sec:cataloguelimitations}
We finish this section by pointing out a few limitations of our results that users of the catalogue should be aware of. 

\subsubsection{Model limitations}
All of our models are axially symmetric and monotonically decreasing in intensity from the centre. We are unable to capture asymmetries such as spiral arms, offset bulges, tidal tails, mergers, star-forming regions etc.; or disk features such as rings, bumps, truncations or flares. If such features are present in the data, they may bias or skew the model parameters. We also remind the reader that when we talk about ``bulges", what we really mean are the central components. This could be a classical bulge, a pseudo-bulge, an AGN, a bar, or any combination (sometimes resulting in the model trying to fit a mixture between e.g. a bar and a bulge). We make no attempt to distinguish between these cases. 

\subsubsection{Model selection caveats}
\label{sec:modelselectioncaveats}
Model selection is accurate to $>90\%$ compared to what could be achieved
by visual classification (Section~\ref{sec:modelselection}). However, it is important to note that our aim in the model selection is to determine which one of our three models is most appropriate to use for the given data; and not how many physically distinct components an object consists of. The reason for this is that for a given galaxy, the data quality will strongly influence how many fitting parameters can be meaningfully constrained and using more model parameters will inevitably overfit the data and lead to unphysical results. Hence, even in the joint model selection, we base our visual classification on the fit and residuals in individual bands (which is what we fit to), rather than e.g. colour images. Due to the different depths and resolutions of the KiDS bands, it is hence common for the same galaxy to be classified as double component in one band, but single component in another.  

In an attempt to make fitting parameters more directly comparable across bands, we introduced the joint model selection (Section~\ref{sec:modelselection}), yet this is necessarily a compromise between the different bands. For example, we lose bulges that are resolved in the $r$-band but not in $g$ and $i$ due to the larger PSFs; or there may be some ill-constrained $i$-band fitting parameters for an extended low-surface brightness disk that is visible in $r$- and $g$- but not in the shallower $i$-band image. There are also more skipped fits and outliers in the joint model selection than in the band-specific ones because all objects that are skipped or flagged in at least one of the three bands is skipped or flagged in the joint model selection. \\

Irrespective of the result of the model selection, we provide all fitted parameters of all models in the catalogue (along with the postage stamps of all fits and a flag indicating the preferred model). This allows users to perform their own selection if desired; but also requires care as not all provided parameters will be meaningful. While single S\'ersic fits to double component objects are mostly reasonable; double component fits to true single component galaxies will have unconstrained and hence potentially unphysical parameters for at least one of the components.

We are also aware of a population of objects that are classified as double component fits but have the bulge component dominating both the centre and the outskirts, with the disk only dominating at intermediate radii or even staying ``below" the bulge at all radii. We believe these are essentially single component systems that do not follow a S\'ersic law (e.g. S\'ersic index would be higher at centre than outskirts); and so the freedom of the disk is used to offset this. This population is easily identifyable by the high bulge-to-total ratio (B/T\,$\gtrsim$\,0.6 or 0.7). The single S\'ersic fits may be more appropriate to use in these cases \citep[see also the discussion of this issue in][]{Allen2006}. An example is shown in Figure~\ref{fig:examplefithighbt} in the appendix. 

\subsubsection{Drawbacks of tight fitting segments}
As detailed in Section~\ref{sec:tightsegments}, we use relatively tight segments around the galaxies for fitting, which results in the best possible fit of the inner regions of the galaxy but can lead to large, unphysical wings. Hence we recommend using only integrated properties, i.e. the summed flux/magnitude within the region that was fitted and the corresponding effective radii and bulge-to-total ratios as given by the corresponding *\_SEGRAD properties in the catalogue. For comparisons to other catalogues using larger fitting segments, their profiles should also be appropriately truncated (see details in Sections~\ref{sec:tightsegments} and \ref{sec:comparelee}). 

\subsubsection{Sources of systematic uncertainties}
We provide errors for each fitted parameter in the catalogue including our best estimate of systematic uncertainties taken from Table~\ref{tab:errorunderestimate}. However, we do not apply the (small) bias corrections given in the same table, since they are only applicable to large random samples of our galaxies and not to individual objects. In addition, we would like to point out that the systematic errors were estimated from single S\'ersic $r$-band fits. We expect that individual components as well as the $g$ and $i$ bands are affected by similar systematics, but we did not test for this. Also, there are some systematic uncertainties that we do not account for in our simulations, most obviously galaxy features that cannot be captured by our models. For these reasons, the given errors should still be considered as lower limits of the true errors. 

\subsubsection{GAMA-KiDS RA/Dec offset}
We observed an average offset between the input and output (fitted) positions of galaxies in both RA and Dec of approx. 0.4 pix ($0\farcs$08). This is due to an offset between the GAMA (SDSS) and KiDS positions; the same offset can
be seen when comparing the KiDS source catalogue with the Gaia catalogue; see also Figure~15 in \citet{Kuijken2019}. We correct for this during the outlier rejection, but give the original (uncorrected) fitted values for position otherwise.

\subsubsection{Completeness limits}
Due to our sample selection (Section~\ref{sec:sampleselection}), our spectroscopic completeness is 100\% and even the faintest objects in our sample are well-resolved and bright enough in KiDS data to allow for robust single S\'ersic fits. However, there is a systemic limit to the component magnitude in that the samples of bulges and disks with magnitudes fainter than the GAMA limit ($r<$19.8\,mag) are incomplete. For example, a bulge with a magnitude of 22\,mag in the $r$-band will only be contained in our sample if the corresponding disk is bright enough such that the total magnitude is below 19.8\,mag; hence the sample of bulges with 22\,mag is incomplete. This applies almost exclusively to the faint bulges from the 1.5-component fits as can be seen in the first column of Figure~\ref{fig:resultshists}.

\section{Comparison to previous works}
\label{sec:prevwork}
After presenting some of the contents of our main catalogue, we now turn towards demonstrating its robustness. We start with a comparison to previous works in this section (including work on the same galaxy sample); then describe additional internal consistency checks and a detailed study of biases and systematic errors with bespoke simulations in Section~\ref{sec:simulations}.

\subsection{Comparison of catalogue statistics}
As a first check, we compare our model selection statistics to those of other bulge-disk decomposition works, although care must be taken in judging these results since they will depend on the sample selection, data quality and observational band. 

\begin{table*}
	\centering
	\caption{Comparison of our catalogue statistics to previous works (in the $r$-band unless stated otherwise).}
	\label{tab:stats}
	\setlength{\tabcolsep}{4pt}
	\begin{tabu}{lcccp{.48\textwidth}} 
	    \hline
	    reference & single (\%) & double (\%) & unsuitable (\%) & notes
	    \\
	    \hline
	    this work & 47 & 23 & 30 & double including 1.5-comp.; unsuitable including masking (20\%)\\
	    \citet{Barsanti2021} & 47 & 28 & 25 & cluster S0 galaxies\\
	    \citet{Meert2015} & 44 & 39 & 17 & larger sample up to higher redshift but smaller magnitude range\\
	    \citet{Head2014} & 19 & 35 & 46 & $g$-band; early-type sample; more stringent criteria for ``good" fits\\
	    \citet{Lackner2012} & 35 & 29 & 36 & single corresponding to pure exponential or de Vaucouleurs; unsuitable corresponding to their ``S\'ersic" category\\
	    \citet{Simard2011} & 73 & 26 & 1 & unsuitable corresponding to failure rate of fitting routine; no selection of ``good" fits given\\
	    \citet{Allen2006} & 43 & 34 & 23 & $B$-band; unsuitable galaxies excluded through cuts in redshift, galaxy size and surface brightness \\
	    \hline	   
	\end{tabu}
 \end{table*}

Table~\ref{tab:stats} summarises the corresponding percentages including a few notes on the most important differences of the quoted works to ours (more details on the majority of these studies are given in Section~\ref{sec:prevcols}). In short, for the automated decomposition of large samples of galaxies in the $r$-band, most authors - including ourselves - class roughly half of all galaxies as being well-represented by a single S\'ersic model, with the other half split approximately evenly into double component fits and objects unsuitable for fitting with such simple models. 

\subsection{Comparison of component colours to literature}
\label{sec:prevcols}
$g-r$ colours of galaxy components, such as those we present in Figure~\ref{fig:colourplots} and Section~\ref{sec:colours}, are not found frequently in the literature, although a number of authors have presented bulge-disk decompositions in several bands. For example, \citet{Simard2011} perform bulge-disk decompositions for a large sample of galaxies in the SDSS $g$ and $r$ bands but only present colour-magnitude diagrams for the total galaxies (their figures 9 and 10). These are visually comparable to our total galaxy colours as indicated by the dot-dashed green contours in the right panel of Figure~\ref{fig:colourplots}. \citet{Mendel2014} add the SDSS $u$, $i$ and $z$ bands to the analysis of \citet{Simard2011} and present component masses in $ugriz$ but also do not study component colours. 

Similarly, \citet{Meert2015} present a large $r$-band catalogue which is extended to include the $g$ and $i$ bands in \citet{Meert2016}. Colour-magnitude diagrams, however, are again only presented for total galaxies, with the authors noting that component colours can be calculated from their catalogue but should be used with care since they are subject to large uncertainites. 

More recently, \citet{Dimauro2018} provide (UVJ) component colours in their catalogue but defer their study to future work; while \citet{Bottrell2019} present $ugriz$ colour-magnitude diagrams for total galaxies (colour-coded by B/T); but again not for individual components. \\

Among the first to show component colours for a large sample of galaxies were \citet{Lackner2012} in their study of $\sim$70000 $z<0.05$ SDSS galaxies in the $g$, $r$ and $i$ bands. However, in contrast to our fits, their $g$- and $i$-band fits are not independent. Instead, in order to decrease the noisyness of the colours, the structural parameters are taken from the $r$-band and only the magnitude is adjusted. Additionally, \citet{Lackner2012} (along with e.g. \citealt{Mendel2014}) fix the S\'ersic index of the bulge to either 1 or 4 for their double component fits to limit the number of free parameters since the data is insufficient to constrain the bulge light profile. Keeping these differences in mind, their figure~32 showing the $g-r$ vs. $M_r$ colour-magnitude diagram for bulges and disks as contours superimposed on the greyscale background for all galaxies can be compared to the right panel of our Figure~\ref{fig:colourplots} (for a more detailed description of Figure~\ref{fig:colourplots}, see Section~\ref{sec:colours}). In general, both plots are very similar: the grey background shows a large blue cloud and a well-separated red sequence. The double component fits populate the red sequence, green valley and the brighter part of the blue cloud. The bulges tend to be slightly redder than the red sequence but with a large scatter especially at the faint end. Disks spread from the red sequence towards the green valley with a smaller population also in the blue cloud.\footnote{For reference, their cyan contours represent 6684 galaxies with a bulge S\'ersic index of 1 and the magenta contours show 14042 objects with a bulge S\'ersic index of 4. Also note that their $x$-axis is reversed with respect to ours.} \citet{Lackner2012} also note the large scatter in colour for bulges in particular, despite their fitting constraints and lower reshift limit. Hence it is not surprising that even with the higher quality KiDS data and our new fitting routines, we get a large scatter in component colours, especially since we leave the bulge S\'ersic index free and perform independent fits in both bands. The latter in particular can lead to very extreme colours since it is not guaranteed that the ``bulge" and ``disk" models actually fit the same features in both images (in particular when there are additional features present that are not fully captured by the models; see also Section~\ref{sec:colours}). 

\citet{Kim2016} found similar difficulties when performing $g$-band and $r$-band decompositions on $\sim$10000 large bright and approximately face-on SDSS galaxies. While they leave the S\'ersic index free as we do, the $g$-band structural parameters are again taken from the $r$-band fits with only the magnitudes adjusted. Despite this, they find it necessary to remove almost 40\% of their sample after fitting because they show excessively red bulge colours (combined with low B/T values in the $r$-band). After this cut, the $g-r$ vs. $M_r$ colour-magnitude diagram for bulges shown in their figure 7 is slightly less noisy than ours (Figure~\ref{fig:colourplots}), although still comparable. \citet{Kim2016} did not study the properties of the disks in their sample. 

One of the most direct comparisons to make is with \citet{Kennedy2016} who study GAMA galaxies in the G09 region (a subset of our sample) in the $ugrizYJHK$ bands from the SDSS \citep{York2000} and the United Kingdom Infra-red Telescope Infrared Deep Sky Survey (UKIDSS, \citealt{Lawrence2007}). They use the \texttt{MEGAMORPH} multi-band fitting method with \texttt{GALAPAGOS-2} and \texttt{GALFITM} \citep{Haeussler2013, Vika2013} to perform simultaneous S\'ersic plus exponential fits across all 9 bands. The structural parameters are constrained to be the same in all bands, with only the component magnitudes allowed to vary freely, therefore providing robust colours. While the paper focusses on studying $u-r$ colours, the corresponding catalogue on the GAMA database (\texttt{MegaMorph:MegaMorphCatv01}) contains the information for the fits in all 9 bands such that $g-r$ colours can easily be derived. 

\begin{figure}
    \includegraphics[width=\columnwidth]{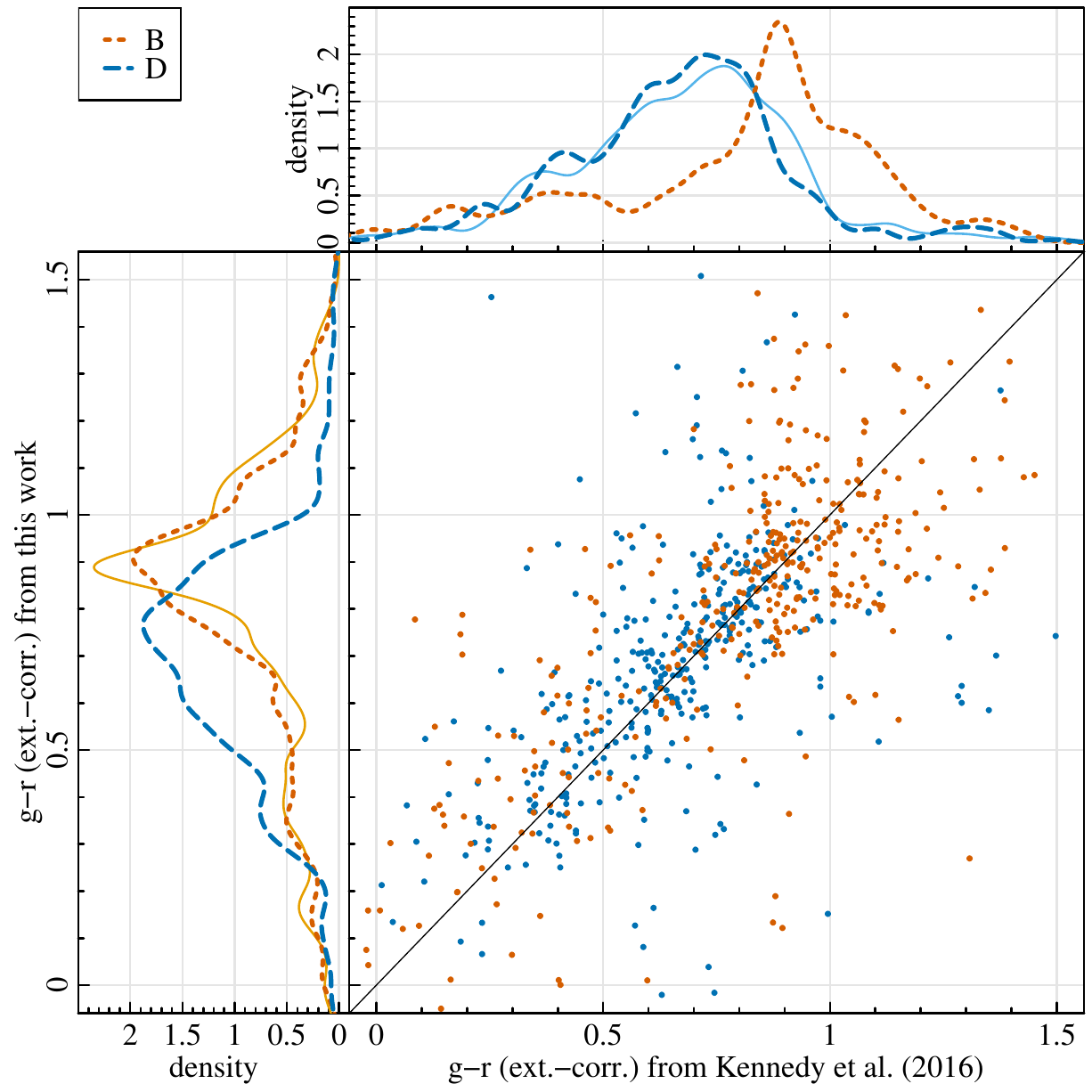}
    \caption{Our Galactic extinction-corrected $g-r$ component colours (limited to segment radii) compared against those from \citet{Kennedy2016} for a subsample of 390 objects that appear in both catalogues and were classified as double component fits in the joint model selection. The scatter plot shows the direct comparison, while the density plots show the respective distributions in both catalogues (ours on the left, \citealt{Kennedy2016} on the top). Bulges are again shown in red with dotted lines and disks in blue with dashed lines. To aid the direct comparison of the distributions, the lighter solid lines also show the \citet{Kennedy2016} bulge distribution on the left and our disk distribution on the top.}	
    \label{fig:colourvskennedy}
\end{figure}

This comparison is shown in Figure~\ref{fig:colourvskennedy} for those galaxies that were present in both catalogues and classified as double component fits (\texttt{NCOMP}\,=\,2) in the joint model selection of our fits (\citealt{Kennedy2016} do not perform model selection nor outlier rejection). In addition to the scatter plot directly comparing the component colours, we show the corresponding distributions in the left (this work) and top \citep{Kennedy2016} panels of Figure~\ref{fig:colourvskennedy}. As always, bulges are shown in red (points and dotted lines) and disks in blue (points and dashed lines). To aid the direct comparison of the distributions, we additionally show the \citet{Kennedy2016} bulge colour distribution in the left panel as a solid orange line and the disk distribution from this work in the top panel as a light blue solid line. Component colours are all corrected for Galactic extinction and limited to segment radii for our fits.

Despite the large scatter, it can be seen that our component colours are generally in agreement with those from \citet{Kennedy2016} with no systematic differences. The scatter in both catalogues is also comparable, although \citet{Kennedy2016} perform multi-band simultaneous fits with fixed structural parameters that should lead to more robust component colours. This advantage of their work seems to be balanced by advantages of our work, such as the improved data quality of KiDS, the robustness of the fitting procedure with \texttt{ProFound} and \texttt{ProFit} and our post-processing steps (in particular outlier rejection and model selection). \\

In addition to the large $g-r$ component colour studies discussed above, there are a number of publications focussing on the $g-i$ colours of bulges and disks for samples ranging between $\sim$100 and $\sim$1000 objects (i.e. roughly a factor of 10 smaller than ours), namely \citet{Gadotti2009, Head2014, Vika2014, Fernandez-Lorenzo2014, Cook2019, Barsanti2021}. We briefly compare our work to their results here, noting that all above authors have more stringent constraints on their fits than we do and also report problems in deriving bulge colours. For example, \citet{Fernandez-Lorenzo2014}, although they fit the galaxies in both bands, use fixed aperture photometry to derive more stable bulge colours. \citet{Vika2014}, while performing \emph{simultaneous} multi-band fits, do not allow for any variation of structural parameters (except magnitudes) with wavelength. \citet{Head2014}, in addition to varying magnitudes, allow for a trend in disk sizes with wavelength in approximately 30\% of their sample, noting that this leads to increased scatter. \citet{Cook2019}, who use \texttt{ProFit} like this work, allow disks to deviate slightly from the exponential profile but fix all bulges to be exactly round (axial ratio of 1), again only allow magnitudes and disk sizes to vary between bands and employ a sophisticated, visually-guided re-fitting procedure to obtain physically meaningful fits for ``difficult" objects. \citet{Barsanti2021}, also employing \texttt{ProFit}, additionally allow for differing bulge sizes and S\'ersic indices in the different bands (but fixing bulge and disk axial ratios and position angles and performing model selection in the $r$-band only), but class approximately half of their double component fits as ``unreliable". \citet{Gadotti2009}, fitting bulges, bars and disks to a sample of face-on, visually-selected ``well-behaved" galaxies refrain from automated fitting and instead treat each galaxy individually. 

After these notes on the inherent difficulties associated with deriving component colours, we can now turn to the corresponding results: \citet{Head2014}, in their study of early-type red-sequence galaxies in the Coma cluster, measure an average $g-i$ difference between bulges and disks of 0.09$\pm$0.01\,mag. Similarly, \citet{Barsanti2021} find a bulge-disk $g-i$ difference of 0.11$\pm$0.02\,mag for their sample of S0 cluster galaxies. \citet{Fernandez-Lorenzo2014}, on the other hand, have a sample of mostly late-type spirals (with B/T$<$0.1 for $\sim$66\% of their objects) and find a difference of 0.29\,mag in the median $g-i$ bulge and disk colours, i.e. a factor of $\sim$3 larger. In line with this, \citet{Vika2014} report that the bulge and disk colours are similar for early-type galaxies but differ significantly for late-types. The $g-i$ differences for the different morphological classes given in their table 2 range from 0.03$\pm$0.04\,mag for ellipticals to 0.28$\pm$0.06\,mag for late-type spirals; with the overall average (comprising approximately two thirds late-types) being 0.19$\pm$0.04. Similarly, the average $g-i$ colour difference of the \citet{Gadotti2009} sample of varying galaxy types amounts to 0.18$\pm$0.04 (from the online-version of their table~2). 

Our results are perfectly in line with this: the median bulge-disk $g-i$ colour difference for our 1.5- or double component fits is 0.17$\pm$0.01\,mag, consistent with the \citet{Vika2014} and \citet{Gadotti2009} results. Limiting to objects with a total $g-i>1$ (red-sequence galaxies) reduces the value to 0.14$\pm$0.01\,mag; while focussing on 2-component fits only (excluding 1.5-component fits) yields 0.10$\pm$0.02\,mag, in agreement with \citet{Head2014} and \citet{Barsanti2021}. This is because our double component galaxies lie predominantly on the red sequence, as can be seen in Figure~\ref{fig:colourplots}. 1.5-component fits on the other hand, have very small (namely unresolved) bulges by definition and hence belong to the class of late-type spirals. In fact, 87\% of the 1.5-component objects have a ($g$-band) B/T ratio less than 0.1, with the median value as low as 0.02. Computing the bulge-disk $g-i$ colour difference for this sample of objects yields a value of 0.46$\pm$0.02, suggesting that the trend described in \citet{Vika2014} continues at very low B/T. \\

From all of these comparisons we conclude that our component colours - although noisy - are in line with previous work. In order to increase the colour robustness while preserving the ability to capture physical trends with wavelength (i.e. not fixing the structural parameters to be the same in all bands), a simultaneous fit in all bands is needed. This has many advantages as shown by the \texttt{MEGAMORPH} project team using \texttt{GALAPAGOS} and \texttt{GALFITM} (\citealt{Haeussler2013, Vika2013, Haeussler2022}), especially for automated analyses, since it naturally ensures smooth wavelength trends while preserving physical variation and additionally allows more robust fits to fainter magnitudes. With \texttt{ProFit} v2.0.0, released in February 2021, now supporting a multi-band fitting mode, this is certainly an interesting avenue to explore in future work and could provide a valuable alternative.

\subsection{Comparison to size-stellar mass relations of \citet{Lange2015}}
\label{sec:sizemass}
\begin{figure}
    \includegraphics[width=\columnwidth]{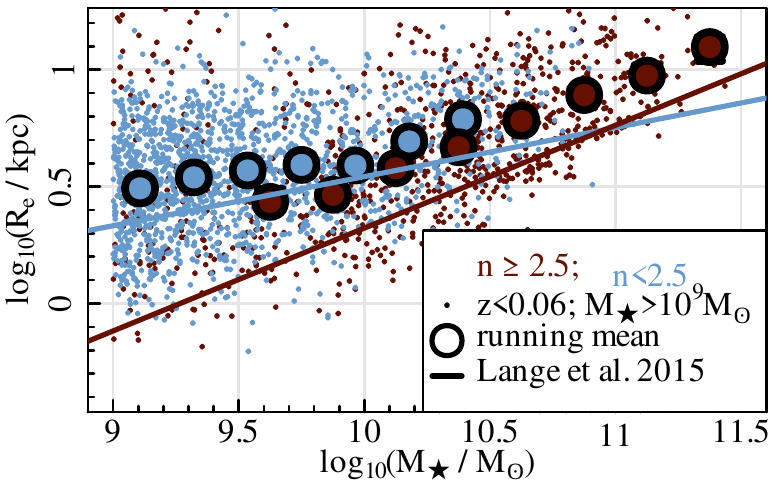}
    \includegraphics[width=\columnwidth]{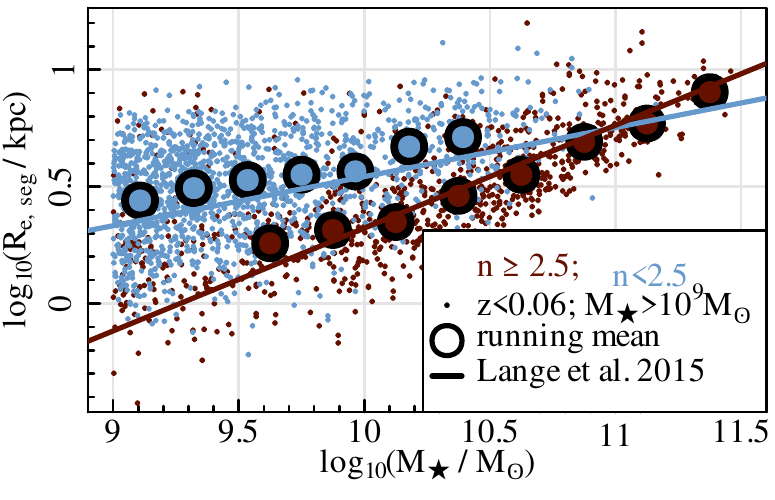}
    \caption{The size-stellar mass relation for our $r$-band fits (dots) compared to the \citet{Lange2015} fits (lines). The sizes are obtained from our single S\'ersic effective radii (top panel: extrapolated to infinity; bottom panel: limited to segment radius) and the distance moduli provided in the \texttt{DistancesFrames} catalogue originally described by \citet{Baldry2012}. The stellar masses are taken from the most recent version of the \texttt{StellarMasses} catalogue initially presented in \citet{Taylor2011}. The sample is limited to the redshift range $0.0001<z<0.06$ (redshifts also from \texttt{DistancesFrames}) and the stellar mass range $M_*>10^9\,M_\odot$. Large circles with error bars indicate the running median with its error (usually smaller than the data point). Solid lines show the single exponential $M_*-R_e$ relation fits obtained by \citet{Lange2015} for their single component $r$-band sample, split by a S\'ersic index cut at $n=2.5$ (taken from their tables~2 and 3).}	
    \label{fig:sizemass}
\end{figure}

Figure \ref{fig:sizemass} shows the size-stellar mass relation obtained from our $r$-band single S\'ersic fits in combination with the redshifts and distance moduli of v14 of the \texttt{DistancesFrames} catalogue \citep{Baldry2012} and v19 of the \texttt{StellarMasses} catalogue \citep{Taylor2011}; both from the GAMA database. The aperture-derived stellar masses have been scaled to match the S\'ersic total flux using the \texttt{fluxscale} keyword provided in the \texttt{StellarMasses} catalogue. The $g$-band and $i$-band results are very similar to those from the $r$-band and hence we do not show them here. 

The sample is limited to objects which were not flagged during our outlier rejection (Section~\ref{sec:flaggingofbadfits}) and split into early- and late-type galaxies according to our fitted S\'ersic index ($n\lessgtr2.5$; analogous to \citealt{Lange2015}). We also limit the redshift range to $0.0001<z<0.06$ and the stellar mass range to $M_*>10^9\,M_\odot$, thus avoiding the need for volume corrections. For comparison, we show the $M_*-R_e$ relations obtained by \citet{Lange2015} by fitting a single power law to the single component $r$-band fits of \citet{Kelvin2012} (pre-release of \texttt{SersicPhotometry:SersicCatSDSSv09}) combined with an earlier version of the stellar masses catalogue of \citet{Taylor2011} (\texttt{StellarMasses:StellarMassesv16}). We note that the stellar masses did not change much between v16 and v19: the mean and standard deviation of $\Delta\log_{10}(M_*/M_\odot)$ are 0.006 and 0.07 respectively for our sample. The two panels show the results obtained with effective radii taken directly from the S\'ersic fits (top; extrapolated to infinity by definition) or limited to the segment radius within which they were fitted (bottom; see Section~\ref{sec:tightsegments} for further explanation). To guide the eye, we also show the running median and its error for our data; where the error is taken as the $1\sigma$-quantile divided by the square-root of data points within that bin (usually smaller than the size of the data points). 

In both cases, the slope of the mass-size relation obtained from our data agrees well with the fit results of \citet{Lange2015}. There is a clear offset in the absolute sizes, but those will inherently depend on the exact definition of the size measurement at hand as well as the (depth of the) data used. Already calculating effective radii within the segments within which we fitted for them (bottom panel) brings our results much closer to those of \citet{Lange2015}; although the measurements are then not directly comparable to their fits anymore since they use S\'ersic values extrapolated to infinity (which will, in turn, depend on the segment size used for fitting). We now discuss these issues further by directly comparing our fits to those of \citet{Kelvin2012}, which the \citet{Lange2015} results were based on.

\subsection{Comparison to single S\'ersic fits of \citet{Kelvin2012}}
\label{sec:comparelee} 
\begin{figure}
    \includegraphics[width=\columnwidth]{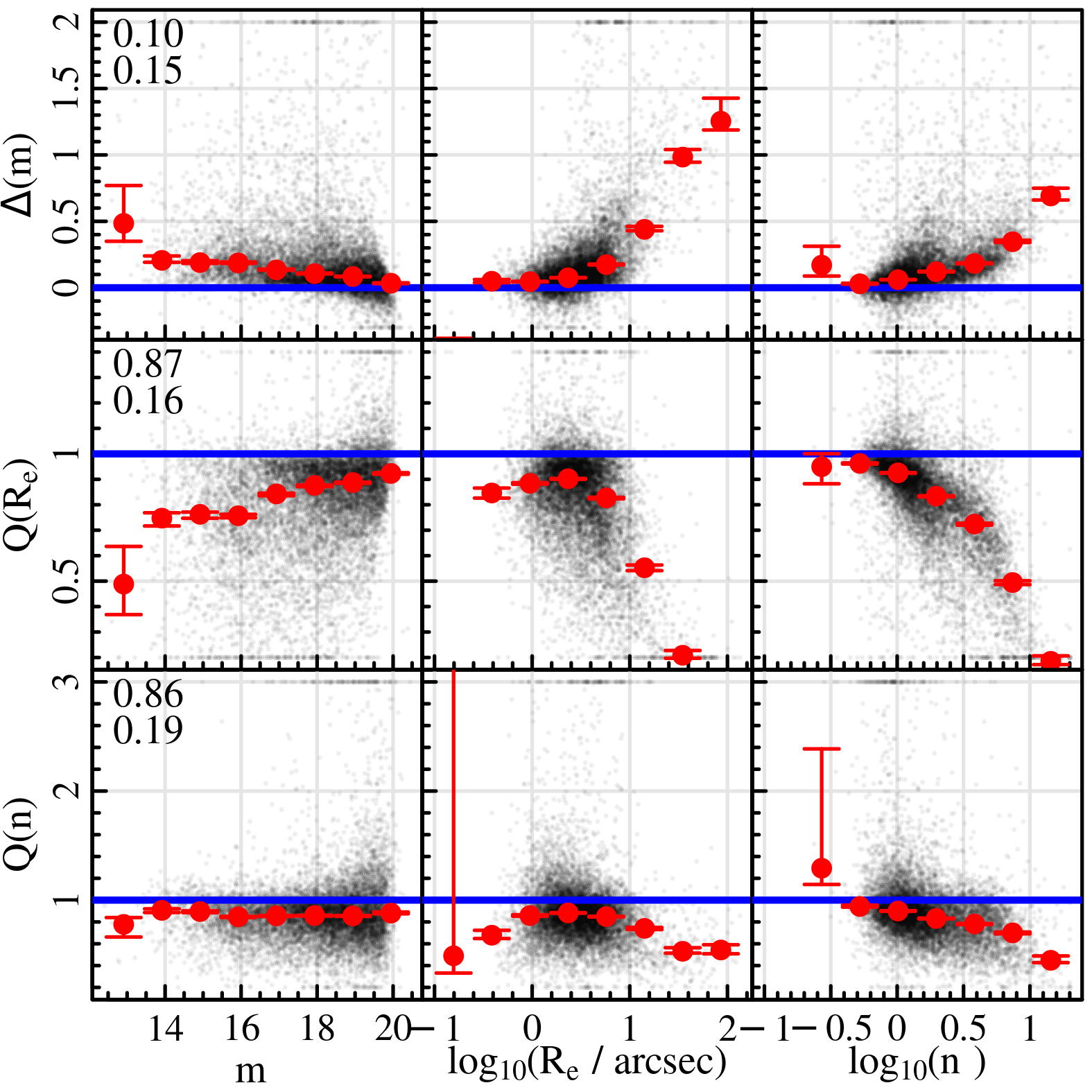}
	\includegraphics[width=\columnwidth]{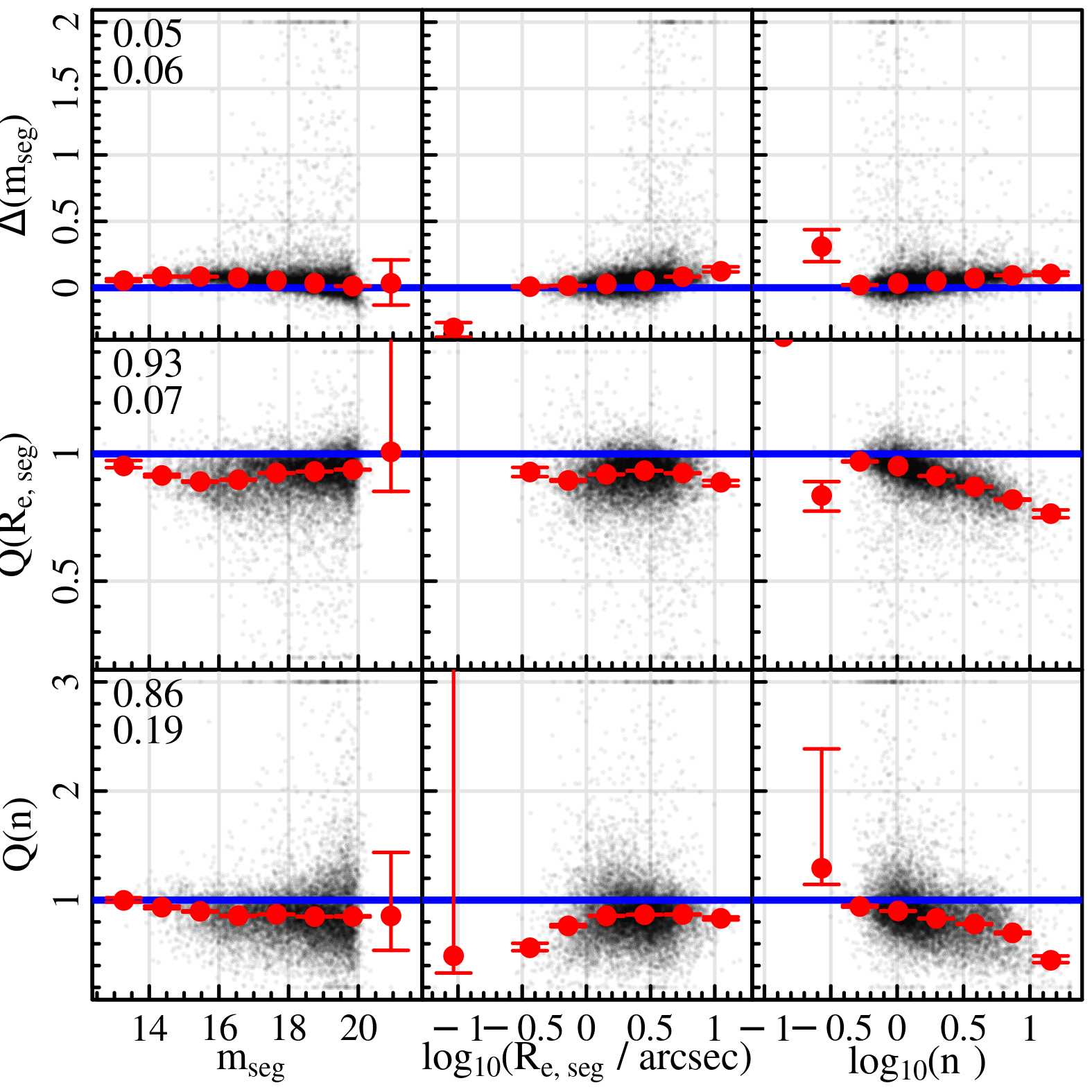}
    \caption{The difference $\Delta$ or quotient Q (for scale parameters) between the \citet{Kelvin2012} fits and our fits plotted against our fits for the three most important single-S\'ersic parameters magnitude, effective radius and S\'ersic index in the $r$-band. The top panels show the S\'ersic parameters extrapolated to infinity, while for the bottom panels we calculated the magnitude and radius within the segment radius for both our and the \citet{Kelvin2012} fits. Outliers are clipped to the plotting interval; which is the same in both cases. Black dots show all fits, red dots with error bars show the running median and its error in evenly spaced bins and horizontal blue lines indicate no difference between the fits. The numbers in the top left corners of the first column of panels show the median and 1$\sigma$-quantile of the respective distribution in the $y$-direction (which is identical for all panels of a row). The sample is limited to fits that are available in the \citet{Kelvin2012} catalogue and were classified as single component in our model selection.}	
    \label{fig:comparelee}
\end{figure}

To further investigate the size offset observed in Section~\ref{sec:sizemass}, we directly compared our fits to those of \citet{Kelvin2012} (\texttt{SersicPhotometry:SersicCatSDSSv09} on the GAMA database). Since \citet{Kelvin2012} do not provide double component fits, the analysis is limited to single S\'ersic fits. We again use the $r$-band as an example for the discussion, but note that results are very similar for the $g$ and $i$ bands. 

The \citet{Kelvin2012} fits are based on the Structural Investigation of Galaxies via Model Analysis (\texttt{SIGMA}) code applied to data from SDSS DR7. \texttt{SIGMA} is a wrapper around \texttt{Source Extractor}, \texttt{PSF Extractor} and \texttt{GALFIT 3} performing similar steps to what we do in our pipeline (Section~\ref{sec:pipeline}), i.e. source identification, background subtraction, PSF estimation and 2D model fits to the surface brightness profile of the galaxies. The differences lie in the data and code used, where we upgrade SDSS to KiDS, \texttt{Source Extractor} to \texttt{ProFound}, \texttt{PSF Extractor} to a combination of \texttt{ProFound and ProFit} and \texttt{GALFIT} to \texttt{ProFit}; with all the advantages described in Sections~\ref{sec:kids}, \ref{sec:profit} and \ref{sec:profound}. In addition, we also perform multi-component fits and model selection. For the comparison to the \citet{Kelvin2012} results, we focus on the three most important single S\'ersic fit parameters: magnitude $m$, S\'ersic index $n$ and effective radius $R_e$, which tend to be the least ``well-behaved" (position, axial ratio and angle are generally more easily constrained and uncorrelated). 

Figure~\ref{fig:comparelee} shows the difference between our fits and the \citet{Kelvin2012} fits for these three parameters. For the magnitudes, we show the difference (\citet{Kelvin2012} fits - our fits), while for the effective radius and S\'ersic index (scale parameters) we show the quotient (\citet{Kelvin2012} fits / our fits) on the $y$-axis; always plotted against our fitted values on the $x$-axis (in logarithmic space for scale parameters). Again, we show the results for S\'ersic parameters extrapolated to infinity (top) and for the magnitude and effective radii calculated within the segment radius (see Section~\ref{sec:tightsegments}), where we limit both our fits and those of \citet{Kelvin2012} to our fitting segment (which are generally smaller than the fitting regions used in \citet{Kelvin2012}) to obtain directly comparable results. \\

For the S\'ersic parameters extrapolated to infinity (top panels), large differences can be seen in all fitted parameters, including clear systematic trends across the parameter space. This shows once again that fitted S\'ersic parameters are not directly comparable given the differences in the data, code and processing steps with a wealth of potentially different systematic uncertainties (Section~\ref{sec:tightsegments}). However, when we limit the analysis to our segment sizes (bottom panels), the fits become much more comparable. On average, now, our fits are $\sim0.03$\,mag brighter and approximately 7\% larger than the \citet{Kelvin2012} fits to the same galaxies, which is not surprising given the increased depth and resolution of KiDS compared to SDSS and the numerous sources of different systematic uncertainties (e.g., differing sky subtraction and PSF estimation). Also, there are fewer trends across the parameter space, indicating that systematic differences arise mainly from the extrapolation to infinity. 

The slight exception to this is the S\'ersic index, which still shows some trends. The reason for this is that the S\'ersic index, unlike the magnitude and effective radius, cannot easily be corrected for different fitting segment sizes. The \citet{Kelvin2012} fits, which were performed within larger fitting segments than our fits, will hence inevitably have to compromise more between the inner and outer regions of the galaxy to be fitted (unless the light profile truly follows a single S\'ersic profile with no deviations out to very large radii, which is rarely the case). Our tight fitting segments, on the other hand, will result in better fits to the inner regions of the galaxy at the expense of producing unphysical wings when extrapolated beyond the fitting region (Section~\ref{sec:tightsegments}). In a sense, fitted S\'ersic indices are hence always a weighted average (or compromise) across a range of radii and their absolute values will never be directly comparable between catalogues unless the fitting regions are exactly the same (or the galaxies studied happen to follow perfect S\'ersic profiles).

\subsection{Comparison to \citet{Lange2016} double component fits}
\begin{figure*}
    \includegraphics[width=1.67\columnwidth]{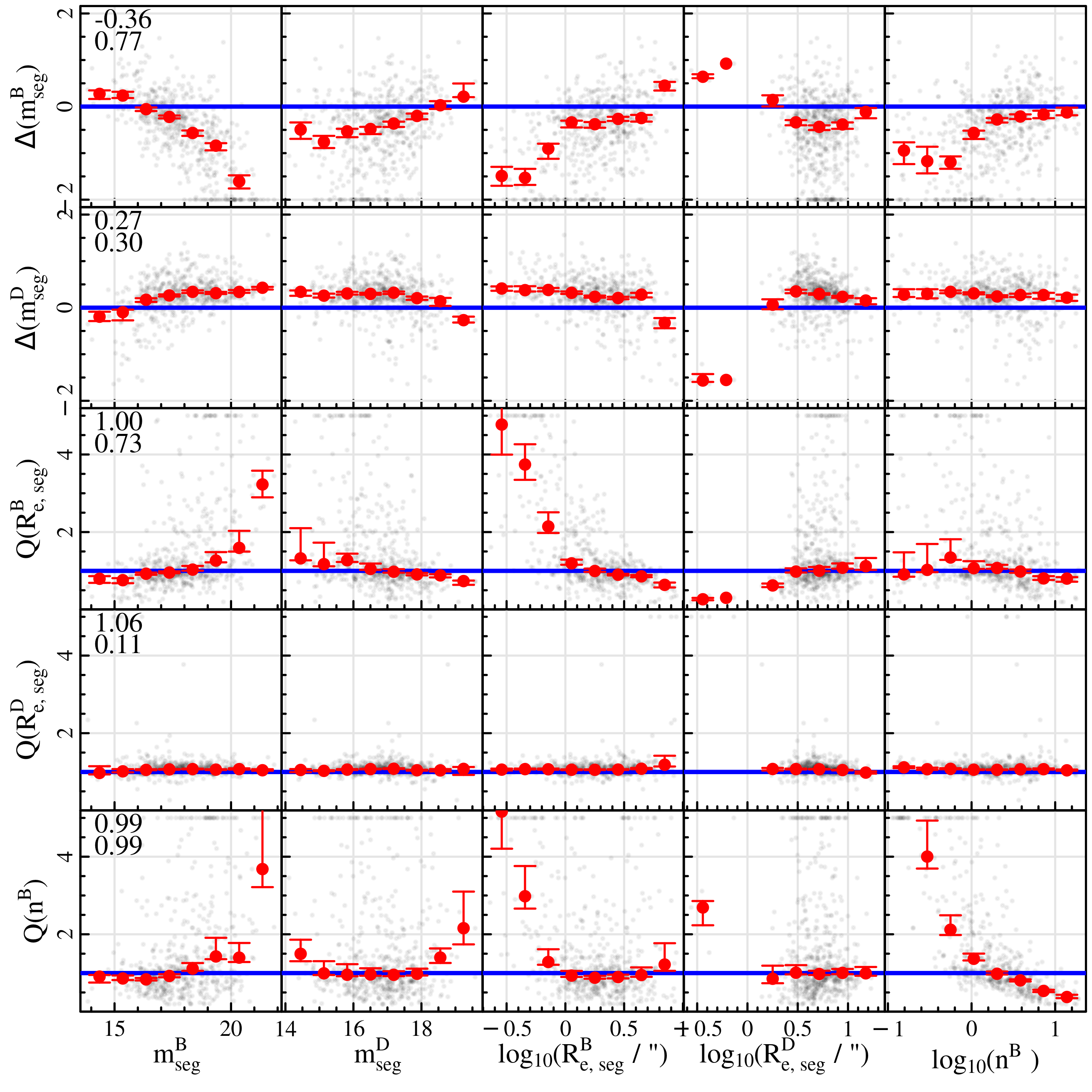}
    \caption{The difference $\Delta$ or quotient Q (for scale parameters) between the \citet{Lange2016} fits and our fits plotted against our fits for the five most important double component parameters bulge magnitude $m^B$, disk magnitude $m^D$, bulge effective radius $R_e^B$, disk effective radius $R_e^D$ and bulge S\'ersic index $n^B$ in the $r$-band. Magnitudes and effective radii are limited to segment radii and all values are clipped to the plotting intervals. Black dots show all fits, red dots with error bars show the running median and its error in evenly spaced bins and horizontal blue lines indicate no difference between the fits. The numbers in the top left corners of the first column of panels show the median and 1$\sigma$-quantile of the respective distribution in the $y$-direction (which is identical for all panels of a row). The sample is limited to fits for which both bulge and disk measurements were available in the \citet{Lange2016} catalogue, that were classified as double component object in our model selection and that were not flagged as unreliable in either catalogue.}	
    \label{fig:comparelangedoub}
\end{figure*}

Double component fits for samples of galaxies overlapping with our sample have been performed by \citet{Lange2016}, \citet{Haeussler2022} and \citet{Robotham2022}. The latter two include a comparison to the fits presented in this work (figures~19 and~20 in \citealt{Robotham2022} and figure~B2 in \citealt{Haeussler2022}), including a discussion of observed differences. We refer the reader to these works and do not repeat the comparisons here. Instead we focus on a comparison to \citet{Lange2016}, for which the fits can be obtained from the GAMA wiki.

The \citet{Lange2016} fits are performed on SDSS $r$-band data for GAMA galaxies with redshifts $z$\,<\,0.06, which is a subsample of our sample. They use \texttt{SIGMA} (Section~\ref{sec:comparelee}) to fit each galaxy with a single S\'ersic and a double S\'ersic function, starting from a grid of starting values in both cases to overcome the susceptibility of \texttt{GALFIT} to initial guesses and to obtain more realistic values for the fit uncertainties. Note that the disk S\'ersic index is left as a free fitting parameter in contrast to our work where we fix it to 1 (exponential).

Figure~\ref{fig:comparelangedoub} shows the direct comparison of our double component fits to those obtained by \citet{Lange2016} for the five most important double component parameters (limited to segment radii), similar to the single S\'ersic comparison against \citet{Kelvin2012} in Figure~\ref{fig:comparelee} and Section~\ref{sec:comparelee}. We note that the comparison to the \citet{Lange2016} single S\'ersic fits is very similar to that for \citet{Kelvin2012} except for the smaller sample size, so we do not show it here. The scatter is larger for all parameters in Figure~\ref{fig:comparelangedoub} compared to Figure~\ref{fig:comparelee}. This is expected since fitting two components will intrinsically show higher uncertainties and more degeneracies than fitting a single S\'ersic function. There is an average offset and clear trends across the parameter space for the bulge and disk magnitudes (first two rows of Figure~\ref{fig:comparelangedoub}). On average, our bulges are $\sim$\,0.3\,mag fainter than those of \citet{Lange2016}, with disks being correspondingly brighter. Small and faint bulges are most severely affected. This is most likely caused by the lower resolution of the SDSS data compared to KiDS: the same trend is observed by \citet{Haeussler2022} when comparing our fits to their fits performed on SDSS data (left two panels of the second row of their figure~B2), but not when comparing our fits to their fits performed on KiDS data (right two panels of the second row of their figure~B2).

The effective radii and bulge S\'ersic index show reduced overall offsets in Figure~\ref{fig:comparelangedoub} as compared to Figure~\ref{fig:comparelee}. The median values are consistent with no difference for most bins with reasonable numbers of data points. The bulge effective radii show an average offset of less than 1\% (albeit with large scatter), while our disks tend to be slightly larger (by around 6\%) than those of \citet{Lange2016}. The bulge S\'ersic indices are also consistent within 1\% on average, despite their large scatter. This reduction of systematic differences compared to the single S\'ersic fits in Figure~\ref{fig:comparelee} is most likely due to the generally better model fit of the double component model, which reduces the effects of the different fitting segment sizes (since those are only relevant for galaxies which cannot be accurately represented by the model, cf. Sections~\ref{sec:tightsegments} and~\ref{sec:comparelee}).

\section{Systematic uncertainties and biases from simulations and the overlap sample}
\label{sec:simulations}
The MCMC chain errors returned by the fitting procedure do not include systematic uncertainties which arise due to galaxy features not accounted for in the models, nearby other objects, imperfect PSF estimation, background subtraction inaccuracies and similar effects. For an individual galaxy, the presence of such ``features" will systematically shift the fitted parameters away from the true values, thus introducing a bias. For a statistically large enough sample of galaxies, however, most of these effects are expected to cancel out on average since they are random from one galaxy to the next (e.g. nearby other sources shifting the fitted positions). These ``random systematics" can - for statistical samples - be accounted for by simply increasing the given parameter errors such that in most cases, the true values are included in the credible intervals again. Such systematics can be studied using overlap sample galaxies, i.e. those that appeared in more than one KiDS tile (cf. Section~\ref{sec:sampleselection}). In addition, there can be ``one-sided effects" that lead to an overall bias across the sample, e.g. due to excess flux from nearby objects. These can only be detected using simulations. In the following, we study both of these effects using our bespoke simulations, the overlap sample of real galaxies and the overlap sample of simulated galaxies; where we refer to the random systematics as ``error underestimates" and to the one-sided effects as ``biases". 

The final corrections for both of these effects are listed in Table~\ref{tab:errorunderestimate}. In short, biases are very small ($\lesssim1\%$), while systematic errors are a factor of 2-3 larger than the random MCMC errors alone. The error underestimate corrections are also applied to the released catalogues, while the bias corrections are not since they are only valid for a large random subset of our galaxies and not for individual objects. Note that the systematic error studies were carried out on \texttt{v03} of the \texttt{BDDecomp} DMU, while the remainder of this work describes \texttt{v04} (both on the GAMA database). However, since \texttt{v04} is statistically identical to \texttt{v03}, the results can directly be transferred. We would also like to point out that we focus on single S\'ersic $r$-band fits in this section. We expect individual components in the 1.5- and double component fits as well as the $g$ and $i$ bands to be affected by similar systematics. Effects are likely to become worse for fainter and/or less well-resolved objects (i.e. bulges in particular). A more detailed investigation into the systematics of double component fits will be included in future work.

\subsection{Overlap sample comparison}
\label{sec:overlapstudies}
\begin{figure}
    \includegraphics[width=\columnwidth]{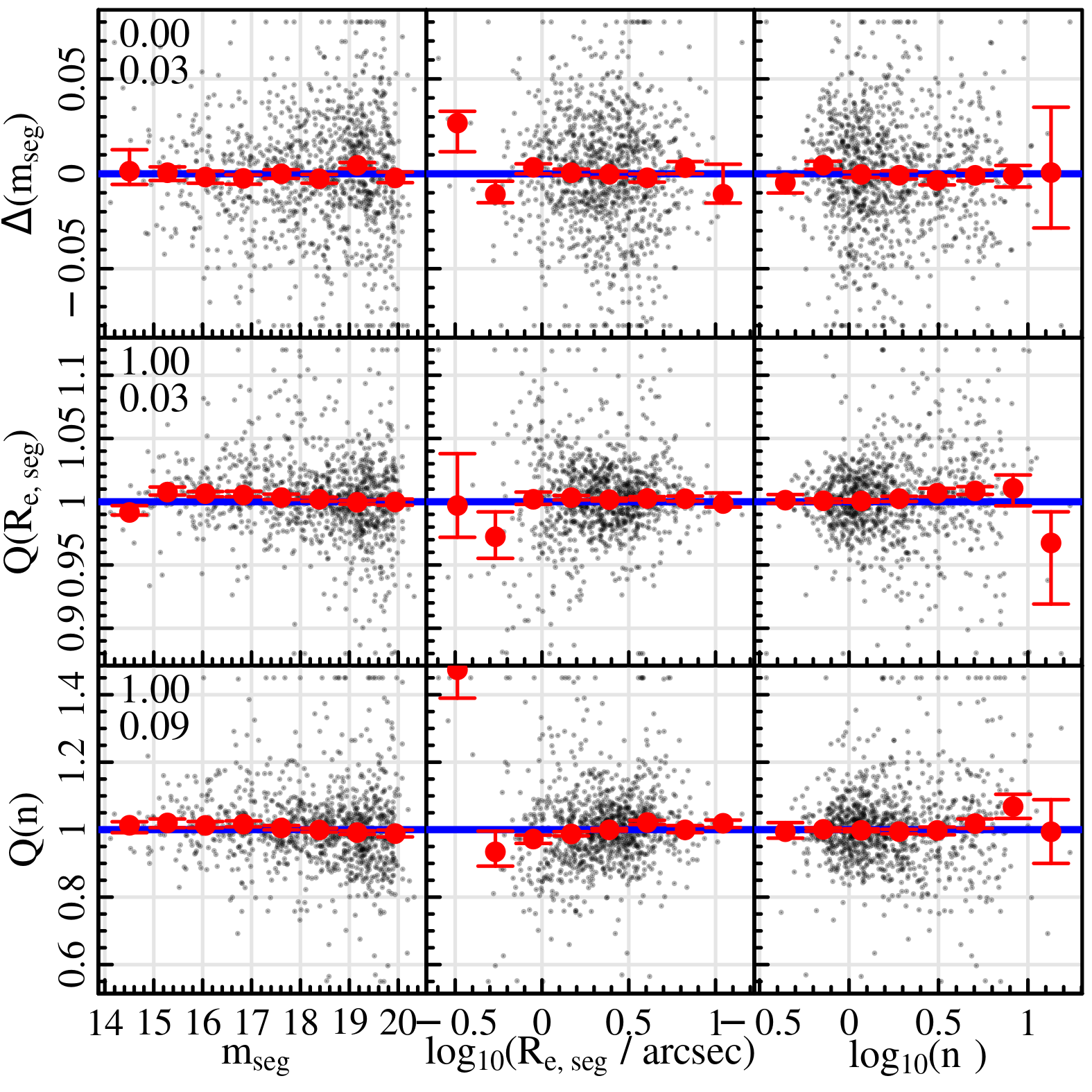}
    \caption{Similar to the bottom panels of Figure~\ref{fig:comparelee} but now showing an internal consistency check of our catalogue using galaxies that were imaged (and successfully fitted) at least twice in the KiDS $r$-band. We pass these duplicate observations of the same physical objects through our pipeline independently and then compare the fit to the shallower image with the fit to the deeper (higher signal-to-noise ratio) image. Note the different plotting ranges relative to Figure~\ref{fig:comparelee} (on the $y$-axis in particular).}	
    \label{fig:compareoverlap}
\end{figure}
As an internal consistency check, we compared the fit results obtained from multiple observations of the same physical object (Section~\ref{sec:sampleselection}) in Figure~\ref{fig:compareoverlap}. The plots are very similar to the ones in the bottom panels of Figure~\ref{fig:comparelee} (see description in Section~\ref{sec:comparelee}; but note the different $y$-axis scale), except that we now show the differences between two of our own fits to different KiDS images of the same galaxy (in the overlap region between the KiDS tiles). Hence all fits shown in Figure~\ref{fig:compareoverlap} are based on KiDS data and use the exact same pipeline for analysis, though the different observations are treated entirely independently. We always use the deeper image as the reference image (the image depth at the edge of KiDS tiles can vary greatly depending on the number of dithers - between 1 and 5 - that cover the area), but evaluate the magnitude and effective radius within whichever of the two segment radii is smaller to avoid extrapolation and obtain consistent results. 

As expected, there are very little differences between the two fits to the same galaxy; and there are no systematic trends across the parameter space. The running median is consistent with 0 (or 1, for scale parameters) in almost all bins, which shows that there are no inherent systematic differences in our fits related to image depth. This holds true despite the segment radii being systematically larger for the deeper images (as expected), as long as both fits are evaluated within the same region before comparison.

\subsection{Simulations: parameter recovery}
\label{sec:parameterrecovery}
As a final test of our pipeline, we ran simulations where we insert single S\'ersic model galaxies convolved with an appropriate PSF at random locations in the KiDS data. To obtain a realistic distribution of parameters for the model galaxies (including correlations), we use the fitted single S\'ersic parameters of a random sample of 1000 $r$-band galaxies that were not classified as outliers. The PSF to convolve with is taken as the model PSF that was fitted to the nearest real galaxy (at the position where the model galaxy is inserted), which is close to the real PSF at that image location. We then simply add the PSF-convolved galaxy to the KiDS data and run the resulting image through our entire pipeline (segmentation, sky subtraction, PSF estimation, galaxy fitting, outlier flagging, model selection). 

In this way we are able to check for intrinsic biases in our entire pipleline, with 3 exceptions: issues due to galaxy features not represented by our models (bars, spiral arms, disk breaks, mergers, etc.), problems in the data processing performed by the KiDS team (if any), and deviations of the true PSF from a Moffat function.\\

\begin{figure}
    \includegraphics[width=\columnwidth]{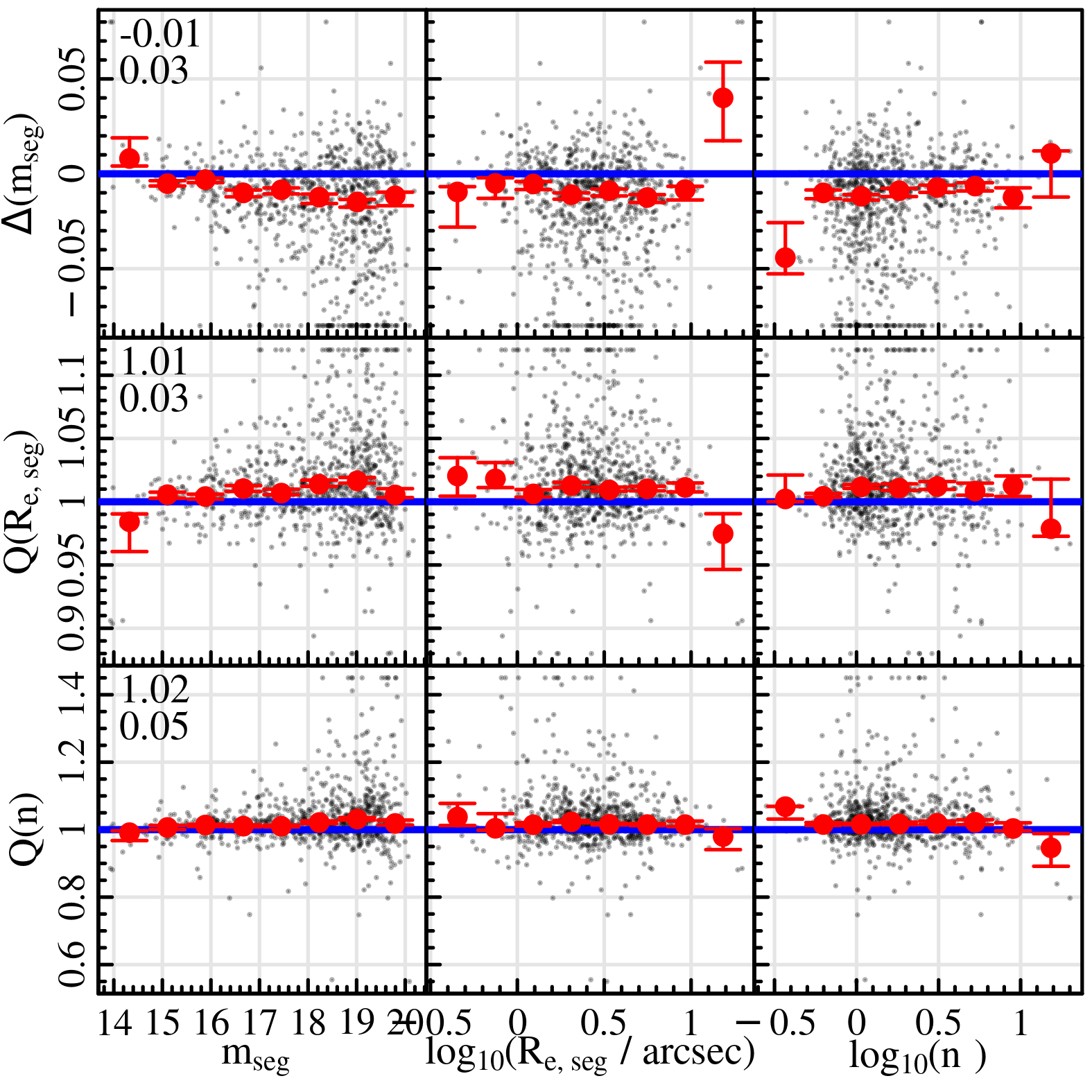}
    \caption{Similar to Figure~\ref{fig:compareoverlap} but comparing the fitted parameters of simulated images to the true (input) parameters; both limited to the segment radius.}	
    \label{fig:comparesimulations}
\end{figure}

In Figure~\ref{fig:comparesimulations} we show the corresponding plots to Figures~\ref{fig:comparelee}~and~\ref{fig:compareoverlap}; where on the $x$-axis we now have the true (input) parameters of our simulated galaxies and on the $y$-axis the difference between the fitted and the true values; both limited to segment radii. 

\begin{figure}
    \includegraphics[width=\columnwidth]{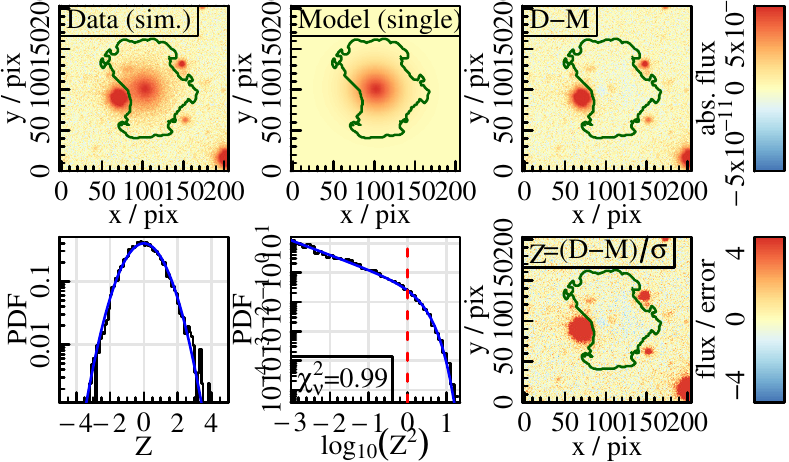}
    \caption{An example fit to a simulated galaxy where the difference between the true and the fitted magnitude is large due to the wings of a nearby bright object and a small faint object included in the segmentation map. Panels are the same as top and middle row in Figure~\ref{fig:examplefit}.}	
    \label{fig:examplebadsim}
\end{figure}

Generally, all parameters are recovered well, although the magnitudes show a slight offset of $\sim$~0.01\,mag (with corresponding trends in effective radius and S\'ersic index since these parameters are correlated); worsening for faint objects. This offset is driven by a number of galaxies scattering to very low values, i.e. where the fit attributes significantly more flux to the galaxy than what we put into the simulation. Visual inspection of these simulated objects revealed that all of them have additional flux from other objects included in the segmentation maps. 
Figure~\ref{fig:examplebadsim} shows an example, where the difference between the fitted and the true magnitude is -0.17. 

Nearby objects affect approximately 5-10\% of our simulated fits in this way. Since it is a one-sided effect (there are no sources with negative flux), it results in a slight overall bias across the sample. This is expected to occur at a similar level also in the fits to real galaxies and could only be improved by simultaneously fitting nearby sources (see also the discussion of this issue in \citealt{Haeussler2007}). However, for this work we decided against this option as explained in Section~\ref{sec:fitandconvergence}. We may revisit this decision in future work. 

\subsection{Simulations: model selection accuracy}
\label{sec:simulationsmodelselection}
Since we know that all of our input galaxies were perfect single S\'ersic systems, the model selection and outlier rejection statistics can be used to judge the failure rate of these routines (cf. also Sections~\ref{sec:postprocessing}~and~\ref{sec:statistics}). We simulated 1000 galaxies at random locations; which resulted in 1126 objects to be fit (due to the overlap regions between KiDS tiles). Of these, 262 (23\%) were skipped; which is similar to the fraction of skipped fits for real galaxies, as expected since the main reason for this are the KiDS masks. Of the remaining objects, 94\% are classified as single component fits, 3\% are 1.5- or double component fits and 3\% are flagged as outliers. 

\begin{figure}
    \includegraphics[width=\columnwidth]{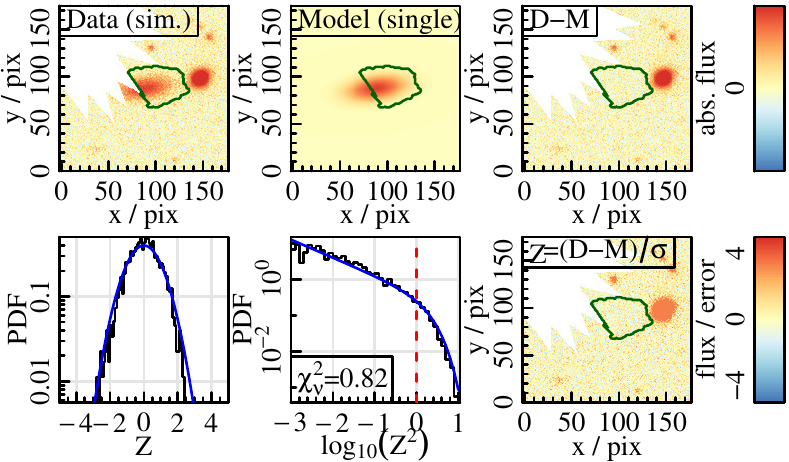}
    \caption{An example fit to a simulated galaxy which was flagged as a bad fit due to a nearby masked area from a bright star chopping up the segmentation map. Panels are the same as top and middle row in Figure~\ref{fig:examplefit}.}	
    \label{fig:exampleoutliersim}
\end{figure}

The number of outliers is significantly less than the 11\% of real $r$-band galaxies flagged (Section \ref{sec:flaggingofbadfits}) because in the simulations all galaxies are intrinsically ``well-behaved". Figure~\ref{fig:exampleoutliersim} shows an example of the most commonly occuring reason for being flagged as an outlier (in the simulations), namely the mask of a nearby bright star chopping up the segmentation map. 

\begin{figure}
    \includegraphics[width=\columnwidth]{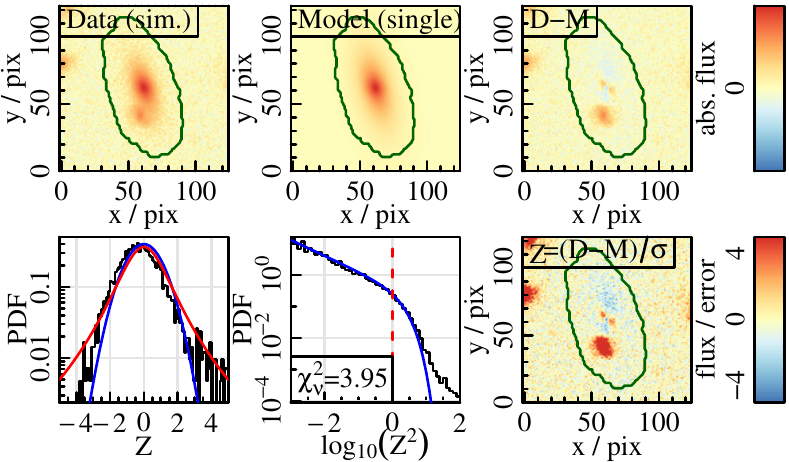}
    \includegraphics[width=\columnwidth]{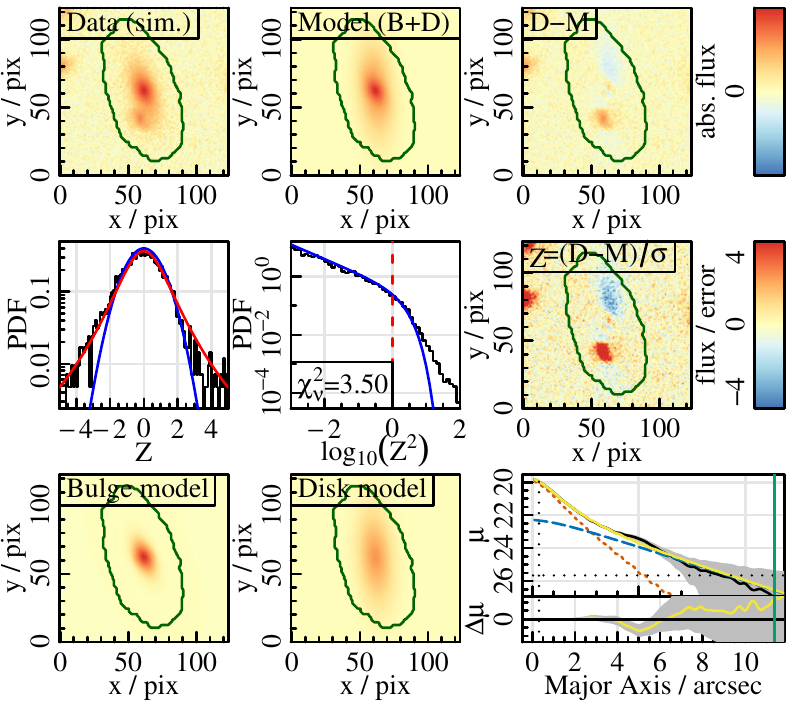}
    \caption{An example fit to a simulated galaxy which was classified as a double component fit. Single component fit at the top; double component fit at the bottom. Panels are the same as in Figure~\ref{fig:examplefit}.} 
    \label{fig:exampledoublesim}
\end{figure}

The fact that 97\% of non-outlier simulated galaxies are correctly classified as single component fits confirms that model selection is accurate provided the galaxy can be unambiguously assigned to the single S\'ersic model (cf. Section~\ref{sec:modelselection}). We visually inspected the 3\% 1.5- and double component fits and found nearby interfering objects in all of them. Figure~\ref{fig:exampledoublesim} shows an example, where the fit attempts to capture the additional ``features" with the freedom of a second component. Note that since we only simulated single S\'ersic objects, we cannot comment on the accuracy of the model selection procedure for double component objects here. However, our model selection procedure was optimised on all types of real galaxies (not just single S\'ersic objects), see Section~\ref{sec:modelselection} for details. We also point out that the confusion rate of the model selection derived from the idealised simulations is a strictly lower limit. Realistic confusion rates for real galaxies, as calibrated against visual inspection, are quantified in Table~\ref{tab:modelselconfusionr} (Section~\ref{sec:modelselection}) for the $r$-band as well as Tables~\ref{tab:modelselconfusiong}, \ref{tab:modelselconfusioni} and~\ref{tab:modelselconfusionjoint} (Appendix~\ref{app:modelselconfusion}) for the $g$, $i$ and joint model selection.

\subsection{Systematic uncertainties}
\label{sec:systematics}

\begin{figure*}
    \includegraphics[width=\textwidth]{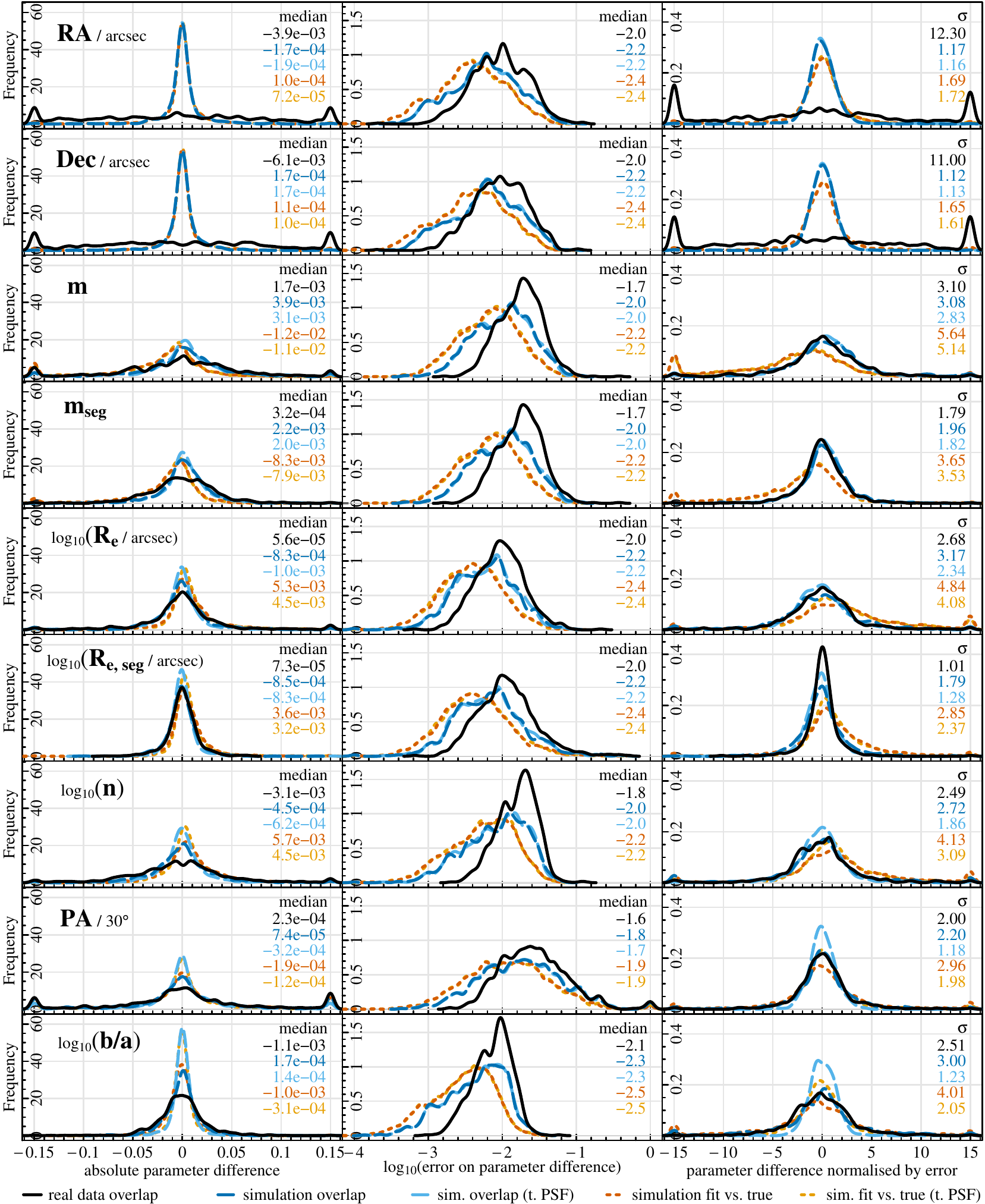}
    \caption{For all single S\'ersic parameters as labelled top to bottom: \textbf{Left column:} The distribution of the absolute difference between the fitted and true values for simulated galaxies; or between the fitted values to two versions of the same (simulated or real) galaxy in the overlap sample. The legend at the bottom indicates which difference is shown; scale parameters are treated in logarithmic space throughout. See text for details. \textbf{Middle column:} The error on the parameter difference shown in the left column. \textbf{Right column:} The parameter difference normalised by its error (i.e. left column divided by middle column).}
    \label{fig:differrnorm}
\end{figure*}

Figure~\ref{fig:differrnorm} shows the results of our systematic error study, which we will now discuss in detail. Going from left to right we show three different plot types as labelled on the $x$-axis and described in more detail in the caption and below. Going from top to bottom, each plot type is shown for each single S\'ersic parameter as labelled in the left panels; and the colours of the lines in the plot indicate which sample was used according to the legend at the bottom of the figure. The left panels show the distribution of absolute differences between the values fitted to both versions of a galaxy (in the indicated units); the middle panels show the corresponding error distribution (errors added in quadrature for the two fits; in the same units as the fit); and the right panels show the distribution of the absolute difference divided by its error (unitless). The solid black lines labelled ``real data overlap" are using the overlap sample. The dashed dark blue lines labelled ``simulation overlap" show the same for the overlap sample in simulated galaxies, where we have run more simulations specifically to boost the number of simulated overlap galaxies to a similar value as we have for the real overlap sample ($\sim700$). The dashed light blue line labelled ``sim. overlap (t. PSF)" is the same as the dashed dark blue line, just that when fitting the galaxy, instead of the estimated PSF we passed in the true PSF (i.e. the one we used to convolve the model galaxy with originally). The dotted dark orange line labelled ``simulation fit vs. true" shows the difference between the fitted value and the true value (instead of between the two fitted values in the overlap region) for the same sample of galaxies. Note the errors here now are just the errors of the fit since the true values do not have errors. The dotted light orange line (``sim. fit vs. true (t. PSF)") is the same for the run that used the true PSFs.
 
All values are clipped to the plotting intervals (for plotting only). For scale parameters, all distributions are shown in logarithmic space (which the parameters were also fitted in). To make the scales comparable to the other parameters, the angle is shown in units of 30\degr (which it was also fitted in to make the MCMC step size comparable to the other parameters). For the magnitude and effective radius, we show both the fitted S\'ersic values and the segment truncated values (cf. Section~\ref{sec:tightsegments}). Comparing rows 3 and 4 (S\'ersic and segment magnitudes) or rows 5 and 6 (S\'ersic and segment effective radii) against each other, it becomes clear immediately that the distributions for the segment values are narrower, i.e. limiting to segment sizes increases the stability and reduces the scatter in those parameters as already observed in many previous sections. Note, though, that our simulated galaxies do follow perfect single S\'ersic profiles, so the differences between the segment and S\'ersic values are generally expected to be much smaller in the simulations than in real data (see Section~\ref{sec:tightsegments} for details). 

For each distribution, we also give the median of the absolute difference and the errors and the 1$\sigma$-quantile (half of the range containing the central 68\% of data) of the normalised difference. These values (with errors) are also given in Table~\ref{tab:errorunderestimate}.

\subsubsection{Overlap sample: real vs. simulated}

Focussing on the real data and simulation overlap samples (solid black and dashed dark blue lines), which are most directly comparable, it can be seen that the distributions of the parameter differences (left column) are broader for the real galaxies than for the simulated galaxies for all parameters; i.e. for two versions of the same galaxy in the overlap sample, the fitted values are on average closer to each other in the simulation than in the real data. 
This could be due to two reasons: either irregular galaxy features in combination with noise (i.e. perfect single S\'ersic objects are just more easily constrained/less easily influenced by noise fluctuations); or differences in the KiDS data processing between tiles (e.g. inaccuracies in their background subtraction procedure) that affect the real galaxies but not the simulated ones since those were added later. In reality, it is probably a combination of these two effects (with the first one presumably dominating). All steps of our own analysis affect both the simulated and the real galaxies and hence on average will have the same effect on both. 

The errors (second column) do reflect this additional uncertainty in real galaxies in that they are larger by 0.2-0.3\,dex for all parameters. In fact, the errors on the simulated galaxies seem to be more severly underestimated than those on the real galaxies, which becomes clear when looking at the parameter differences normalised by the respective errors (third column). In an ideal world, these would all be Gaussians centred on zero with a standard deviation of 1. As there will always be a few outliers due to interfering objects or image artifacts, instead of the mean and standard deviation we will consider their more robust equivalents, the median and 1$\sigma$-quantile (shown in plots). Indeed, all overlap sample distributions (simulated and real, i.e. black, dark blue and light blue lines) are centred on zero, as already expected from the results shown in Figure~\ref{fig:compareoverlap}. However, it can be seen that for all parameters except segment effective radius, the 1$\sigma$-quantile is larger than 1 both for the simulated and the real sample: values generally range between 2 and 3 with simulated galaxies performing slightly worse due to the underestimated errors (which are most likely caused by the PSFs, see discussion below). 

The exception to this is position (RA and Dec, top two rows), for which the normalised distribution is a factor of approximately 10 broader for real galaxies than for simulated ones. In fact, a considerable fraction of these distributions fall outside of the plotting range, such that the clipping to these intervals results in prominent peaks at the plot edges (top right two panels of Figure~\ref{fig:differrnorm}). We believe this to be mainly due to the accuracy of the astrometric solution of the KiDS data, which shows a scatter of approximately $0\farcs$04 in DR4.0 in both RA and Dec (\citealt{Kuijken2019}; and we also confirmed this using the KiDS $r$-band source catalogues). This is a factor of 4 larger than the median MCMC error on position (top two panels in the middle row of Figure~\ref{fig:differrnorm}). Accounting for this additional source of scatter between the tiles (which only affects real objects but not the simulations since those were inserted after the astrometric calibration) would bring the normalised error distribution for the real data overlap sample into much closer agreement with the simulated version. The remaining factor of $\sim$2-3 difference could also be due to the astrometry, considering that the overlap sample by definition sits at tile edges, where the astrometric solution is the most uncertain; or - as for all other parameters - due to irregular galaxy features in combination with noise (see discussion above). As a last point we would like to note that the absolute differences in position are usually still within 1 pixel ($0\farcs$2), i.e. although it stands out from the plot, this is a sub-pixel effect.

\subsubsection{Simulated overlap: imperfect vs. true PSFs}

These distributions of the simulated overlap sample can be compared to their equivalent distributions using the true PSFs (dashed light blue lines in Figure~\ref{fig:differrnorm}). This allows us to determine which parameters are affected by imperfect PSF estimates. However, we note that we can only make qualitative and relative statements here since we do not know how close to the truth our estimated PSFs for the real galaxies are. When simulating our galaxies, we convolve it with the model PSF fitted to the nearest real galaxy, i.e. this is the true PSF (cf. Section~\ref{sec:parameterrecovery}). When processing the simulated galaxy through our pipeline, the estimated PSF is then obtained by fitting nearby stars in the usual way. The nearest galaxy (which the true PSF is based on) is typically around 200\arcsec away, with the distribution ranging between $\sim0$ and $\sim500$\arcsec. This is close enough to provide a realistic PSF for the position of interest since KiDS tiles are much larger than this ($\sim1\deg^2$) and the PSF varies only slowly across the tiles. However, it is further away than the stars used to obtain the estimated PSF, which are typically within $\sim100-200\arcsec$ and can at the very maximum be $\sqrt{2}*200\arcsec$ away since the large cutouts used for PSF fitting are 400\arcsec\ on each side. This is expected since the density of stars is much higher than that of GAMA galaxies. However, it implies that the deviations of the estimated to the true PSFs in the simulations will on average be larger than for real data, hence leading to the errors on simulated galaxies being more severely underestimated as noted in the previous sections. 

Comparing the simulations with true PSFs to the real data, it can be seen that the simulations now perform better than the real data for all parameters except segment effective radius (1$\sigma$-quantiles between 1 and 2.8). In addition, comparing the simulations with the true and the false PSFs against each other, we can assess which parameters are most affected (relatively speaking): position angle and axial ratio are most severely influenced; followed by S\'ersic index, effective radius and magnitude, while the position is nearly unaffected. This makes sense: the axial ratio and position angle are very sensitive to mistakes in the ellipticity and orientation of the PSF; while the fitted S\'ersic index, effective radius and magnitude depend on the concentration and FWHM of the PSF. The position is only very weakly affected since the PSF is always centred and symmetric. 

Note that all 1$\sigma$-quantiles are still larger than 1 (ranging from 1.1 to 2.8) even for the simulations with true PSFs. This indicates that the error underestimates of these parameters are not exclusively caused by the effects studied so far (galaxy and/or image processing features not accounted for in the simulations and PSF uncertainties) but there is an additional contribution from features that are also present in the simulations such as nearby objects, noise fluctuations, background subtraction inaccuracies or image artifacts.

\subsubsection{Simulated sample: fitted vs. true values}

Finally, for the simulated samples we can compare the fitted values to the true values (instead of the overlap sample comparison), which is shown as dotted orange lines in Figure~\ref{fig:differrnorm}. This allows us to detect biases, but is less directly comparable to the sample of real galaxies where the true value is unknown. Note the errors are generally slightly smaller compared to the overlap studies since for those, the errors for both fits were added in quadrature while the true values now obviously do not have errors. Correspondingly, the normalised distributions are slightly broader even though the absolute differences between parameters are comparable. Most notably, however, the median of the distribution is now shifted away from zero for magnitude, effective radius and S\'ersic index (see also Table~\ref{tab:errorunderestimate}).
This is due to the bias caused by nearby objects already described in Section~\ref{sec:parameterrecovery}: magnitudes are too bright by $\sim$~0.01\,mag; effective radii and S\'ersic indices too large by approximately 1\% (always a bit better for segment values and/or simulations using the true PSFs). All other parameters still have their distributions centred on zero, i.e. do not show any bias - at least not one that we can test with our simulations. This makes sense since position, axial ratio and angle will be influenced by nearby objects (and other effects) as well, but without any preferred direction and hence on average this leads to an error underestimate rather than an overall bias. Using the true PSFs (dotted light orange lines) narrows all distributions slightly as expected; but again there is only an error underestimate rather than an overall bias introduced by the wrong PSFs since they are ``randomly wrong". 

One source of potential bias that we cannot test with the simulations are galaxy features not accounted for in the models. If, for example, there is a large population of galaxies that have bars; and these bars lead to the bulge axial ratios being systematically underestimated, this is again a one-sided effect that could lead to an overall bias. In addition, such features could further increase the error underestimate because they will obviously tend to influence both fits to a galaxy in the overlap sample in the same way and hence are difficult to detect in the above analysis. If there are systematic one-sided deviations of the true PSFs from Moffat functions, these could lead to one-sided systematically wrong PSF estimates which could in turn also introduce an additional bias that cannot be tested by the simulations which use Moffat model PSFs.

\subsubsection{Corrections for systematics and their validity}

\begin{table}
	\centering
	\caption{Biases and error underestimates for all single S\'ersic parameters derived from our systematic error studies (Section~\ref{sec:systematics}). The bias is additive (indicated with $\pm$) for those parameters that were treated in linear space and multiplicative (indicated with $\divideontimes$) for those treated in logarithmic space. Error underestimates are always multiplicative. The column ``bias/$\sigma$" gives the significance of each bias.}
	\label{tab:errorunderestimate}
	\begin{tabular}{llrr} 
		\hline
		\multirow{2}{*}{param.} & \multicolumn{1}{c}{bias} & \multirow{2}{*}{bias/$\sigma$} & \multicolumn{1}{c}{error}\\
		 & \multicolumn{1}{c}{(using true PSFs)} & & \multicolumn{1}{c}{underest.}\\
		\hline
RA & $\pm\,\, (7 \pm 18) \times 10^{-5}$\,arcsec &   0.39 & $ 12.27 \pm 0.71 $ \\ 
 Dec & $\pm\,\, (10 \pm 19) \times 10^{-5}$\,arcsec &   0.56 & $ 10.98 \pm 0.53 $ \\ 
 $m$ & $\pm\,\, (-11.4 \pm 0.8) \times 10^{-3}$  & -13.83 & $  3.10 \pm 0.16 $ \\ 
 $m_{seg}$ & $\pm\,\, (-8 \pm 0.5) \times 10^{-3}$  & -14.92 & $  1.79 \pm 0.09 $ \\ 
 $R_{e}$ & $\divideontimes\,\, (1.0105 \pm 0.0009)$  &  11.61 & $  2.68 \pm 0.12 $ \\ 
 $R_{e, seg}$ & $\divideontimes\,\, (1.0074 \pm 0.0005)$  &  14.39 & $  1.01 \pm 0.05 $ \\ 
 $n$ & $\divideontimes\,\, (1.010 \pm 0.001)$  &  10.93 & $  2.49 \pm 0.10 $ \\ 
 PA & $\pm\,\, (-4 \pm 19) \times 10^{-3}$\,deg &  -0.19 & $  2.00 \pm 0.10 $ \\ 
 b/a & $\divideontimes\,\, (0.9993 \pm 0.0004)$  &  -2.07 & $  2.51 \pm 0.09 $ \\ 
        \hline
	\end{tabular}
\end{table}

Table~\ref{tab:errorunderestimate} summarises the results of the systematic error studies: for all single S\'ersic parameters (plus the segment magnitude and segment effective radius), we give the average bias and error underestimates with errors. The bias is estimated from the median of the offset between fitted and true values in the simulation using the true PSFs (light orange numbers in the first column of Figure~\ref{fig:differrnorm}). The errors on the median are taken as the $1\sigma$-quantiles of these distributions divided by the square-root of the number of data points ($\sim2000$). Logarithmic parameters are converted back into linear space to simplify bias correction in the catalogue. Nonetheless, scale parameters have multiplicative correction factors while location parameters have additive corrections (in given units). In other words: to correct for the bias, subtract the values in Table~\ref{tab:errorunderestimate} from the catalogue values for position, magnitude and position angle; and divide by the given values for effective radius, S\'ersic index and axial ratio. Users should note, however, that due to the way these biases were estimated, they do not include all sources of potential bias (e.g. galaxy features such as bars and spiral arms; or systematically wrong PSFs). Also, we recommend to apply the bias correction only to statistically large and random samples; they are average values not applicable to individual galaxies as evident from Figure~\ref{fig:comparesimulations}. 

The next column in Table~\ref{tab:errorunderestimate} gives the significance of each bias, which is the deviation of the median from 0 (or 1 for scale parameters) divided by its error. It can be seen that position, position angle and axial ratio are not biased (consistent with 0/1 within $2\sigma$), while magnitude, effective radius and S\'ersic index are biased (deviation from 0/1 of $>5\sigma$); as found and discussed before. 

Finally, the last column in Table~\ref{tab:errorunderestimate} gives the error underestimates estimated from the width of the distribution of the normalised difference between fits to two versions of the same galaxy in the overlap sample (black numbers in the last column of Figure~\ref{fig:differrnorm}). The errors in this case were estimated by bootstrapping the distributions 1000 times each to get an estimate of the variation of the distribution width. Since these distributions were normalised (by the respective errors), there is no need to convert between linear and logarithmic space and the given values can hence directly be used as correction factors for the MCMC errors. We have applied the relevant correction to all quoted errors in the catalogue, but also give the original (purely random) errors for completeness. Also note that since these values are now based on real data, they do include PSF uncertainties (in contrast to the biases). Also, since the overlap sample is in many ways the worst in terms of data quality (sitting at tile edges), these are likely upper limits. However, they still do not include error underestimates caused by (galaxy) features not accounted for in the models such as bars, rings, spiral arms or similar, as well as nearby objects. Since these are physical (rather than related to the data taking or image processing), they will be present in both versions of the overlap sample galaxy and influence both fits in similar ways - leading to a (random) bias on individual galaxy fits; and hence an error underestimate for a large enough sample. These issues should be kept in mind when using the catalogue.

\section{Conclusions}
\label{sec:conclusions}

In this work we presented our pipeline for the single S\'ersic fits and bulge-disk decompositions of 13096 galaxies at redshifts $z<0.08$ in the GAMA II equatorial fields in the KiDS $g$, $r$ and $i$ bands. The galaxy modelling is done using \texttt{ProFit}, the Bayesian two-dimensional surface profile fitting code of \citet{Robotham2017}, fitting three models to each galaxy:
\begin{enumerate} 
\item{a single S\'ersic component,}
\item{a two-component model consisting of a S\'ersic bulge plus exponential disk and }
\item{a two-component model consisting of a point source bulge plus exponential disk (for unresolved bulges).}
\end{enumerate} 
The preparatory work (image segmentation, background subtraction and obtaining initial parameter guesses) is carried out using the sister package \texttt{ProFound} \citep{Robotham2018}; with the PSF estimated by fitting nearby stars using a combination of \texttt{ProFound} and \texttt{ProFit}. Segmentation maps are defined on joint $gri$-images, while the remaining analysis is performed individually in each band except for the model selection, for which we offer both a per-band and a joint version. The analysis is fully automated and self-contained with no dependency on additional tools. 

In addition to the galaxy fitting, we perform a number of post-processing steps including flagging of bad fits and model selection. An overview of the number of galaxies successfully fitted in each band as well as the number classified in each category is given in Table~\ref{tab:results} and Figure~\ref{fig:ncompstats}. For our planned applications of the catalogue, which involves the statistical study of dust attenuation effects, we need fits that are most directly comparable to each other. Hence, we choose to model a maximum of two components for each galaxy even if more features may be present; and focus on achieving good fits in the high signal-to-noise regions of the galaxies by choosing relatively small segments for fitting. Consequently, we recommend using truncated magnitudes and effective radii for all analyses instead of the S\'ersic values which are extrapolated to infinity by definition. The quality of the fits is tested and ensured by visual inspection, comparing to previous works \citep{Kelvin2012, Lange2015}, studying independent fits of galaxies in the overlap regions of KiDS tiles and bespoke simulations. The latter two are also used for a detailed analysis of how systematic uncertainties affect our fits. \\

We find that the combination of \texttt{ProFound} and \texttt{ProFit} is well-suited to the automated analysis of large datasets. The fully Bayesian MCMC treatment enabled by \texttt{ProFit} is able to overcome the main shortcomings of traditionally engaged downhill-gradient based optimisers, namely their susceptibility to initial guesses and their inability to easily derive realistic error estimates. The watershed deblending algorithm used by \texttt{ProFound} is less prone to catastrophic segmentation failures and allows us to extract more complex object shapes than other commonly used algorithms based on elliptical apertures; while still preserving the total flux well. With its wealth of utility functions, it not only facilitates the robust segmentation of large sets of images but also provides sky background estimates and reasonable initial guesses for the MCMC fitting. 

These characteristics, in combination with our own routines for quality assurance, led to results that are robust across a variety of galaxy types and image qualities and in reasonable agreement with previous studies given the different data, code and focus of the analysis. The outlier rejection routine efficiently identifies objects for which none of our models is appropriate such as irregular galaxies or those compromised by masked areas. Model selection is based on a DIC cut and accurate to $>90\%$ compared to what could be achieved by visual inspection. There is a minimal bias in the fitted magnitude, effective radius and S\'ersic index of approximately 0.01\,mag, 1\% and 1\% respectively (on average across the full sample) caused by excess flux from nearby other objects. The errors obtained from the MCMC chains are underestimated with respect to the true errors by factors typically between 2 and 3 (see Table~\ref{tab:errorunderestimate}) and can easily be corrected for statistically large samples of galaxies.\\

All results are integrated into the GAMA database as \texttt{v04} of the \texttt{BDDecomp} DMU. This consists of a total of 8 catalogues giving the results of the preparatory work, the 2D surface brightness distribution fits and the post-processing of all 13096 galaxies in our full sample ($z<0.08$ in the GAMA II equatorial survey regions) in the KiDS $g$, $r$ and $i$ bands (see Section~\ref{sec:results} for details). 

The full DMU is currently available to GAMA team members with a version restricted to SAMI galaxies available to the SAMI team. It will be made publicly available in one of the forthcoming GAMA data releases. Readers interested in using (parts of) the catalogue before it is publicly released are encouraged to contact the authors to explore the possibilities for a collaboration\footnote{\url{http://www.gama-survey.org/collaborate/}}.\\

We plan to extend this work to include the KiDS $u$ and VIKING $Z, Y, J, H, K_s$ bands, ideally making use of \texttt{ProFit}'s new multi-frame and multi-band fitting functionality. The decompositions will then be used to derive the stellar mass functions of bulges and disks and constrain the nature and distribution of dust in galaxy disks. This is achieved by comparing the distribution of bulges and disks in the luminosity-size plane to the dust radiative transfer models of Tuffs and Popescu \citep[and preceding papers of this series]{Popescu2011}; essentially expanding the work of \citet{Driver2007} with more and better data and at several wavelengths.

\section*{Acknowledgements}
GAMA is a joint European-Australasian project based around a spectroscopic campaign using the Anglo-Australian Telescope. The GAMA input catalogue is based on data taken from the Sloan Digital Sky Survey and the UKIRT Infrared Deep Sky Survey. Complementary imaging of the GAMA regions is being obtained by a number of independent survey programmes including GALEX MIS, VST KiDS, VISTA VIKING, WISE, Herschel-ATLAS, GMRT and ASKAP providing UV to radio coverage. GAMA is funded by the STFC (UK), the ARC (Australia), the AAO, and the participating institutions. The GAMA website is http://www.gama-survey.org/ .

Based on observations made with ESO Telescopes at the La Silla Paranal Observatory under programme IDs 177.A-3016, 177.A-3017, 177.A-3018 and 179.A-2004, and on data products produced by the KiDS consortium. The KiDS production team acknowledges support from: Deutsche Forschungsgemeinschaft, ERC, NOVA and NWO-M grants; Target; the University of Padova, and the University Federico II (Naples).

\section*{Data availability} 
The data underlying this article are integrated into the GAMA database. For access options, see \url{http://www.gama-survey.org/collaborate/}. 




\bibliographystyle{mnras}
\bibliography{references} 

\providecommand{\noopsort}[1]{}
\begin{thebibliography}{}
\makeatletter
\relax
\def\mn@urlcharsother{\let\do\@makeother \do\$\do\&\do\#\do\^\do\_\do\%\do\~}
\def\mn@doi{\begingroup\mn@urlcharsother \@ifnextchar [ {\mn@doi@}
  {\mn@doi@[]}}
\def\mn@doi@[#1]#2{\def\@tempa{#1}\ifx\@tempa\@empty \href
  {http://dx.doi.org/#2} {doi:#2}\else \href {http://dx.doi.org/#2} {#1}\fi
  \endgroup}
\def\mn@eprint#1#2{\mn@eprint@#1:#2::\@nil}
\def\mn@eprint@arXiv#1{\href {http://arxiv.org/abs/#1} {{\tt arXiv:#1}}}
\def\mn@eprint@dblp#1{\href {http://dblp.uni-trier.de/rec/bibtex/#1.xml}
  {dblp:#1}}
\def\mn@eprint@#1:#2:#3:#4\@nil{\def\@tempa {#1}\def\@tempb {#2}\def\@tempc
  {#3}\ifx \@tempc \@empty \let \@tempc \@tempb \let \@tempb \@tempa \fi \ifx
  \@tempb \@empty \def\@tempb {arXiv}\fi \@ifundefined
  {mn@eprint@\@tempb}{\@tempb:\@tempc}{\expandafter \expandafter \csname
  mn@eprint@\@tempb\endcsname \expandafter{\@tempc}}}

\bibitem[\protect\citeauthoryear{{Akaike}}{{Akaike}}{1974}]{Akaike1974}
{Akaike} H.,  1974, IEEE Transactions on Automatic Control, \href
  {https://ui.adsabs.harvard.edu/abs/1974ITAC...19..716A} {19, 716}

\bibitem[\protect\citeauthoryear{{Allen}, {Driver}, {Graham}, {Cameron},
  {Liske}  \& {de Propris}}{{Allen} et~al.}{2006}]{Allen2006}
{Allen} P.~D.,  {Driver} S.~P.,  {Graham} A.~W.,  {Cameron} E.,  {Liske} J.,
  {de Propris} R.,  2006, \mn@doi [\mnras] {10.1111/j.1365-2966.2006.10586.x},
  \href {http://adsabs.harvard.edu/abs/2006MNRAS.371....2A} {371, 2}

\bibitem[\protect\citeauthoryear{{Argyle}, {M{\'e}ndez-Abreu}, {Wild}  \&
  {Mortlock}}{{Argyle} et~al.}{2018}]{Argyle2018}
{Argyle} J.~J.,  {M{\'e}ndez-Abreu} J.,  {Wild} V.,   {Mortlock} D.~J.,  2018,
  \mn@doi [\mnras] {10.1093/mnras/sty1691}, \href
  {https://ui.adsabs.harvard.edu/abs/2018MNRAS.479.3076A} {479, 3076}

\bibitem[\protect\citeauthoryear{{Baldry} et~al.,}{{Baldry}
  et~al.}{2010}]{Baldry2010}
{Baldry} I.~K.,  et~al., 2010, \mn@doi [\mnras]
  {10.1111/j.1365-2966.2010.16282.x}, \href
  {http://adsabs.harvard.edu/abs/2010MNRAS.404...86B} {404, 86}

\bibitem[\protect\citeauthoryear{{Baldry} et~al.,}{{Baldry}
  et~al.}{2012}]{Baldry2012}
{Baldry} I.~K.,  et~al., 2012, \mn@doi [\mnras]
  {10.1111/j.1365-2966.2012.20340.x}, \href
  {https://ui.adsabs.harvard.edu/abs/2012MNRAS.421..621B} {421, 621}

\bibitem[\protect\citeauthoryear{{Baldry} et~al.,}{{Baldry}
  et~al.}{2014}]{Baldry2014}
{Baldry} I.~K.,  et~al., 2014, \mn@doi [\mnras] {10.1093/mnras/stu727}, \href
  {http://adsabs.harvard.edu/abs/2014MNRAS.441.2440B} {441, 2440}

\bibitem[\protect\citeauthoryear{{Barden} et~al.,}{{Barden}
  et~al.}{2005}]{Barden2005}
{Barden} M.,  et~al., 2005, \mn@doi [\apj] {10.1086/497679}, \href
  {https://ui.adsabs.harvard.edu/abs/2005ApJ...635..959B} {635, 959}

\bibitem[\protect\citeauthoryear{{Barsanti} et~al.,}{{Barsanti}
  et~al.}{2021}]{Barsanti2021}
{Barsanti} S.,  et~al., 2021, \mn@doi [\apj] {10.3847/1538-4357/abe5ac}, \href
  {https://ui.adsabs.harvard.edu/abs/2021ApJ...911...21B} {911, 21}

\bibitem[\protect\citeauthoryear{{Bellstedt} et~al.,}{{Bellstedt}
  et~al.}{2020}]{Bellstedt2020}
{Bellstedt} S.,  et~al., 2020, \mn@doi [\mnras] {10.1093/mnras/staa1466}, \href
  {https://ui.adsabs.harvard.edu/abs/2020MNRAS.496.3235B} {496, 3235}

\bibitem[\protect\citeauthoryear{{Benson}, {D{\v{z}}anovi{\'c}}, {Frenk}  \&
  {Sharples}}{{Benson} et~al.}{2007}]{Benson2007}
{Benson} A.~J.,  {D{\v{z}}anovi{\'c}} D.,  {Frenk} C.~S.,   {Sharples} R.,
  2007, \mn@doi [\mnras] {10.1111/j.1365-2966.2007.11923.x}, \href
  {https://ui.adsabs.harvard.edu/abs/2007MNRAS.379..841B} {379, 841}

\bibitem[\protect\citeauthoryear{{Bertin} \& {Arnouts}}{{Bertin} \&
  {Arnouts}}{1996}]{Bertin1996}
{Bertin} E.,  {Arnouts} S.,  1996, \mn@doi [\aaps] {10.1051/aas:1996164}, \href
  {https://ui.adsabs.harvard.edu/abs/1996A&AS..117..393B} {117, 393}

\bibitem[\protect\citeauthoryear{{Blanton} et~al.,}{{Blanton}
  et~al.}{2003}]{Blanton2003}
{Blanton} M.~R.,  et~al., 2003, \mn@doi [\apj] {10.1086/375528}, \href
  {https://ui.adsabs.harvard.edu/abs/2003ApJ...594..186B} {594, 186}

\bibitem[\protect\citeauthoryear{{Blanton}, {Eisenstein}, {Hogg}, {Schlegel}
  \& {Brinkmann}}{{Blanton} et~al.}{2005}]{Blanton2005}
{Blanton} M.~R.,  {Eisenstein} D.,  {Hogg} D.~W.,  {Schlegel} D.~J.,
  {Brinkmann} J.,  2005, \mn@doi [\apj] {10.1086/422897}, \href
  {https://ui.adsabs.harvard.edu/abs/2005ApJ...629..143B} {629, 143}

\bibitem[\protect\citeauthoryear{{Bluck} et~al.,}{{Bluck}
  et~al.}{2019}]{Bluck2019}
{Bluck} A. F.~L.,  et~al., 2019, \mn@doi [\mnras] {10.1093/mnras/stz363}, \href
  {https://ui.adsabs.harvard.edu/abs/2019MNRAS.485..666B} {485, 666}

\bibitem[\protect\citeauthoryear{{Bottrell}, {Torrey}, {Simard}  \&
  {Ellison}}{{Bottrell} et~al.}{2017}]{Bottrell2017b}
{Bottrell} C.,  {Torrey} P.,  {Simard} L.,   {Ellison} S.~L.,  2017, \mn@doi
  [\mnras] {10.1093/mnras/stx276}, \href
  {https://ui.adsabs.harvard.edu/abs/2017MNRAS.467.2879B} {467, 2879}

\bibitem[\protect\citeauthoryear{{Bottrell}, {Simard}, {Mendel}  \&
  {Ellison}}{{Bottrell} et~al.}{2019}]{Bottrell2019}
{Bottrell} C.,  {Simard} L.,  {Mendel} J.~T.,   {Ellison} S.~L.,  2019, \mn@doi
  [\mnras] {10.1093/mnras/stz855}, \href
  {https://ui.adsabs.harvard.edu/abs/2019MNRAS.486..390B} {486, 390}

\bibitem[\protect\citeauthoryear{{Broyden}}{{Broyden}}{1970}]{Broyden1970}
{Broyden} C.~G.,  1970, \mn@doi [Journal of the Institute of Mathematics and
  Its Applications] {10.1093/imamat/6.1.76}, 6, 76

\bibitem[\protect\citeauthoryear{{Bryant} et~al.,}{{Bryant}
  et~al.}{2015}]{Bryant2015}
{Bryant} J.~J.,  et~al., 2015, \mn@doi [\mnras] {10.1093/mnras/stu2635}, \href
  {https://ui.adsabs.harvard.edu/abs/2015MNRAS.447.2857B} {447, 2857}

\bibitem[\protect\citeauthoryear{{Cluver} et~al.,}{{Cluver}
  et~al.}{2020}]{Cluver2020}
{Cluver} M.~E.,  et~al., 2020, \mn@doi [\apj] {10.3847/1538-4357/ab9cb8}, \href
  {https://ui.adsabs.harvard.edu/abs/2020ApJ...898...20C} {898, 20}

\bibitem[\protect\citeauthoryear{{Cole}, {Lacey}, {Baugh}  \& {Frenk}}{{Cole}
  et~al.}{2000}]{Cole2000}
{Cole} S.,  {Lacey} C.~G.,  {Baugh} C.~M.,   {Frenk} C.~S.,  2000, \mn@doi
  [\mnras] {10.1046/j.1365-8711.2000.03879.x}, \href
  {https://ui.adsabs.harvard.edu/abs/2000MNRAS.319..168C} {319, 168}

\bibitem[\protect\citeauthoryear{{Cook}, {Lapi}  \& {Granato}}{{Cook}
  et~al.}{2009}]{Cook2009}
{Cook} M.,  {Lapi} A.,   {Granato} G.~L.,  2009, \mn@doi [\mnras]
  {10.1111/j.1365-2966.2009.14962.x}, \href
  {https://ui.adsabs.harvard.edu/abs/2009MNRAS.397..534C} {397, 534}

\bibitem[\protect\citeauthoryear{{Cook}, {Cortese}, {Catinella}  \&
  {Robotham}}{{Cook} et~al.}{2019}]{Cook2019}
{Cook} R. H.~W.,  {Cortese} L.,  {Catinella} B.,   {Robotham} A.,  2019,
  \mn@doi [\mnras] {10.1093/mnras/stz2789}, \href
  {https://ui.adsabs.harvard.edu/abs/2019MNRAS.490.4060C} {490, 4060}

\bibitem[\protect\citeauthoryear{{Davies} et~al.,}{{Davies}
  et~al.}{2018}]{Davies2018}
{Davies} L.~J.~M.,  et~al., 2018, \mn@doi [\mnras] {10.1093/mnras/sty1553},
  \href {https://ui.adsabs.harvard.edu/abs/2018MNRAS.480..768D} {480, 768}

\bibitem[\protect\citeauthoryear{{Dimauro} et~al.,}{{Dimauro}
  et~al.}{2018}]{Dimauro2018}
{Dimauro} P.,  et~al., 2018, \mn@doi [\mnras] {10.1093/mnras/sty1379}, \href
  {https://ui.adsabs.harvard.edu/abs/2018MNRAS.478.5410D} {478, 5410}

\bibitem[\protect\citeauthoryear{{Dimauro} et~al.,}{{Dimauro}
  et~al.}{2019}]{Dimauro2019}
{Dimauro} P.,  et~al., 2019, \mn@doi [\mnras] {10.1093/mnras/stz2421}, \href
  {https://ui.adsabs.harvard.edu/abs/2019MNRAS.489.4135D} {489, 4135}

\bibitem[\protect\citeauthoryear{{Driver}, {Popescu}, {Tuffs}, {Liske},
  {Graham}, {Allen}  \& {de Propris}}{{Driver} et~al.}{2007a}]{Driver2007}
{Driver} S.~P.,  {Popescu} C.~C.,  {Tuffs} R.~J.,  {Liske} J.,  {Graham} A.~W.,
   {Allen} P.~D.,   {de Propris} R.,  2007a, \mn@doi [\mnras]
  {10.1111/j.1365-2966.2007.11862.x}, \href
  {http://adsabs.harvard.edu/abs/2007MNRAS.379.1022D} {379, 1022}

\bibitem[\protect\citeauthoryear{{Driver}, {Allen}, {Liske}  \&
  {Graham}}{{Driver} et~al.}{2007b}]{Driver2007b}
{Driver} S.~P.,  {Allen} P.~D.,  {Liske} J.,   {Graham} A.~W.,  2007b, \mn@doi
  [\apjl] {10.1086/513106}, \href
  {http://adsabs.harvard.edu/abs/2007ApJ...657L..85D} {657, L85}

\bibitem[\protect\citeauthoryear{{Driver}, {Popescu}, {Tuffs}, {Graham},
  {Liske}  \& {Baldry}}{{Driver} et~al.}{2008}]{Driver2008}
{Driver} S.~P.,  {Popescu} C.~C.,  {Tuffs} R.~J.,  {Graham} A.~W.,  {Liske} J.,
    {Baldry} I.,  2008, \mn@doi [\apjl] {10.1086/588582}, \href
  {http://adsabs.harvard.edu/abs/2008ApJ...678L.101D} {678, L101}

\bibitem[\protect\citeauthoryear{{Driver} et~al.,}{{Driver}
  et~al.}{2009}]{Driver2009}
{Driver} S.~P.,  et~al., 2009, \mn@doi [Astronomy and Geophysics]
  {10.1111/j.1468-4004.2009.50512.x}, \href
  {http://adsabs.harvard.edu/abs/2009A%26G....50e..12D} {50, 5.12}

\bibitem[\protect\citeauthoryear{{Driver} et~al.,}{{Driver}
  et~al.}{2011}]{Driver2011}
{Driver} S.~P.,  et~al., 2011, \mn@doi [\mnras]
  {10.1111/j.1365-2966.2010.18188.x}, \href
  {http://adsabs.harvard.edu/abs/2011MNRAS.413..971D} {413, 971}

\bibitem[\protect\citeauthoryear{{Driver}, {Robotham}, {Bland-Hawthorn},
  {Brown}, {Hopkins}, {Liske}, {Phillipps}  \& {Wilkins}}{{Driver}
  et~al.}{2013}]{Driver2013}
{Driver} S.~P.,  {Robotham} A.~S.~G.,  {Bland-Hawthorn} J.,  {Brown} M.,
  {Hopkins} A.,  {Liske} J.,  {Phillipps} S.,   {Wilkins} S.,  2013, \mn@doi
  [\mnras] {10.1093/mnras/sts717}, \href
  {https://ui.adsabs.harvard.edu/abs/2013MNRAS.430.2622D} {430, 2622}

\bibitem[\protect\citeauthoryear{{Driver} et~al.,}{{Driver}
  et~al.}{2016}]{Driver2016}
{Driver} S.~P.,  et~al., 2016, \mn@doi [\mnras] {10.1093/mnras/stv2505}, \href
  {http://adsabs.harvard.edu/abs/2016MNRAS.455.3911D} {455, 3911}

\bibitem[\protect\citeauthoryear{{Driver} et~al.,}{{Driver}
  et~al.}{2022}]{Driver2022}
{Driver} S.~P.,  et~al., 2022, \mn@doi [\mnras] {10.1093/mnras/stac472}, \href
  {https://ui.adsabs.harvard.edu/abs/2022MNRAS.tmp..552D} {}

\bibitem[\protect\citeauthoryear{{Dutton} et~al.,}{{Dutton}
  et~al.}{2011}]{Dutton2011}
{Dutton} A.~A.,  et~al., 2011, \mn@doi [\mnras]
  {10.1111/j.1365-2966.2010.17555.x}, \href
  {https://ui.adsabs.harvard.edu/abs/2011MNRAS.410.1660D} {410, 1660}

\bibitem[\protect\citeauthoryear{{Edge}, {Sutherland}, {Kuijken}, {Driver},
  {McMahon}, {Eales}  \& {Emerson}}{{Edge} et~al.}{2013}]{Edge2013}
{Edge} A.,  {Sutherland} W.,  {Kuijken} K.,  {Driver} S.,  {McMahon} R.,
  {Eales} S.,   {Emerson} J.~P.,  2013, The Messenger, \href
  {https://ui.adsabs.harvard.edu/abs/2013Msngr.154...32E} {154, 32}

\bibitem[\protect\citeauthoryear{{Erwin}}{{Erwin}}{2015}]{Erwin2015}
{Erwin} P.,  2015, \mn@doi [\apj] {10.1088/0004-637X/799/2/226}, \href
  {https://ui.adsabs.harvard.edu/abs/2015ApJ...799..226E} {799, 226}

\bibitem[\protect\citeauthoryear{{Fern{\'a}ndez Lorenzo}
  et~al.,}{{Fern{\'a}ndez Lorenzo} et~al.}{2014}]{Fernandez-Lorenzo2014}
{Fern{\'a}ndez Lorenzo} M.,  et~al., 2014, \mn@doi [\apjl]
  {10.1088/2041-8205/788/2/L39}, \href
  {https://ui.adsabs.harvard.edu/abs/2014ApJ...788L..39F} {788, L39}

\bibitem[\protect\citeauthoryear{{Fletcher}}{{Fletcher}}{1970}]{Fletcher1970}
{Fletcher} R.,  1970, \mn@doi [Computer Journal] {10.1093/comjnl/13.3.317}, 13,
  317

\bibitem[\protect\citeauthoryear{{Gadotti}}{{Gadotti}}{2009}]{Gadotti2009}
{Gadotti} D.~A.,  2009, \mn@doi [\mnras] {10.1111/j.1365-2966.2008.14257.x},
  \href {https://ui.adsabs.harvard.edu/abs/2009MNRAS.393.1531G} {393, 1531}

\bibitem[\protect\citeauthoryear{{Gao} \& {Ho}}{{Gao} \& {Ho}}{2017}]{Gao2017}
{Gao} H.,  {Ho} L.~C.,  2017, \mn@doi [\apj] {10.3847/1538-4357/aa7da4}, \href
  {https://ui.adsabs.harvard.edu/abs/2017ApJ...845..114G} {845, 114}

\bibitem[\protect\citeauthoryear{{Goldfarb}}{{Goldfarb}}{1970}]{Goldfarb1970}
{Goldfarb} D.,  1970, \mn@doi [Mathematics of Computation]
  {10.1090/S0025-5718-1970-0258249-6}, 24, 23

\bibitem[\protect\citeauthoryear{{\noopsort{Graaff}}{de Graaff}, {Trayford},
  {Franx}, {Schaller}, {Schaye}  \& {van der Wel}}{{\noopsort{Graaff}}{de
  Graaff} et~al.}{2022}]{deGraaff2022}
{\noopsort{Graaff}}{de Graaff} A.,  {Trayford} J.,  {Franx} M.,  {Schaller} M.,
   {Schaye} J.,   {van der Wel} A.,  2022, \mn@doi [\mnras]
  {10.1093/mnras/stab3510}, \href
  {https://ui.adsabs.harvard.edu/abs/2022MNRAS.511.2544D} {511, 2544}

\bibitem[\protect\citeauthoryear{{Graham}}{{Graham}}{2013}]{Graham2013b}
{Graham} A.~W.,  2013, {Elliptical and Disk Galaxy Structure and Modern Scaling
  Laws}.
pp 91--140, \mn@doi{10.1007/978-94-007-5609-0\_2}

\bibitem[\protect\citeauthoryear{{Graham} \& {Driver}}{{Graham} \&
  {Driver}}{2005}]{Graham2005}
{Graham} A.~W.,  {Driver} S.~P.,  2005, \mn@doi [\pasa] {10.1071/AS05001},
  \href {https://ui.adsabs.harvard.edu/abs/2005PASA...22..118G} {22, 118}

\bibitem[\protect\citeauthoryear{{H{\"a}ussler} et~al.,}{{H{\"a}ussler}
  et~al.}{2007}]{Haeussler2007}
{H{\"a}ussler} B.,  et~al., 2007, \mn@doi [\apjs] {10.1086/518836}, \href
  {https://ui.adsabs.harvard.edu/abs/2007ApJS..172..615H} {172, 615}

\bibitem[\protect\citeauthoryear{{H{\"a}u{\ss}ler} et~al.,}{{H{\"a}u{\ss}ler}
  et~al.}{2013}]{Haeussler2013}
{H{\"a}u{\ss}ler} B.,  et~al., 2013, \mn@doi [\mnras] {10.1093/mnras/sts633},
  \href {https://ui.adsabs.harvard.edu/abs/2013MNRAS.430..330H} {430, 330}

\bibitem[\protect\citeauthoryear{{H{\"a}u{\ss}ler} et~al.,}{{H{\"a}u{\ss}ler}
  et~al.}{2022}]{Haeussler2022}
{H{\"a}u{\ss}ler} B.,  et~al., 2022, arXiv e-prints, \href
  {https://ui.adsabs.harvard.edu/abs/2022arXiv220405907H} {p. arXiv:2204.05907}

\bibitem[\protect\citeauthoryear{{Head}, {Lucey}, {Hudson}  \& {Smith}}{{Head}
  et~al.}{2014}]{Head2014}
{Head} J. T.~C.~G.,  {Lucey} J.~R.,  {Hudson} M.~J.,   {Smith} R.~J.,  2014,
  \mn@doi [\mnras] {10.1093/mnras/stu325}, \href
  {https://ui.adsabs.harvard.edu/abs/2014MNRAS.440.1690H} {440, 1690}

\bibitem[\protect\citeauthoryear{{Hilbe}, {de Souza}  \& {Ishida}}{{Hilbe}
  et~al.}{2017}]{Hilbe2017}
{Hilbe} J.~M.,  {de Souza} R.~S.,   {Ishida} E.~E.~O.,  2017, Bayesian Models
  for Astrophysical Data: Using R, JAGS, Python, and Stan.
Cambridge University Press, Cambridge

\bibitem[\protect\citeauthoryear{{Hopkins} et~al.,}{{Hopkins}
  et~al.}{2013}]{Hopkins2013}
{Hopkins} A.~M.,  et~al., 2013, \mn@doi [\mnras] {10.1093/mnras/stt030}, \href
  {http://adsabs.harvard.edu/abs/2013MNRAS.430.2047H} {430, 2047}

\bibitem[\protect\citeauthoryear{{Hyde} \& {Bernardi}}{{Hyde} \&
  {Bernardi}}{2009}]{Hyde2009}
{Hyde} J.~B.,  {Bernardi} M.,  2009, \mn@doi [\mnras]
  {10.1111/j.1365-2966.2009.14445.x}, \href
  {https://ui.adsabs.harvard.edu/abs/2009MNRAS.394.1978H} {394, 1978}

\bibitem[\protect\citeauthoryear{{Ivezi{\'c}} et~al.,}{{Ivezi{\'c}}
  et~al.}{2019}]{Ivezic2019}
{Ivezi{\'c}} {\v{Z}}.,  et~al., 2019, \mn@doi [\apj]
  {10.3847/1538-4357/ab042c}, \href
  {https://ui.adsabs.harvard.edu/abs/2019ApJ...873..111I} {873, 111}

\bibitem[\protect\citeauthoryear{{Jeffreys}}{{Jeffreys}}{1946}]{Jeffreys1946}
{Jeffreys} H.,  1946, \mn@doi [Proceedings of the Royal Society of London.
  Series A, Mathematical and Physical Sciences] {10.1098/rspa.1946.0056}, 186,
  453

\bibitem[\protect\citeauthoryear{Johnson}{Johnson}{2017}]{nloptr}
Johnson S.~G.,  2017, The NLopt nonlinear-optimization package.
\url {http://ab-initio.mit.edu/nlopt}

\bibitem[\protect\citeauthoryear{{\noopsort{Jong}}{de Jong}, {Verdoes Kleijn},
  {Kuijken}  \& {Valentijn}}{{\noopsort{Jong}}{de Jong}
  et~al.}{2013}]{deJong2013}
{\noopsort{Jong}}{de Jong} J. T.~A.,  {Verdoes Kleijn} G.~A.,  {Kuijken} K.~H.,
    {Valentijn} E.~A.,  2013, \mn@doi [Experimental Astronomy]
  {10.1007/s10686-012-9306-1}, \href
  {https://ui.adsabs.harvard.edu/abs/2013ExA....35...25D} {35, 25}

\bibitem[\protect\citeauthoryear{{\noopsort{Jong}}{de Jong}
  et~al.,}{{\noopsort{Jong}}{de Jong} et~al.}{2015}]{deJong2015}
{\noopsort{Jong}}{de Jong} J.~T.~A.,  et~al., 2015, \mn@doi [\aap]
  {10.1051/0004-6361/201526601}, \href
  {http://adsabs.harvard.edu/abs/2015A%26A...582A..62D} {582, A62}

\bibitem[\protect\citeauthoryear{{Kawinwanichakij} et~al.,}{{Kawinwanichakij}
  et~al.}{2021}]{Kawinwanichakij2021}
{Kawinwanichakij} L.,  et~al., 2021, arXiv e-prints, \href
  {https://ui.adsabs.harvard.edu/abs/2021arXiv210909766K} {p. arXiv:2109.09766}

\bibitem[\protect\citeauthoryear{{Kelvin} et~al.,}{{Kelvin}
  et~al.}{2012}]{Kelvin2012}
{Kelvin} L.~S.,  et~al., 2012, \mn@doi [\mnras]
  {10.1111/j.1365-2966.2012.20355.x}, \href
  {http://adsabs.harvard.edu/abs/2012MNRAS.421.1007K} {421, 1007}

\bibitem[\protect\citeauthoryear{{Kelvin} et~al.,}{{Kelvin}
  et~al.}{2014}]{Kelvin2014}
{Kelvin} L.~S.,  et~al., 2014, \mn@doi [\mnras] {10.1093/mnras/stu1507}, \href
  {https://ui.adsabs.harvard.edu/abs/2014MNRAS.444.1647K} {444, 1647}

\bibitem[\protect\citeauthoryear{{Kennedy} et~al.,}{{Kennedy}
  et~al.}{2016}]{Kennedy2016}
{Kennedy} R.,  et~al., 2016, \mn@doi [\mnras] {10.1093/mnras/stw1176}, \href
  {https://ui.adsabs.harvard.edu/abs/2016MNRAS.460.3458K} {460, 3458}

\bibitem[\protect\citeauthoryear{{Kim}, {Oh}, {Jeong}, {Arag{\'o}n-Salamanca},
  {Smith}  \& {Yi}}{{Kim} et~al.}{2016}]{Kim2016}
{Kim} K.,  {Oh} S.,  {Jeong} H.,  {Arag{\'o}n-Salamanca} A.,  {Smith} R.,
  {Yi} S.~K.,  2016, \mn@doi [\apjs] {10.3847/0067-0049/225/1/6}, \href
  {https://ui.adsabs.harvard.edu/abs/2016ApJS..225....6K} {225, 6}

\bibitem[\protect\citeauthoryear{{Kormendy}}{{Kormendy}}{1977}]{Kormendy1977}
{Kormendy} J.,  1977, \mn@doi [\apj] {10.1086/155687}, \href
  {https://ui.adsabs.harvard.edu/abs/1977ApJ...218..333K} {218, 333}

\bibitem[\protect\citeauthoryear{{\noopsort{Kruit}}{van der Kruit} \&
  {Searle}}{{\noopsort{Kruit}}{van der Kruit} \&
  {Searle}}{1981}]{vanderKruit1981}
{\noopsort{Kruit}}{van der Kruit} P.~C.,  {Searle} L.,  1981, \aap, \href
  {https://ui.adsabs.harvard.edu/abs/1981A&A....95..105V} {95, 105}

\bibitem[\protect\citeauthoryear{{Kruk} et~al.,}{{Kruk}
  et~al.}{2018}]{Kruk2018}
{Kruk} S.~J.,  et~al., 2018, \mn@doi [\mnras] {10.1093/mnras/stx2605}, \href
  {https://ui.adsabs.harvard.edu/abs/2018MNRAS.473.4731K} {473, 4731}

\bibitem[\protect\citeauthoryear{{Kuijken} et~al.,}{{Kuijken}
  et~al.}{2019}]{Kuijken2019}
{Kuijken} K.,  et~al., 2019, \mn@doi [\aap] {10.1051/0004-6361/201834918},
  \href {https://ui.adsabs.harvard.edu/abs/2019A&A...625A...2K} {625, A2}

\bibitem[\protect\citeauthoryear{{La Barbera}, {de Carvalho}, {de La Rosa},
  {Lopes}, {Kohl-Moreira}  \& {Capelato}}{{La Barbera}
  et~al.}{2010}]{LaBarbera2010}
{La Barbera} F.,  {de Carvalho} R.~R.,  {de La Rosa} I.~G.,  {Lopes} P.~A.~A.,
  {Kohl-Moreira} J.~L.,   {Capelato} H.~V.,  2010, \mn@doi [\mnras]
  {10.1111/j.1365-2966.2010.16850.x}, \href
  {https://ui.adsabs.harvard.edu/abs/2010MNRAS.408.1313L} {408, 1313}

\bibitem[\protect\citeauthoryear{{Lackner} \& {Gunn}}{{Lackner} \&
  {Gunn}}{2012}]{Lackner2012}
{Lackner} C.~N.,  {Gunn} J.~E.,  2012, \mn@doi [\mnras]
  {10.1111/j.1365-2966.2012.20450.x}, \href
  {https://ui.adsabs.harvard.edu/abs/2012MNRAS.421.2277L} {421, 2277}

\bibitem[\protect\citeauthoryear{{Lagos} et~al.,}{{Lagos}
  et~al.}{2018}]{Lagos2018}
{Lagos} C. d.~P.,  et~al., 2018, \mn@doi [\mnras] {10.1093/mnras/stx2667},
  \href {https://ui.adsabs.harvard.edu/abs/2018MNRAS.473.4956L} {473, 4956}

\bibitem[\protect\citeauthoryear{{Lange} et~al.,}{{Lange}
  et~al.}{2015}]{Lange2015}
{Lange} R.,  et~al., 2015, \mn@doi [\mnras] {10.1093/mnras/stu2467}, \href
  {http://adsabs.harvard.edu/abs/2015MNRAS.447.2603L} {447, 2603}

\bibitem[\protect\citeauthoryear{{Lange} et~al.,}{{Lange}
  et~al.}{2016}]{Lange2016}
{Lange} R.,  et~al., 2016, \mn@doi [\mnras] {10.1093/mnras/stw1495}, \href
  {http://adsabs.harvard.edu/abs/2016MNRAS.462.1470L} {462, 1470}

\bibitem[\protect\citeauthoryear{{Lawrence} et~al.,}{{Lawrence}
  et~al.}{2007}]{Lawrence2007}
{Lawrence} A.,  et~al., 2007, \mn@doi [\mnras]
  {10.1111/j.1365-2966.2007.12040.x}, \href
  {https://ui.adsabs.harvard.edu/abs/2007MNRAS.379.1599L} {379, 1599}

\bibitem[\protect\citeauthoryear{{Liske} et~al.,}{{Liske}
  et~al.}{2015}]{Liske2015}
{Liske} J.,  et~al., 2015, \mn@doi [\mnras] {10.1093/mnras/stv1436}, \href
  {http://adsabs.harvard.edu/abs/2015MNRAS.452.2087L} {452, 2087}

\bibitem[\protect\citeauthoryear{{Meert}, {Vikram}  \& {Bernardi}}{{Meert}
  et~al.}{2015}]{Meert2015}
{Meert} A.,  {Vikram} V.,   {Bernardi} M.,  2015, \mn@doi [\mnras]
  {10.1093/mnras/stu2333}, \href
  {https://ui.adsabs.harvard.edu/abs/2015MNRAS.446.3943M} {446, 3943}

\bibitem[\protect\citeauthoryear{{Meert}, {Vikram}  \& {Bernardi}}{{Meert}
  et~al.}{2016}]{Meert2016}
{Meert} A.,  {Vikram} V.,   {Bernardi} M.,  2016, \mn@doi [\mnras]
  {10.1093/mnras/stv2475}, \href
  {https://ui.adsabs.harvard.edu/abs/2016MNRAS.455.2440M} {455, 2440}

\bibitem[\protect\citeauthoryear{{Mendel}, {Simard}, {Palmer}, {Ellison}  \&
  {Patton}}{{Mendel} et~al.}{2014}]{Mendel2014}
{Mendel} J.~T.,  {Simard} L.,  {Palmer} M.,  {Ellison} S.~L.,   {Patton} D.~R.,
   2014, \mn@doi [\apjs] {10.1088/0067-0049/210/1/3}, \href
  {http://adsabs.harvard.edu/abs/2014ApJS..210....3M} {210, 3}

\bibitem[\protect\citeauthoryear{{M{\'e}ndez-Abreu} et~al.,}{{M{\'e}ndez-Abreu}
  et~al.}{2017}]{Mendez-Abreu2017}
{M{\'e}ndez-Abreu} J.,  et~al., 2017, \mn@doi [\aap]
  {10.1051/0004-6361/201629525}, \href
  {https://ui.adsabs.harvard.edu/abs/2017A&A...598A..32M} {598, A32}

\bibitem[\protect\citeauthoryear{{Moffett} et~al.,}{{Moffett}
  et~al.}{2016}]{Moffett2016}
{Moffett} A.~J.,  et~al., 2016, \mn@doi [\mnras] {10.1093/mnras/stw1861}, \href
  {https://ui.adsabs.harvard.edu/abs/2016MNRAS.462.4336M} {462, 4336}

\bibitem[\protect\citeauthoryear{{Nedkova} et~al.,}{{Nedkova}
  et~al.}{2021}]{Nedkova2021}
{Nedkova} K.~V.,  et~al., 2021, \mn@doi [\mnras] {10.1093/mnras/stab1744},
  \href {https://ui.adsabs.harvard.edu/abs/2021MNRAS.506..928N} {506, 928}

\bibitem[\protect\citeauthoryear{{Oh} et~al.,}{{Oh} et~al.}{2020}]{Oh2020}
{Oh} S.,  et~al., 2020, \mn@doi [\mnras] {10.1093/mnras/staa1330}, \href
  {https://ui.adsabs.harvard.edu/abs/2020MNRAS.495.4638O} {495, 4638}

\bibitem[\protect\citeauthoryear{{Peng}, {Ho}, {Impey}  \& {Rix}}{{Peng}
  et~al.}{2010}]{Peng2010}
{Peng} C.~Y.,  {Ho} L.~C.,  {Impey} C.~D.,   {Rix} H.-W.,  2010, \mn@doi [\aj]
  {10.1088/0004-6256/139/6/2097}, \href
  {https://ui.adsabs.harvard.edu/abs/2010AJ....139.2097P} {139, 2097}

\bibitem[\protect\citeauthoryear{{Pillepich} et~al.,}{{Pillepich}
  et~al.}{2018}]{Pillepich2018}
{Pillepich} A.,  et~al., 2018, \mn@doi [\mnras] {10.1093/mnras/stx2656}, \href
  {https://ui.adsabs.harvard.edu/abs/2018MNRAS.473.4077P} {473, 4077}

\bibitem[\protect\citeauthoryear{{Popescu}, {Tuffs}, {Dopita}, {Fischera},
  {Kylafis}  \& {Madore}}{{Popescu} et~al.}{2011}]{Popescu2011}
{Popescu} C.~C.,  {Tuffs} R.~J.,  {Dopita} M.~A.,  {Fischera} J.,  {Kylafis}
  N.~D.,   {Madore} B.~F.,  2011, \mn@doi [\aap] {10.1051/0004-6361/201015217},
  \href {http://adsabs.harvard.edu/abs/2011A%26A...527A.109P} {527, A109}

\bibitem[\protect\citeauthoryear{{R Core Team}}{{R Core Team}}{2020}]{Rv3.6}
{R Core Team} 2020, R: A Language and Environment for Statistical Computing.
R Foundation for Statistical Computing, Vienna, Austria, \url
  {https://www.R-project.org/}

\bibitem[\protect\citeauthoryear{{Robotham} et~al.,}{{Robotham}
  et~al.}{2010}]{Robotham2010}
{Robotham} A.,  et~al., 2010, \mn@doi [\pasa] {10.1071/AS09053}, \href
  {http://adsabs.harvard.edu/abs/2010PASA...27...76R} {27, 76}

\bibitem[\protect\citeauthoryear{{Robotham}, {Taranu}, {Tobar}, {Moffett}  \&
  {Driver}}{{Robotham} et~al.}{2017}]{Robotham2017}
{Robotham} A.~S.~G.,  {Taranu} D.~S.,  {Tobar} R.,  {Moffett} A.,   {Driver}
  S.~P.,  2017, \mn@doi [\mnras] {10.1093/mnras/stw3039}, \href
  {http://adsabs.harvard.edu/abs/2017MNRAS.466.1513R} {466, 1513}

\bibitem[\protect\citeauthoryear{{Robotham}, {Davies}, {Driver}, {Koushan},
  {Taranu}, {Casura}  \& {Liske}}{{Robotham} et~al.}{2018}]{Robotham2018}
{Robotham} A.~S.~G.,  {Davies} L.~J.~M.,  {Driver} S.~P.,  {Koushan} S.,
  {Taranu} D.~S.,  {Casura} S.,   {Liske} J.,  2018, \mn@doi [\mnras]
  {10.1093/mnras/sty440}, \href
  {http://adsabs.harvard.edu/abs/2018MNRAS.476.3137R} {476, 3137}

\bibitem[\protect\citeauthoryear{{Robotham}, {Bellstedt}  \&
  {Driver}}{{Robotham} et~al.}{2022}]{Robotham2022}
{Robotham} A.~S.~G.,  {Bellstedt} S.,   {Driver} S.~P.,  2022, \mn@doi [\mnras]
  {10.1093/mnras/stac1032}, \href
  {https://ui.adsabs.harvard.edu/abs/2022MNRAS.tmp.1018R} {}

\bibitem[\protect\citeauthoryear{{Rodriguez-Gomez} et~al.,}{{Rodriguez-Gomez}
  et~al.}{2019}]{Rodriguez-Gomez2019}
{Rodriguez-Gomez} V.,  et~al., 2019, \mn@doi [\mnras] {10.1093/mnras/sty3345},
  \href {https://ui.adsabs.harvard.edu/abs/2019MNRAS.483.4140R} {483, 4140}

\bibitem[\protect\citeauthoryear{{Salo} et~al.,}{{Salo}
  et~al.}{2015}]{Salo2015}
{Salo} H.,  et~al., 2015, \mn@doi [\apjs] {10.1088/0067-0049/219/1/4}, \href
  {https://ui.adsabs.harvard.edu/abs/2015ApJS..219....4S} {219, 4}

\bibitem[\protect\citeauthoryear{{S{\'a}nchez-Janssen}
  et~al.,}{{S{\'a}nchez-Janssen} et~al.}{2016}]{Sanchez-Janssen2016}
{S{\'a}nchez-Janssen} R.,  et~al., 2016, \mn@doi [\apj]
  {10.3847/0004-637X/820/1/69}, \href
  {https://ui.adsabs.harvard.edu/abs/2016ApJ...820...69S} {820, 69}

\bibitem[\protect\citeauthoryear{{Schaye} et~al.,}{{Schaye}
  et~al.}{2015}]{Schaye2015}
{Schaye} J.,  et~al., 2015, \mn@doi [\mnras] {10.1093/mnras/stu2058}, \href
  {https://ui.adsabs.harvard.edu/abs/2015MNRAS.446..521S} {446, 521}

\bibitem[\protect\citeauthoryear{{Schwarz}}{{Schwarz}}{1978}]{Schwarz1978}
{Schwarz} G.,  1978, Annals of Statistics, \href
  {https://ui.adsabs.harvard.edu/abs/1978AnSta...6..461S} {6, 461}

\bibitem[\protect\citeauthoryear{{S{\'e}rsic}}{{S{\'e}rsic}}{1963}]{Sersic1963}
{S{\'e}rsic} J.~L.,  1963, Boletin de la Asociacion Argentina de Astronomia La
  Plata Argentina, \href
  {https://ui.adsabs.harvard.edu/abs/1963BAAA....6...41S} {6, 41}

\bibitem[\protect\citeauthoryear{{Shanno}}{{Shanno}}{1970}]{Shanno1970}
{Shanno} D.~F.,  1970, \mn@doi [Mathematics of Computation]
  {10.1090/S0025-5718-1970-0274029-X}, 24, 647

\bibitem[\protect\citeauthoryear{{Shen}, {Mo}, {White}, {Blanton}, {Kauffmann},
  {Voges}, {Brinkmann}  \& {Csabai}}{{Shen} et~al.}{2003}]{Shen2003}
{Shen} S.,  {Mo} H.~J.,  {White} S. D.~M.,  {Blanton} M.~R.,  {Kauffmann} G.,
  {Voges} W.,  {Brinkmann} J.,   {Csabai} I.,  2003, \mn@doi [\mnras]
  {10.1046/j.1365-8711.2003.06740.x}, \href
  {https://ui.adsabs.harvard.edu/abs/2003MNRAS.343..978S} {343, 978}

\bibitem[\protect\citeauthoryear{{Shibuya}, {Ouchi}  \& {Harikane}}{{Shibuya}
  et~al.}{2015}]{Shibuya2015}
{Shibuya} T.,  {Ouchi} M.,   {Harikane} Y.,  2015, \mn@doi [\apjs]
  {10.1088/0067-0049/219/2/15}, \href
  {https://ui.adsabs.harvard.edu/abs/2015ApJS..219...15S} {219, 15}

\bibitem[\protect\citeauthoryear{{Simard} et~al.,}{{Simard}
  et~al.}{2002}]{Simard2002}
{Simard} L.,  et~al., 2002, \mn@doi [\apjs] {10.1086/341399}, \href
  {https://ui.adsabs.harvard.edu/abs/2002ApJS..142....1S} {142, 1}

\bibitem[\protect\citeauthoryear{{Simard}, {Mendel}, {Patton}, {Ellison}  \&
  {McConnachie}}{{Simard} et~al.}{2011}]{Simard2011}
{Simard} L.,  {Mendel} J.~T.,  {Patton} D.~R.,  {Ellison} S.~L.,
  {McConnachie} A.~W.,  2011, \mn@doi [\apjs] {10.1088/0067-0049/196/1/11},
  \href {http://adsabs.harvard.edu/abs/2011ApJS..196...11S} {196, 11}

\bibitem[\protect\citeauthoryear{{Sivia} \& {Skilling}}{{Sivia} \&
  {Skilling}}{2006}]{Sivia2006}
{Sivia} D.~S.,  {Skilling} J.,  2006, Data Analysis: A Bayesian Tutorial, 2nd
  edn.
Oxford Science Publications, Oxford University Press

\bibitem[\protect\citeauthoryear{{\noopsort{Souza}}{de Souza}, {Gadotti}  \&
  {dos Anjos}}{{\noopsort{Souza}}{de Souza} et~al.}{2004}]{deSouza2004}
{\noopsort{Souza}}{de Souza} R.~E.,  {Gadotti} D.~A.,   {dos Anjos} S.,  2004,
  \mn@doi [\apjs] {10.1086/421554}, \href
  {https://ui.adsabs.harvard.edu/abs/2004ApJS..153..411D} {153, 411}

\bibitem[\protect\citeauthoryear{Spiegelhalter, Best, Carlin  \& {Van Der
  Linde}}{Spiegelhalter et~al.}{2002}]{Spiegelhalter2002}
Spiegelhalter D.,  Best N.,  Carlin B.,   {Van Der Linde} A.,  2002, \mn@doi
  [Journal of the Royal Statistical Society. Series B: Statistical Methodology]
  {10.1111/1467-9868.00353}, 64, 583

\bibitem[\protect\citeauthoryear{{Statisticat} \& {LLC.}}{{Statisticat} \&
  {LLC.}}{2018}]{LaplacesDemon}
{Statisticat} {LLC.} 2018, LaplacesDemon: Complete Environment for Bayesian
  Inference.
\url
  {https://web.archive.org/web/20150206004624/http://www.bayesian-inference.com/software}

\bibitem[\protect\citeauthoryear{{Taranu}}{{Taranu}}{2022}]{AllStarFit}
{Taranu} D.~S.,  2022, {AllStarFit: R package for source detection, PSF and
  multi-component galaxy fitting}, Astrophysics Source Code Library (\mn@eprint
  {ascl} {2201.005})

\bibitem[\protect\citeauthoryear{{Tasca} et~al.,}{{Tasca}
  et~al.}{2014}]{Tasca2014}
{Tasca} L.~A.~M.,  et~al., 2014, \mn@doi [\aap] {10.1051/0004-6361/201423699},
  \href {https://ui.adsabs.harvard.edu/abs/2014A&A...564L..12T} {564, L12}

\bibitem[\protect\citeauthoryear{{Taylor} et~al.,}{{Taylor}
  et~al.}{2011}]{Taylor2011}
{Taylor} E.~N.,  et~al., 2011, \mn@doi [\mnras]
  {10.1111/j.1365-2966.2011.19536.x}, \href
  {http://adsabs.harvard.edu/abs/2011MNRAS.418.1587T} {418, 1587}

\bibitem[\protect\citeauthoryear{{Trujillo} et~al.,}{{Trujillo}
  et~al.}{2006}]{Trujillo2006}
{Trujillo} I.,  et~al., 2006, \mn@doi [\apj] {10.1086/506464}, \href
  {https://ui.adsabs.harvard.edu/abs/2006ApJ...650...18T} {650, 18}

\bibitem[\protect\citeauthoryear{{\noopsort{Vaucouleurs}}{de
  Vaucouleurs}}{{\noopsort{Vaucouleurs}}{de
  Vaucouleurs}}{1948}]{deVaucouleurs1948}
{\noopsort{Vaucouleurs}}{de Vaucouleurs} G.,  1948, Annales d'Astrophysique,
  \href {https://ui.adsabs.harvard.edu/abs/1948AnAp...11..247D} {11, 247}

\bibitem[\protect\citeauthoryear{{Vika}, {Bamford}, {H{\"a}u{\ss}ler}, {Rojas},
  {Borch}  \& {Nichol}}{{Vika} et~al.}{2013}]{Vika2013}
{Vika} M.,  {Bamford} S.~P.,  {H{\"a}u{\ss}ler} B.,  {Rojas} A.~L.,  {Borch}
  A.,   {Nichol} R.~C.,  2013, \mn@doi [\mnras] {10.1093/mnras/stt1320}, \href
  {https://ui.adsabs.harvard.edu/abs/2013MNRAS.435..623V} {435, 623}

\bibitem[\protect\citeauthoryear{Vika, Bamford, Häußler  \& Rojas}{Vika
  et~al.}{2014}]{Vika2014}
Vika M.,  Bamford S.~P.,  Häußler B.,   Rojas A.~L.,  2014, \mn@doi [Monthly
  Notices of the Royal Astronomical Society] {10.1093/mnras/stu1696}, 444, 3603

\bibitem[\protect\citeauthoryear{{Vogelsberger} et~al.,}{{Vogelsberger}
  et~al.}{2014}]{Vogelsberger2014}
{Vogelsberger} M.,  et~al., 2014, \mn@doi [\mnras] {10.1093/mnras/stu1536},
  \href {https://ui.adsabs.harvard.edu/abs/2014MNRAS.444.1518V} {444, 1518}

\bibitem[\protect\citeauthoryear{{\noopsort{Wel}}{van der Wel}
  et~al.,}{{\noopsort{Wel}}{van der Wel} et~al.}{2012}]{vanderWel2012}
{\noopsort{Wel}}{van der Wel} A.,  et~al., 2012, \mn@doi [\apjs]
  {10.1088/0067-0049/203/2/24}, \href
  {https://ui.adsabs.harvard.edu/abs/2012ApJS..203...24V} {203, 24}

\bibitem[\protect\citeauthoryear{{\noopsort{Wel}}{van der Wel}
  et~al.,}{{\noopsort{Wel}}{van der Wel} et~al.}{2014}]{vanderWel2014}
{\noopsort{Wel}}{van der Wel} A.,  et~al., 2014, \mn@doi [\apj]
  {10.1088/0004-637X/788/1/28}, \href
  {https://ui.adsabs.harvard.edu/abs/2014ApJ...788...28V} {788, 28}

\bibitem[\protect\citeauthoryear{{York} et~al.,}{{York}
  et~al.}{2000}]{York2000}
{York} D.~G.,  et~al., 2000, \mn@doi [\aj] {10.1086/301513}, \href
  {https://ui.adsabs.harvard.edu/abs/2000AJ....120.1579Y} {120, 1579}

\makeatother
\end{thebibliography}




\appendix
\section{PSF estimation details}
\label{app:psfdetails}
The selection of star candidates for PSF estimation proceeds in three steps.

From the segmentation statistics returned by \texttt{profoundProFound}, we select relatively round and isolated objects as follows: 
\begin{itemize}
\item objects that do not touch other segments, masked regions or image edges (edge fraction~=~1)
\item objects with a regular boundary geometry (edge excess $<$ 1) 
\item objects with an axial ratio (minor/major axis) larger than 0.5
\item objects which were not flagged as possibly spurious
\end{itemize}

Of these relatively round and isolated objects, a given fraction (depending on the depth of the image and source extraction, 4\% for the $gri$ stacks and a \texttt{skycut} value of 2) are identified as star candidates via a joint cut in \texttt{R50} (semi-major axis containing half the flux) and the concentration (\texttt{R50/R90}, where \texttt{R90} is the semi-major axis containing 90\% of the total flux). This selection is based on the notion that we would expect stars to be small (i.e. low \texttt{R50}) and highly concentrated (i.e. low \texttt{R50/R90}) and was calibrated empirically.

Around each of these star candidates, a smaller cutout is taken and a subsample selected:
\begin{itemize}
\item objects brighter than the 5-sigma point source detection limit and fainter than the saturation limit (both taken from the headers of the corresponding KiDS tile)
\item objects where less than 10\% of pixels in the cutout belong to other segments
\item objects where the star cutout does not overlap with the edge of the large cutout
\item objects with a positive sum of the cutout (excluding poorly background-subtracted and/or purely noise-dominated objects)
\end{itemize}

In the next step, the star candidate cutouts are normalised to a magnitude of 0, masked appropriately and fitted with a Moffat function using \texttt{ProFit} (see fitting details in Section~\ref{sec:psfestimation}).\\

After fitting, suitable stars for PSF estimation are determined as follows:
\begin{itemize}
\item The fitted centre in x and y must be within $\pm$\,1 pixel of the centre of the cutout (fixed since the cutout was centred on the star)
\item The fitted magnitude must be within $\pm$\,0.1\,mag of 0 (fixed since the cutout was normalised to a magnitude of zero according to the segmentation statistics, so any deviation indicates a difference between the magnitudes estimated by \texttt{ProFound} and \texttt{ProFit}, which likely points to bad segmentation, additional objects in the segment or a bad model fit)
\item The reduced chi-square ($\chi^2_\nu$) of the fit must be smaller than 3 (calibrated by inspection to exclude visually bad fits)
\item FWHM, concentration index, angle and axial ratio must not be equal to the fit limits (except for the axial ratio, which is allowed to be exactly 1 although this is the upper limit of the fit).
\item Outliers in any of FWHM, concentration index, angle, axial ratio or background are rejected via an iterative $2\sigma$ clip (in logarithmic space where appropriate). 
\end{itemize}

The stars fulfilling these criteria are classified as suitable, from which the selection is made:
\begin{itemize} 
\item The closest two from each quadrant (8 in total) are selected to make sure they are roughly evenly distributed around the galaxy. 
\item If one or more quadrant contains less than two stars, the closest stars from any other quadrant (which are not already used) are taken instead to give 8 stars in total. 
\item If there are less than 8 stars in total, all of them are used.
\item If there are no stars classified as suitable, the object is flagged as having a failed PSF and consequently skipped during the galaxy fitting.
\end{itemize}

The stars selected in this way are then used to create the model PSF as described in Section~\ref{sec:psfestimation}.

\section{Model selection accuracy}
\label{app:modelselconfusion}
Tables~\ref{tab:modelselconfusiong}, \ref{tab:modelselconfusioni} and \ref{tab:modelselconfusionjoint} show the confusion matrices for the model selection in the $g$ and $i$ bands and the joint model selection. The confusion matrix for the $r$ band is already shown in Table~\ref{tab:modelselconfusionr}.
\begin{table}
	\centering
	\caption{The confusion matrix for our model selection based on a DIC difference cut compared against visual inspection for the $g$-band. All values are in percent of the total number of visually inspected $g$-band galaxies. Bold font highlights galaxies classified correctly, while grey shows those that were ignored during the calibration.}
	\label{tab:modelselconfusiong}
	\begin{tabu}{lcrrrc} 
		\hline
		 & \multicolumn{5}{r}{number of components} \\
		visual classification & & 1 & 1.5 & 2 &\\
		\hline
		``single" && \textbf{48.9} & 0.4 & 1.0 &\\
		``1.5" && 2.1 & \textbf{3.7} & 0.6 &\\
		``double" && 1.8 & 0.1 & \textbf{6.0} &\\
		``1.5 or double" && 0.4 & \textbf{1.5} & \textbf{1.9} &\\
		\rowfont{\color{gray}}
		``unsure" && 19.5 & 0.9 & 7.3 &\\
		\rowfont{\color{gray}}
		``unfittable" && 1.3 & 0.3 & 2.1 & \\
        \hline
	\end{tabu}
\end{table}

\begin{table}
	\centering
	\caption{The confusion matrix for our model selection based on a DIC difference cut compared against visual inspection for the $i$-band. All values are in percent of the total number of visually inspected $i$-band galaxies. Bold font highlights galaxies classified correctly, while grey shows those that were ignored during the calibration.}
	\label{tab:modelselconfusioni}
	\begin{tabu}{lcrrrc} 
		\hline
		 & \multicolumn{5}{r}{number of components} \\
		visual classification & & 1 & 1.5 & 2 &\\
		\hline
		``single" && \textbf{51.8} & 0.4 & 1.5 &\\
		``1.5" && 1.9 & \textbf{3.0} & 1.5 &\\
		``double" && 0.7 & 0 & \textbf{8.9} &\\
		``1.5 or double" && 0 & \textbf{1.0} & \textbf{1.9} &\\
		\rowfont{\color{gray}}
		``unsure" && 10.0 & 0.7 & 12.4 &\\
		\rowfont{\color{gray}}
		``unfittable" && 1.2 & 0.3 & 2.2 & \\
        \hline
	\end{tabu}
\end{table}

\begin{table}
	\centering
	\caption{The confusion matrix for our model selection based on a DIC difference cut compared against visual inspection for the joint model selection. All values are in percent of the total number of visually inspected $g$, $r$ and $i$ band galaxies. Bold font highlights galaxies classified correctly, while grey shows those that were ignored during the calibration.}
	\label{tab:modelselconfusionjoint}
	\begin{tabu}{lcrrrc} 
		\hline
		 & \multicolumn{5}{r}{number of components} \\
		visual classification & & 1 & 1.5 & 2 &\\
		\hline
		``single" && \textbf{47.4} & 0.5 & 1.5 &\\
		``1.5" && 2.2 & \textbf{2.4} & 1.4 &\\
		``double" && 2.2 & 0.4 & \textbf{7.3} &\\
		``1.5 or double" && 0.3 & \textbf{1.1} & \textbf{2.1} &\\
		\rowfont{\color{gray}}
		``unsure" && 15.0 & 0.7 & 11.1 &\\
		\rowfont{\color{gray}}
		``unfittable" && 1.0 & 0.4 & 2.4 & \\
        \hline
	\end{tabu}
\end{table}

\section{Fit examples}
\label{app:exampleplots}
\begin{figure}
	\includegraphics[width=\columnwidth]{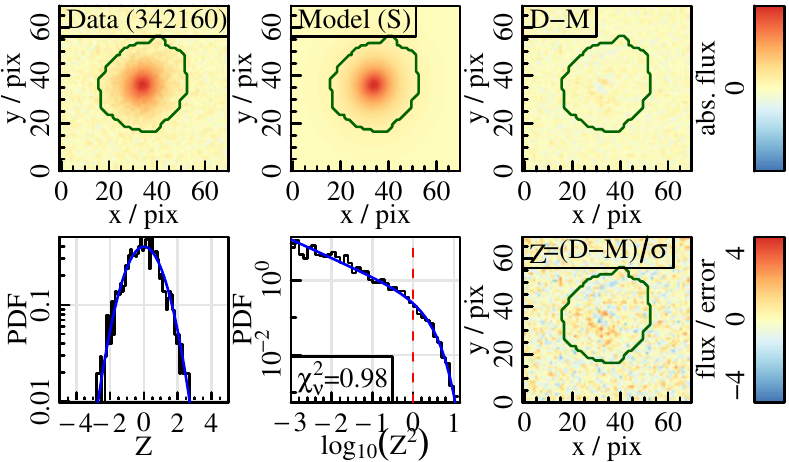}
    \caption{The single S\'ersic fit to galaxy 342160, classified as single component object (\texttt{NCOMP}\,=\,1) in the KiDS $r$-band. Panels are the same as the top two rows in Figure~\ref{fig:examplefit}.} 
    \label{fig:examplefit1}
\end{figure}

\begin{figure}
	\includegraphics[width=\columnwidth]{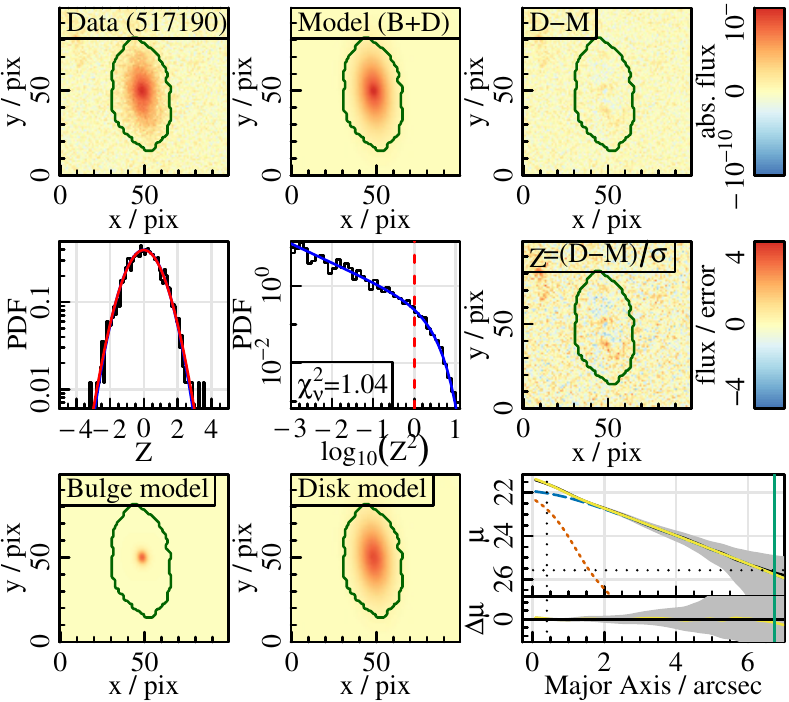}
    \caption{The 1.5-component fit to galaxy 517190, classified as 1.5-component object (\texttt{NCOMP}\,=\,1.5) in the KiDS $r$-band. Panels are the same as those in Figure~\ref{fig:examplefit}.} 
    \label{fig:examplefit1.5}
\end{figure}

\begin{figure}
	\includegraphics[width=\columnwidth]{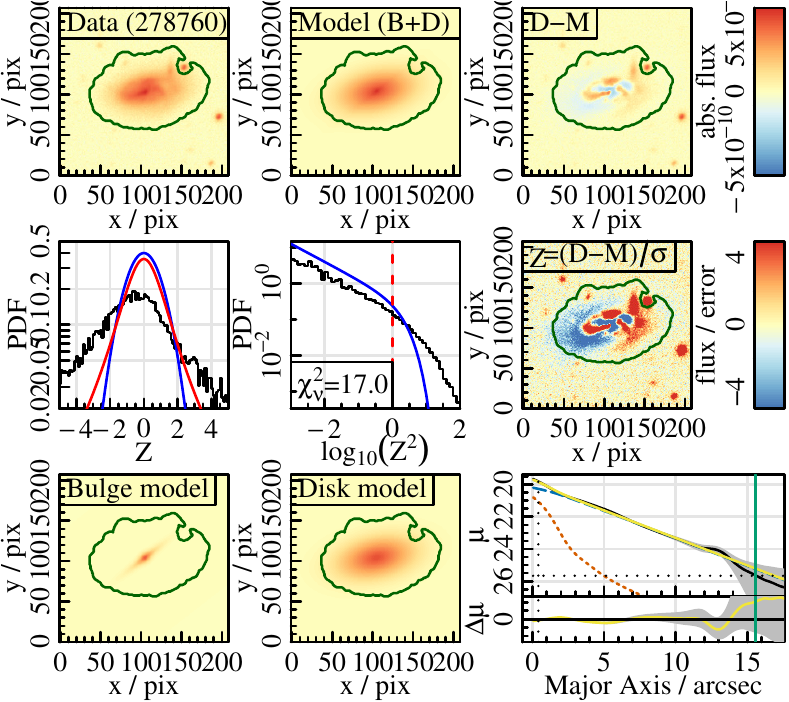}
    \caption{The double component fit to galaxy 278760, classified as outlier (\texttt{NCOMP}\,=\,$-2$) in the KiDS $r$-band. Panels are the same as those in Figure~\ref{fig:examplefit}.} 
    \label{fig:examplefit-2}
\end{figure}

\begin{figure}
	\includegraphics[width=\columnwidth]{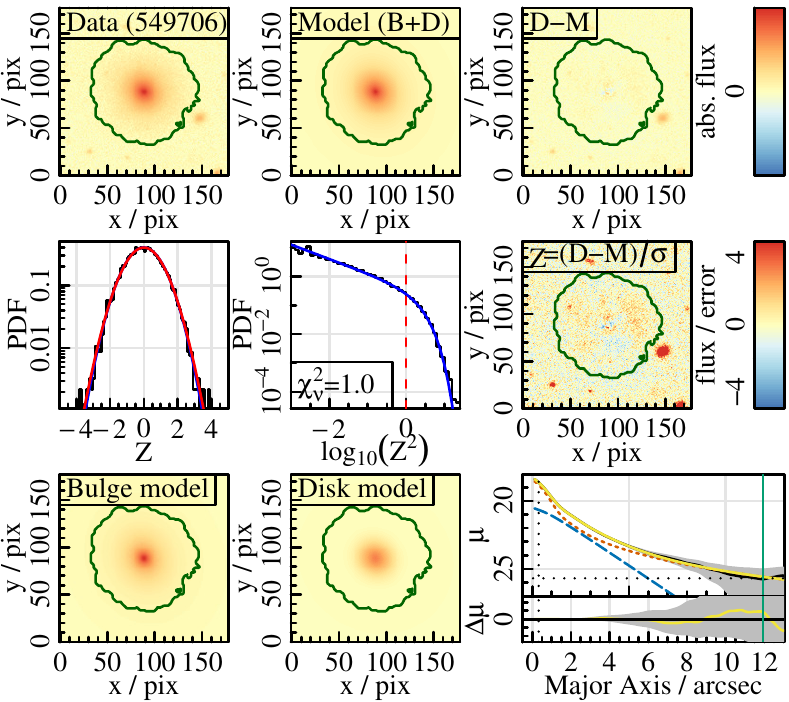}
    \caption{The double component fit to galaxy 549706, classified as double component object but with a very high B/T ratio of 0.71 in the KiDS $r$-band. Panels are the same as those in Figure~\ref{fig:examplefit}.} 
    \label{fig:examplefithighbt}
\end{figure}

Figures~\ref{fig:examplefit1}, \ref{fig:examplefit1.5} and \ref{fig:examplefit-2} show examples of galaxies classified as single S\'ersic, 1.5-component fit and outlier (\texttt{NCOMP}\,=\,1, 1.5 and $-2$) respectively. An example double component fit (\texttt{NCOMP}\,=\,2) is already shown in Figure~\ref{fig:examplefit}. In addition, Figure~\ref{fig:examplefithighbt} shows an example of a very high B/T object where the bulge dominates both the centre and the wings of the object (see Section~\ref{sec:modelselectioncaveats}).  

\bsp	
\label{lastpage}
\end{document}